\documentstyle[12pt,urthesis,epsfig]{report}

\def \lsim
{\raisebox{-3pt}{$\>\stackrel{<}{\scriptstyle\sim}\>$}}
\def \gsim
{\raisebox{-3pt}{$\>\stackrel{>}{\scriptstyle\sim}\>$}}

\def\half{{{1\over 2}}}

\def\as{\alpha_S}
\def\L{\Lambda_{QCD}}
\newcommand{\eq}[1]{(\ref{#1})}
\newcommand{\refsec}[1]{{\rm Section} (\ref{#1})}
\def\tr{{\rm tr}}
\def\MSbar{\overline{\mathrm{MS}}}
\def\lepton{\ell}

\def\MeV{~{\rm MeV}}
\def\GeV{~{\rm GeV}}
\def\TeV{~{\rm TeV}}
\def\pk{(p_t.p_g)}
\def\qk{(p_b.p_g)}
\def\pq{(p_t.p_b)}
\newcommand{\Sp}[1]{{\rm Li}_2{\left( #1 \right)}}
\newcommand{\Ln}[1]{\ln{\left( #1 \right)}}

\begin{document}
\pagenumbering{roman}
\title{Applications of Perturbative Quantum Chromodynamics to Processes with
Heavy Quarks}
\author{Alexander Mitov}
\department{Physics and Astronomy}
\degree{Doctor of Philosophy}
\supervisor{Professor Lynne H. Orr}
\degreeyear{2003}
\maketitle

\pagestyle{headings}

%%%%%%%%%%%%%%%%%%%%%%%%%

\vspace*{4.cm} \hspace*{2.cm}
\parbox{11.3cm}{\Large\it To my wife and true friend Milena Mitova,
in appreciation for all her support.}

\newpage
%%%%%%%%%%%%%%%%%%%%%%%%

\begin{center}
{ \large \bf Curriculum Vitae }
\end{center}

\vspace{1.cm}

The author was born in Sofia, Bulgaria. He attended Sofia
University of St. Kliment Ohridski from 1990 to 1996. He graduated
with a Master of Science degree in Physics in 1996 under the
supervision of Professor~D.~Stoyanov. The author came to the
University of Rochester in the fall of 1999. He pursued his
research in elementary particle physics under the direction of
Professor L.~H.~Orr and received the Master of Arts degree in
2001.

\newpage

{ \large \bf Acknowledgments}

\vspace{1.cm}

First, I would like to thank my advisor Professor Lynne~H.~Orr for
opening the doors of perturbative QCD to me and for her entire
support during my research and study at the University of
Rochester. I would like to thank Dr. Gennaro Corcella for his
collaboration on the three papers that became the backbone of this
thesis and for his consistency and hard work, and for the numerous
important discussions that we have had while working together. I
would also like to thank Dr.~Matteo~Cacciari for his collaboration
on one of the papers that shaped this thesis, for the many useful
discussions and for supplying us with his numerical code for
performing inverse Mellin transformation and fits of hadronization
models to $e^+e^-$ data.

I would also like to thank Dr.~S.~Catani for very useful
discussions on the subject of soft-gluon resummation, as well as
Dr.~A.~Bodek and Dr.~S.~Kretzer for the helpful insights on the
subject of DIS. It is also my pleasure to thank Dr.~A.~Das and
Dr.~D.~Wackeroth for the many discussions on this and related
topics.

\newpage

%%%%%%%%%%%%%%%%%%%%%%%

\abstractpage

{\def\baselinestretch{2.}\normalsize

In this thesis we apply perturbative QCD to make precision
predictions for some observables in high-energy processes
involving heavy quarks.

The first application we consider is a prediction for the spectrum
of $b$-flavored hadrons in top quark decay. For that purpose we
calculate at NLO the QCD corrections for bottom fragmentation in
top decay with the $b$-mass fully taken into account. Using the
perturbative fragmentation function formalism we then resum with
NLL accuracy large collinear logs of the ratio of bottom-to-top
mass, which leads to an essential improvement of the results. Next
we perform the threshold resummation for the coefficient function
for top decay with NLL accuracy. That resummation leads to an
important improvement of the $b$ spectrum in the region where the
produced bottom in top decay carries a large fraction of the
momentum of the parent top. Finally, we extract information for
the non-perturbative $b$-fragmentation into hadrons from $e^+e^-$
data and make a prediction for the spectrum of those $b$-flavored
hadrons produced in top-quark decay.

Our second application is to charm production in charged-current
DIS. We first calculate with NLL accuracy the soft-gluon resummed
coefficient function for heavy quark production (initiated by a
light quark) in inclusive DIS. Our result is applicable for the
case of low momentum transfer that is of the order of the mass of
the heavy quark. We also make a connection of this result to the
known result for massless quark production. We then apply this
result for charm quark production at NuTeV and HERA for a wide
range of the transferred momentum, and present the effect of the
threshold resummation on the charm structure functions.

\par}

\endabstractpage

\addcontentsline{toc}{chapter}{Table of Contents}
\tableofcontents

\newpage
\listoftables \addcontentsline{toc}{chapter}{List of Tables}
\newpage

\addcontentsline{toc}{chapter}{List of Figures} \listoffigures

%%%%%%%%%%%%%%%%%%%%%%%%%%%%%%%%%%%%%%%%%%%%%%%%%%%%%%%%%%%%%%%%%%
\chapter{Introduction}\pagenumbering{arabic}

For the three decades of its existence, the Standard Model (SM) of
elementary particles has proved to be extremely successful. It has
withstood all experimental tests and has become a well established
theory. All predictions based on the SM have been experimentally
verified and most of its parameters have been fixed. The only
sector of the SM that has not been directly experimentally
verified as of today is the Higgs sector and, in particular, the
existence of a neutral massive spin-zero particle often simply
referred to as Higgs. There is still no direct evidence for its
existence and, despite the many constraints from precision
electro-weak physics, the Higgs mass is not known. Not all of the
parameters of the Higgs potential are determined, and its Yukawa
couplings to the fermions are only implicitly tested through the
measurements of the masses of the fermions (quarks and leptons).
Experiments in the near future and most notably the Large Hadron
Collider (LHC) and the Tevatron, will either confirm that
particle's existence and fix the above mentioned parameters, or
will significantly increase the limit on its mass.

The SM describes three of the four presently known interactions:
Electromagnetic, Weak and Strong. The fourth one - Gravity - will
not be considered in this thesis. We will only point out that the
orthodox viewpoint -- that gravity becomes important only at
scales of the order of the Plank one and is thus completely
irrelevant for present day collider experiments -- has been
recently challenged. New ideas have emerged (Arkani-Hamed,
Dimopoulos and Dvali \cite{add}, Randall and Sundrum \cite{rs})
suggesting that it may be possible for gravity to play an
important role at energies as low as a $\TeV$. However, at least
at presently accessible energies, the experimental data do not
favor any gravitational effects (usually through so-called
Kaluza-Klein modes of gravity or other higher dimensional matter)
and most of the models dealing with those effects typically assume
that the lowest masses of the new particles are close to or above
$1\TeV$.

The SM is a gauge-field theoretical model based on the following
non-abelian gauge group:
\begin{equation}
G_{SM} = SU(3)_C\times SU(2)_I\times U(1)_Y. \label{gauge group}
\end{equation}

The gauge group is non-simple and involves three different
dimensionless coupling constants corresponding to each of the
three group factors above. The $SU(2)_I\times U(1)_Y$ part
corresponds to the Weinberg-Salam model \cite{WS} and provides a
unified description
\footnote{ Strictly speaking it is not truly unified since the
group is not simple and we introduce {\it a priori} more than one
dimensionless coupling. }
of the electromagnetic and weak interactions. There are four gauge
bosons associated with that group: two neutral ones $\gamma$ and
$Z$ and two charged ones $W^\pm$. The photon $\gamma$ is exactly
massless due to the unbroken $U(1)$ subgroup that is identified
with electro-magnetism. The other three are heavy, with masses
approximately
\footnote{ The values of the various constants used throughout
this thesis are taken from the Particle Data Group \cite{pdg}. }
$m_Z=91\GeV$ and $m_{W^\pm}=80\GeV$; their masses arise as a
result of electroweak symmetry breaking via the Higgs mechanism.
The same mechanism also supplies the tree-level masses of the
fermions through their Yukawa couplings to the Higgs doublet. The
electroweak vector bosons couple to all fermions. The magnitude of
those couplings is relatively small; the electromagnetic
interactions are suppressed by powers of $\alpha = e^2/4\pi
\approx 1/137$, while the effects of the weak interactions are
typically proportional to powers of the Fermi constant $G_F$. For
that reason the quantum corrections introduced by electro-weak
interactions are much smaller than the ones due to strong
interactions
\footnote{This is also the case in the applications considered in
this thesis as will be detailed in the subsequent sections.}.

The dependence of the dimensionless electroweak gauge couplings
$g$ and $g'$ on energy is ``intuitive" i.e. the strength of the
electroweak interactions increase with the energy scale. For that
reason the $SU(2)\times U(1)$ theory of the electroweak
interactions is a true perturbation theory formulated directly in
terms of observable fields
\footnote{ With the exception of the quarks; see below. }.
The situation changes dramatically when one considers the strong
interactions described by the $SU(3)_C$ factor in \eq{gauge
group}. First, the coupling constant is not small. Moreover, as is
well known, a non-abelian gauge theory with such a gauge group and
with sufficiently small number of active fermions (flavors)
exhibits the ``counter-intuitive" behavior known as asymptotic
freedom: the dimensionless coupling constant associated with that
group decreases with the increasing of the energy scale and
effectively such a theory behaves as a free theory at high
energies. However, at small energies the coupling grows and
eventually diverges at some finite value of the energy scale. That
scale, usually denoted as $\L$, has a value of the order of
$200-300\MeV$ and quantifies the borderline between the
perturbative and non-perturbative regimes in such a theory.
Although the detailed description of QCD will be spared for the
next Chapter, from the above discussion it becomes clear that such
a theory exhibits another remarkable property: confinement. The
growing of the coupling at low energies (which corresponds to
large distances) indicates that the particles whose interaction is
described by such a gauge theory may not be able to exist as free
(asymptotic) states at all. Instead, they will form bound states.
That expectation is confirmed by experiment: no free quarks have
ever been observed experimentally. Unlike the quarks that carry an
additional quantum number called color, the observable strongly
interacting particles -- the hadrons -- are colorless objects and
have the quantum numbers of two-particle (mesons) and
three-particle (baryons) bound states of quarks. Strictly
speaking, the property of confinement is an assumption based on
the above mentioned behavior of the strong coupling constant and
the non-observation of free colored particles. The derivation of
the properties of the hadrons from the first principles (i.e. from
QCD) is one of the fundamental problems in theoretical physics
today.

It is a remarkable achievement that we are able to make precise
predictions for the observed hadronic states in high energy
experiments based on a theory formulated in terms of
non-observable constituents - the quarks. One of the main
ingredients of the theory that makes this possible is the
so-called factorization theorem. In essence, it states that in
hard scattering experiments with typical hard scale $Q>>\L$ an
observable (e.g. a cross-section) can be written as a product
(more precisely - convolution) of perturbative and
non-perturbative parts. The former part can be calculated in
perturbation theory based on an expansion in the strong coupling
constant $\as(Q^2)$ while the latter part has to be extracted from
the experiment. It is possible to extend the factorization theorem
even to processes where other scales besides $Q$ are present, e.g.
when heavy quarks are involved. As will be clear from discussions
throughout this thesis, there is no absolute notion of heavy
quarks, i.e. whether a quark is considered heavy or light depends
on the particular problem being studied. However, in general,
light quarks are considered to be those with masses below $\L$
(i.e. $u,d,s$) while the ones with masses above that scale are
usually considered to be heavy ($c,b,t$). The reason behind such a
separation is easy to understand: for a heavy quark with mass $M$,
$\as(M^2)<1$ and it is indeed sensible to apply perturbation
theory. Studies of QCD involving processes with heavy quarks are
at present an important internal test for QCD as well as for
obtaining precision predictions that will be needed to distinguish
signals from new physics (Supersymmetry, Extra Dimensions, etc.).

QCD has another peculiar feature: there are situations where the
convergence of the perturbation series is spoiled because of the
appearance of additional factors that multiply the coupling
constant to any order in perturbation series. The presence of such
terms effectively alters the expansion parameter to a larger value
which in turn spoils the convergence of the series. To be able to
obtain useful information in that case, one needs to resum classes
of such terms to all orders in the coupling constant.

There are many examples of physical processes where the
factorization theorem plays a decisive role in studies of
processes involving heavy quarks and where often the application
of the above mentioned resummations leads to serious improvement
of the perturbative results. This thesis is devoted to the study
of two such processes in perturbative QCD (pQCD) with detailed
phenomenological applications: the spectrum of $b$-flavored
hadrons in top quark decay and neutrino production of charm in
Deep Inelastic Scattering (DIS).

In the SM, it is the mass of the top that uniquely distinguishes
it from the other five flavors. The very large value of this
parameter, however, is sufficient to draw particular attention to
the top quark. Top's large mass is responsible for its small
lifetime, the latter preventing the top from forming bound states
(a process known as hadronization). For that reason, the top quark
behaves like a real particle and one can safely describe its decay
in perturbation theory. Since the only experimental information
about the top is through its decay products, it is very important
to have a precise prediction for the decay products of the top. In
this thesis we make a prediction for the spectrum of the hadrons
resulting from the hadronization of the $b$-quark in top quark
decay and resumming to all orders in $\as$ and with
next-to-leading logarithmic (NLL) accuracy two classes of large
logs: quasi-collinear ones that are due to the large ratio of
top-to-bottom mass and soft ones that are due to soft-gluon
radiation and affect the distribution in particular kinematical
regions. Such results will be very important after the near future
high energy experiments supply enough data on top decay.

The second process studied in this thesis is Deep Inelastic
Scattering (DIS), a process of scattering of a lepton off a
hadronic target. Since the leptons do not interact strongly, that
process can serve as a tool to study the structure of the hadrons
by probing them with a virtual vector boson emitted from the
lepton. DIS is perhaps the most studied high energy process; it
was in this process in the late 1960's that Bjorken scaling was
discovered, and which in turn served as a motivation for the
introduction of the quark-parton model. Using the available
next-to-leading (NLO) calculations for heavy quark production in
charged-current (CC) DIS we calculate the corrections to the charm
structure functions in the region of low momentum transfer and
fully accounting for the mass of the charm quark. Such a result is
important for the precise determination of the parton distribution
functions in the target hadrons as well as the value of the charm
mass.

This thesis is organized as follows: in the next Chapter we
discuss some general features of QCD from the perspective of our
applications. In Chapter 3 we present our original results on
$b$-fragmentation in top decay. In Chapter 4, we discuss our
original results for soft-gluon resummation in the coefficient
function for heavy quark production in CC DIS, which we apply to
study the effect of the threshold resummation on charm production
at low energy transfer. In Chapter 5 are our conclusions. In the
appendix we have listed some useful results.

%
%%%%%%%%%%%%%%%%%%%%%%%%%%%%%%%%%%%%%%%%%%%%%%%%%%%%%%%%%%%%%%%%%%
\chapter{Preliminaries: Perturbative QCD}

In the previous Chapter we made some general remarks about the
relevance of QCD as a theory of the strong interactions as well as
few of its peculiar features. In this Chapter we are going to
systematically review that theory and derive many of its
properties. The organization of the material in the present
Chapter does not follow any particular review on QCD but is
presented in a way that is suitable for our applications. There
are many excellent reviews of QCD; some of those include
\cite{Muta,purple,BCM,Dokshitzer Dyakonov Troyan,pQCDbook}.

\section{QCD as a Fundamental Model for the Strong Interactions}

The strong interactions govern the interactions of hadrons at a
wide range of energies: from the highest energies accessible to
the present day colliders down to energies typical for the nuclear
physics. At the same time the behavior of the strong interaction
is very different in the low and high energy regimes. At low
energies, i.e. energies characterized by a scale $\mu<<\L$, the
hadrons behave as if they are fundamental particles. However, no
successful field--theoretical description in terms of the observed
hadrons was found that was able to describe the high energy regime
and explain the proliferation of observed hadrons at high energy
colliders. Contrary to the early expectations, the breakthrough
towards the understanding of the strong interactions was made in
the study of the high energy behavior of the hadrons.
\subsection{Quark Hypothesis}
In 1964 Gell-Mann and Zweig \cite{quarks} introduced the idea of
quarks: a few elementary particles that are the building blocks of
all hadrons. There are six known types of quarks (quark flavors);
they are spin $1/2$ fermions with rational electric charges (in
units of the charge of the electron): $u,c,t$ have charge $+2/3$
while $d,s,b$ have charge $-1/3$. The quark hypothesis assumes
that the wave function of a hadron is constructed from the
one-particle wave functions of quarks and/or antiquarks. The
mesons have the quantum numbers of a quark-antiquark pair while
the baryons are combinations of three (anti)quarks. Also, in order
to avoid a problem with the spin-statistics theorem, it was
necessary to introduce additional hidden quantum number - {\it
color} \cite{color}. From a comparison with experiment it was
concluded that each quark flavor must have three different
``copies" labelled by an additional color index. Since no colored
particles have been observed it was postulated that the hadrons
can only form ``colorless" combinations of quarks, i.e. color was
introduced as an exact global symmetry. It can be shown that the
above observations plus the requirement that quarks and antiquarks
transform under different (complex-conjugated) irreducible
representations of the color symmetry group uniquely fixes the
group to be $SU(3)_C$. The quarks transform under the fundamental
representation ${\bf 3}$ of that group, while the antiquarks
transform under its conjugated representation $\overline{{\bf
3}}$.

\subsection{Parton Model}\label{sec partmod}

The idea that the hadrons are built from elementary constituents -
the partons \cite{parton model} - was extremely successful not
only in explaining the hadron spectroscopy but also in the
description of the Bjorken scaling \cite{Bj scaling} observed in
Deep Inelastic Scattering (DIS) experiments \cite{SLACDIS}. The
experimental data showed that at large scales the structure
functions of the nucleons are (approximately) independent of the
value of the hard energy scale $Q$ and depend only on the Bjorken
variable $x$ (for a detailed description of DIS we refer the
reader to Section (4.1)). The parton model assumes that in high
energy lepton-nucleon scattering, where the transferred momentum
is large enough so the masses of the partons and their transverse
motion inside the nucleon can be neglected, the virtual
electroweak vector boson emitted from the initial lepton is
scattered incoherently by a single free point-like parton $a$. The
whole information about the structure of the hadron that is
relevant to the high energy process is encoded in a scalar
function $f_a(\xi)$ called the parton distribution function (pdf)
representing the probability distribution for finding the parton
$a$ inside the hadron and carrying a fraction $\xi:~ 0\leq \xi\leq
1$ of the momentum of the parent hadron. The philosophy of the
parton model then suggests the following form of the hadronic
interactions at high energy:
\begin{equation}
d\sigma(h,\dots) = \sum_a \int_0^1 d\xi~ f_a(\xi)~
d\hat\sigma(a,\dots)~, \label{parton model}
\end{equation}
where $d\sigma(h,\dots)$ is the cross-section for scattering of a
hadron $h$ and $\dots$ stands for the other particles in the
scattering process; $d\hat\sigma(a\dots)$ is a parton level
cross-section with the hadron $h$ replaced by a free parton $a$,
and the partonic cross-section is weighted with the distribution
function $f_a(\xi)$. Since the parton model is a free theory, to
lowest order in the electroweak coupling the partonic
cross-section is very simple: $d\hat\sigma \sim \delta(\xi - x)$,
so that the momentum fraction $\xi$ is identified with the Bjorken
variable $x$ (cf. Section (4.1)). Therefore under the parton model
assumption \eq{parton model}, the structure functions are simply
proportional to the pdf $f_a(x)$ and naturally independent of the
hard transferred momentum $Q$.

The success of the parton model goes beyond the description of
Bjorken scaling. It also makes a prediction about various
relations involving the measurable structure functions $F_1,
F_2,\dots$ (that will be properly introduced in Section (4.1)),
among which is the Callan-Gross relation \cite{CallanGross}:
\begin{equation}
F_2(x) = x F_1(x)~,\label{Callan-Gross}
\end{equation}
which is related to the fact that quarks have spin $1/2$. The
parton model can be generalized to other processes as well; one
just needs to measure the pdfs for the various quarks in a
specific process in order to predict a measurable quantity for
another process. In that procedure, the following relations
between the various parton distributions following from isospin
invariance are often assumed:
\begin{equation}
f_u^{proton}(x) = f_d^{neutron}(x)\ ;\ f_d^{proton}(x) =
f_u^{neutron}(x).\label{isospin}
\end{equation}

\subsection{QCD: the Dynamical Theory of Color}

Although the parton model was quite successful in the description
of many high energy processes, it was clear that it is at best a
good hint towards a complete dynamical theory of the strong
interactions. The complete theory would be able to explain one of
the basic assumptions of the parton model - asymptotic freedom.
Such a theory was constructed around 1973 after it was understood
that the Yang-Mills theories do play an important role in high
energy physics; at that time the renormalizeability of those
theories was proved and methods for their quantization were
developed \cite{renormalize}. It was also shown that the
non-abelian theories were the only theories which may exhibit
asymptotic freedom, or technically, have negative first
coefficient in the $\beta$-function (see next section). All these
developments led to the construction of QCD as the dynamical
theory of the strong interactions \cite{QCDth} as follows:

QCD is a non-abelian gauge theory with six quark flavors. The
gauge group can be naturally obtained by gauging the exact global
color symmetry group $SU(3)_C$. The quarks transform under the
fundamental representation of $SU(3)_C$. Since $\dim(SU(3))=8$,
there are eight gauge bosons called {\it gluons} that are
electrically neutral, carry color charge and, as usual, are
hermitian fields that transform under the adjoint representation
of the gauge group $SU(3)_C$.

The lagrangian of QCD has the following form:
\begin{equation}
{\mathcal{L}}_{QCD} = -{1\over 4} F_{\mu\nu}^a F^{a,\mu\nu} +
i\sum_q \overline{\psi^i}_q\gamma^\mu
\left(D_\mu\right)_{ij}\psi^j_q - \sum_q m_q \overline{\psi^i}_q
\psi^i_q, \label{L QCD}
\end{equation}
where the index $i=1,2,3$ runs over the different quark colors and
$q$ over the quark flavors: $q=u,b,s,c,b,t$. The field-strengths
are given by:
\begin{equation}
F^a_{\mu\nu} = \partial_\mu A_\nu^a - \partial_\nu A_\mu^a -g_S
f^{abc}A^b_\mu A^c_\nu, \label{F mn}
\end{equation}
and the gauge-covariant derivative is:
\begin{equation}
\left(D_\mu\right)_{ij} = \delta_{ij}\partial_\mu + ig_S \sum_a
t^a_{ij}A^a_\mu. \label{D mu}
\end{equation}
In the above equations, $g_S$ is the strong coupling constant,
$f^{abc}$ are the structure constants of the gauge group and $t^a$
are the generators of the fundamental representation of the gauge
group. In general, the lagrangian \eq{L QCD} must be supplemented
with gauge-fixing and ghost terms. The quarks have non-zero masses
but their origin is outside QCD; in SM their masses result from
the electroweak symmetry breaking. The only free parameters in QCD
are the six quark masses and the single gauge coupling constant.

The gauge group has the following matrix structure which, for
generality, we present for arbitrary group $SU(N)$: the
fundamental representation has generators $t^a,~ a=1,\dots,N^2-1$
that satisfy
\begin{eqnarray}
\tr(t^at^b)&=&{1\over 2}\delta^{ab}\nonumber\\
\sum_a t^a_{ij}t^a_{jk} &=& C_F\delta_{ik},~ i,j,k=1,\dots,N.
\label{matrix fermions}
\end{eqnarray}
For the group $SU(3)$, the generators $t^a$ are usually given by
the Gell-Mann matrices $\lambda^a$: $t^a=\lambda^a/2$. Similarly,
the adjoint representation has generators $T^a$ that can be
related to the structure constants $f^{abc}$ through:
\begin{equation}
(T^a)_{bc} = f^{abc}, \label{structure const}
\end{equation}
and
\begin{equation}
\tr(T^cT^d) = \sum_{a,b} f^{abc}f^{abd} = C_A \delta^{cd}.
\label{matrix gluon}
\end{equation}
Above, $C_F$ and $C_A$ are the values of the quadratic Casimir of
the gauge algebra in the fundamental and adjoint representations
respectively:
\begin{equation}
C_F = {N^2-1\over 2N}~;~~ C_A = N,  \label{CFCA}
\end{equation}
and for the case of $SU(3)$:
\begin{equation}
C_F= {4\over 3}~;~~ C_A = 3. \label{CACF N=3}
\end{equation}

Once formulated, it must be shown that QCD indeed is capable of
reproducing the success of the parton model as a first step. That
in fact follows since from the very formulation of QCD it is clear
that the parton model corresponds to the Born approximation of
QCD. The real challenge however is to derive the asymptotic
freedom from first principles and also to derive the corrections
to the Bjorken scaling.

\subsection{Strong Coupling Constant}

The running of the renormalized strong coupling $\as=g_S^2/4\pi$
is determined from the following equation:
\begin{equation}
\mu{\partial \as\over \partial\mu} = 2\beta(\as). \label{gs}
\end{equation}
The $\beta$-function $\beta(\as)$
\footnote{In contrast to most of the standard presentations (e.g.
\cite{Muta}), we have introduced $\beta(\as)$ through the
relation: $\beta(\as)= (g_S/4\pi) \beta(g)$.}
has a series expansion in the coupling constant $\as$. It can be
determined up to a fixed order in perturbation theory from
explicit evaluation of the gauge coupling renormalization constant
$Z_g$:
\begin{equation}
\beta(g) = \lim_{\epsilon\to 0} \left( -\epsilon g - {\mu\over
Z_g}{dZ_g\over d\mu} g \right).  \label{equation beta}
\end{equation}
At present, the $\beta$-function of QCD is known to four loops in
the $\MSbar$ scheme \cite{beta function}. However, since for all
applications in this Thesis we need the evolution of the strong
coupling to two loops, we will present only the two loop result:
\begin{equation}
\beta(\as) = -b_0 \as^2 - b_1\as^3 -\dots, \nonumber\\
\end{equation}
with $b_0$ and $b_1$ given by
\begin{equation}
b_0={{33-2n_f}\over {12\pi}},\ \ b_1={{153-19n_f}\over{24\pi^2}}.
\label{beta NLO}
\end{equation}
The first two coefficients of the $\beta$-function are independent
of the renormalization scheme. However, that is no longer true for
the higher order terms. In Eq.\eq{beta NLO}, $n_f$ is the number
of active massless flavors. In the presence of quark masses, its
value becomes scale dependent. If one assumes strong ordering
among the quark masses, then for scale $\mu:~m_n<<\mu<<m_{n+1}$
all flavors with masses below $m_{n+1}$ are effectively massless
while the rest of the flavors are heavy and can be integrated out.
An example is the case of top decay considered in Chapter 3 where
the scale is running between the $b$ and the $t$ masses and
because of their large difference we simply fix $n_f=5$. However,
in practical applications (especially with scales of the order of
the $b$ and the $c$ quark) the strong ordering assumption is not
always valid and therefore the choice of $n_f$ is somehow
ambiguous. The common practice is to change the value of $n_f$ by
one unit when the hard scale crosses the mass of the corresponding
heavy quark. Such a change should be supplemented with an
additional constraint that relates the values of the strong
coupling evaluated in the two schemes at the switching point. In
the $\MSbar$ renormalization scheme, the strong coupling is
continuous at the switching points \cite{continuity as} ( see also
\cite{pdg}), up to negligible corrections of order
${\mathcal{O}}(\as^3)$.

Now it is easy to show that indeed QCD enjoys the property of
asymptotic freedom \cite{asympt free}. In a regime where the
strong coupling is small, from \eq{beta NLO} and \eq{gs}, it is
easy to see that the strong coupling is a decreasing function of
the scale $\mu$ if the number of flavors $n_f<33/2$. That
requirement is satisfied in QCD. The exact solution of Eq.\eq{gs}
to NLO is given by:
\begin{equation}
\alpha_S(\mu^2)={1\over {b_0\ln(\mu^2/\Lambda^2)}} \left\{
1-{{b_1\ln\left[\ln (\mu^2/\Lambda^2)\right]}\over
{b_0^2\ln(\mu^2/\Lambda^2)}}\right\}. \label{alpha s}
\end{equation}
One can use this expression in order to relate the values of the
strong coupling at two different scales with NLO accuracy
\cite{cmn}:
\begin{eqnarray}
\as(k^2) &=& {\as(\mu^2)\over 1+ b_0 \as(\mu^2)\ln(k^2/\mu^2)}
\left( 1-{b_1\over b_0}{\as(\mu^2)\over 1+ b_0
\as(\mu^2)\ln(k^2/\mu^2)}\right.\label{alpha s ii}\\
&\times & \ln(1+b_0\as(\mu^2)\ln(k^2/\mu^2))+{\mathcal{O}}\left(
\as^2(\mu^2)[\as(\mu^2)\ln(k^2/\mu^2)]^n \right)\Bigg).\nonumber
\end{eqnarray}
The constant $\Lambda$ contains all the information about the
boundary condition to which Eq.\eq{gs} must be subjected. It is a
low energy scale where the strong coupling diverges. As we
mentioned in Chapter 1, $\Lambda$ represents the border between
the perturbative and non-perturbative regimes of QCD. In practice
the value of $\Lambda$ is not unambiguous. In high energy
experiments one typically obtains information about the strong
coupling constant at some large scale and only from there the
value of $\Lambda$ is inferred. It is obvious however, that in
this way the determination of $\Lambda$ ``absorbs" all ambiguities
such as the dependence on the order of $\as$ (LO, NLO, etc.), the
choice of the value of $n_f$ and the choice of the renormalization
scheme. In this Thesis we use the following value of the strong
coupling at NLO \cite{pdg}:
\begin{equation}
\alpha_S(M_Z^2)=0.118 \label{alpha s numerical}
\end{equation}
It leads to $\Lambda^{(5)} \approx 200\MeV$ and $\Lambda^{(4)}
\approx 300\MeV$. The precise values used in our applications will
be discussed in the next Chapters.

\section{QCD Factorization Theorem}\label{Fac Th}

QCD is formulated in terms of quarks and gluons while the
experimentally observed states are hadrons. Since at present we
are not able to describe the non-perturbative regime of QCD, a
fruitful application of QCD to the hadronic interactions requires
a universal way of splitting the contributions from short- and
long-distance physics.

Such a separation is possible. It is known as the (QCD)
factorization theorem  \cite{CSS,Collins} and states that for
processes that have initial and/or observed final state hadrons
the differential cross-section has the following form:
\begin{eqnarray}
d\sigma(x,Q^2,m^2) &=& \prod_{h,h'} \sum_{i,f}~
f_{i/h}(x,\mu^2)~\otimes~ d\hat\sigma_{i\to
f}(x,Q^2,m^2,\mu_r^2,\mu_F^2)\nonumber\\
&\otimes & D_{h'/f}(x,\mu^2) + {\mathcal{O}}(\Lambda/Q)
\label{fac}
\end{eqnarray}
The factor $f_{i/h}$ stands for the parton distribution function
of the parton $i$ inside the hadron $h$ present in the initial
state, and $Q$ and $x$ represent the hard scale and some
kinematical variable respectively. Unlike the simple parton model
\eq{parton model}, the parton distributions also depend on the
factorization/renormalization scale $\mu^2$. That difference has
important consequences since it incorporates the violations of
Bjorken scaling. The second factor $d\hat\sigma_{i\to f}$, also
known as the (Wilson) coefficient function, represents the
partonic hard scattering cross section for the reaction $i\to f$
that depends on the unphysical renormalization and factorization
scales $\mu_r^2$ and $\mu_F^2$ and on the masses of the heavy
quarks $m^2$. The last factor in \eq{fac} is the so called
fragmentation function $D$. It contains the information for the
hadronization of the hard parton $f$ (that is produced in the hard
process described by the partonic cross-section $d\hat\sigma$)
into an observed hadron $h'$. The integral convolution appearing
in \eq{fac} is defined as:
\begin{equation}
(f\otimes g) (x) = \int_x^1 {dz\over z} f(z)g(x/z),
\label{convolution}
\end{equation}
where $f$ and $g$ are two functions with argument $x:~0\leq x\leq
1$. The convolution has the property \eq{Mellin fac} which is
important in practical applications.

The real power of the factorization theorem is in the fact that
the distribution/fragmentation functions are universal: they
depend only on the non-perturbative transition they describe and
not on the hard scattering process. That is why once they are
measured in one process, they can be applied to any other process.
At the same time, the coefficient function contains all the
information about the hard scattering process and is independent
of the details of the non-perturbative transitions. Although the
$\otimes$-product of coefficient function and
distribution/fragmentation function is an observable and therefore
free from any ambiguity, the distribution, fragmentation and the
coefficient functions are separately ambiguous. In particular,
they are scheme dependent; the origin of that scheme dependence is
in the treatment of the IR divergences associated with their
computation. Let us describe that in more detail:

The evaluation of the coefficient function proceeds in the
following way: one calculates the hadronic process $d\sigma$ that
is under study by formally replacing each initial hadron $h$ with
an {\it on-shell} parton $a$. To that end, and in accordance with
the factorization theorem \eq{fac} we just described, one
introduces (also formally) new parton distributions $f_{i/a}$
which have the meaning of a distribution of a parton $i$ inside a
parton $a$; we treat the fragmenting partons in a similar fashion.
The main purpose of the fictitious distributions $f_{i/a}$ is to
absorb all the IR sensitive contributions from the calculated
cross-section. As a next step, one simply discards the functions
$f_{i/a}$ and what is left is the needed coefficient function
$d\hat\sigma$. The extraction of the partonic pdfs is physically
equivalent to the absorbtion of the IR sensitive contributions
into the pdfs $f_{i/h}$ (see also \refsec{HQM}).

Clearly, such a procedure is very similar (and is in fact related)
to the UV renormalization where one introduces appropriate
counter-terms to absorb (and thus cancel) the UV divergences
appearing in the Feynman diagrams. In the most often used $\MSbar$
subtraction scheme the partonic pdfs read:
\begin{equation}
f_{i/a}(x) = \delta_{ia}\delta(1-x) +{\as\over 2\pi} \left(
-{1\over\epsilon} +\gamma_E-\ln(4\pi)\right)P^{(0)}_{ia}(x)
+{\mathcal{O}}(\as^2),\label{parton pdf}
\end{equation}
where $P^{(0)}_{ij}(x)$ are the leading order Altarelli-Parisi
splitting functions that will be defined and thoroughly discussed
in the next section, and $\epsilon = (4-D)/2$ (see the appendix
for more details). The subtraction scheme for the IR divergences
emerging in the limit $\epsilon\to 0$ is related to the
renormalization scheme used to remove the UV divergences appearing
in the formal (operator) definition of the parton densities.
Following the approach in \cite{CSS}, the quark distribution can
be defined (in the light-cone gauge $A^+=0$) as:
\begin{equation}
f_{q/h}(x) = {1\over 4\pi} \int dy^- e^{-ixP^+y^-} \langle
h(P)\vert \overline{\psi}(0,y^-,0_{\bot}) \gamma^+
\psi(0,0,0_{\bot})\vert h(P) \rangle , \label{pdf q}
\end{equation}
with similar results for the antiquark and gluon distributions. In
Eq.\eq{pdf q} $P$ is the momentum of the hadronic state $h$,
$V^\pm = 1/\sqrt{2}(V^0\pm V^3)$ is the light-cone co-ordinate of
a four-vector $V$ and $\psi$ is the operator of the quark field
that carries momentum fraction $x:~ p_q^+=xP^+$. For a partonic
initial state, one can evaluate the partonic PDF in powers of
$\as$ by explicit calculation of the matrix element in \eq{pdf q}.
The result to order $\as$ shows the following structure of the
singularities:
\begin{equation}
f_{i/a}(x,\epsilon) = \delta_{ia}\delta(1-x) +{\as\over 2\pi}
\left( {1\over\epsilon_{UV}}-{1\over\epsilon_{IR}}
\right)P^{(0)}_{ia}(x) + {\rm UV}~{\rm counterterm}. \label{parton
pdf reg}
\end{equation}
Since by the UV renormalization the UV counterterm cancels the
$1/\epsilon_{UV}$ pole, we have a remaining IR divergence in the
partonic PDF. The explicit form of that singularity obviously
depends on the chosen UV renormalization scheme; for the $\MSbar$
scheme the result is presented in \eq{parton pdf}. Similar
considerations apply for the distribution functions as well; that
is discussed e.g. in \cite{CS} and \cite{pij}.

The factorization theorem in the presence of massive quarks is
accurate up to terms ${\mathcal{O}}(\Lambda/Q)$; see Eq.\eq{fac}.
The remainder in \eq{fac} contains terms that are power suppressed
and are uniform in $Q$ i.e. no terms with suppression $\sim
(m/Q)^n$ are present \cite{Collins}. The proof of the
factorization theorem presented in \cite{Collins} uses a variable
flavor number scheme (VFNS). The VFNS treats the light quarks
$u,d$ and $s$ as massless and always includes them as active
flavors in the running of the strong coupling $\as$. The treatment
of the heavy flavors $c,b$ and $t$ is, however, process dependent.
If the typical energy scale is below the corresponding quark mass
then that quark is treated as heavy and is integrated out. In
particular it does not contribute to the evolution of the strong
coupling and does not have an associated parton density. The
quarks with masses below the hard scale are treated in a different
way: they contribute to the strong coupling as if they are exactly
massless, and they have their own distribution functions
\footnote{These are introduced in order to systematically resum
large logs of collinear origin that appear to all orders in $\as$.
Schemes without heavy quark densities exist and are called Fixed
Flavor Number Schemes (FFNS). An example is the GRV 98 set of
parton distributions \cite{GRV}.}
which are evolved with the energy scale via evolution equations
with massless kernels (see next section). Such a scheme
corresponds to the renormalization scheme of Collins, Wilczek and
Zee \cite{CWZ} and is also used (through the so-called
Aivazis-Collins-Olness-Tung (ACOT) prescription \cite{ACOT II}) in
the evolution of the CTEQ parton densities \cite{CTEQ}.

The physical picture behind the factorization theorem is quite
simple. One formally introduces a scale $\mu_F$ which separates
the short- from long-distance physics involved in the process. It
is intuitively clear that such separation must indeed occur in the
limit of large values of the hard scale $Q$. The time scale for
the hard interaction is of the order of $Q^{-1}$ and therefore
quite small, while the typical time for the hadronization effects
is not smaller than $\L^{-1}\gg Q^{-1}$. As a result, in the limit
$Q\to\infty$, short- and long-distance effects cannot interfere
with each other and therefore factorize. More formally, the
separation between small and large scales means that all
contributing Feynman diagrams that have lines with small
virtuality can be separated from the lines with large (of the
order of the hard scale $Q$) virtuality. The former diagrams
constitute the distribution functions while the latter give the
hard coefficient function. Such non-trivial factorization for the
terms with leading power in $1/Q^2$ (the so-called leading twist
terms) was proved by Libby and Sterman \cite{Libby Sterman}. If
one works in a covariant gauge then there are present additional
gluon lines that connect the small- and large-scale factors.
However, those additional contributions cancel when summed up in a
gauge invariant subset of diagrams. These contributions disappear,
however, if one works in a physical gauge $n.A=0$, where $n$ is
some fixed four-vector. The physical gauge also has the important
property that each diagram that contributes to the physical
cross-section can be given a direct partonic interpretation
\cite{Dokshitzer Dyakonov Troyan}.

\section{Perturbative Evolution: DGLAP Equations}

In this section we will devote our attention to the dependence of
the various factors in Eq.\eq{fac} on the renormalization and
factorization scales which incorporate the scaling-violation
effects. To better illuminate our point, we are going to make the
following two simplifications throughout this section: first, we
will set $\mu_r=\mu_F\equiv\mu$. This is a standard choice in the
studies of pQCD which, however, will not restrict the generality
of our discussion. If needed, the separate dependence on both
scales can be easily restored with the use of the running of the
strong coupling (see Eq.\eq{alpha s ii}). The second
simplification is that we will consider Eq.\eq{fac} as having a
single fragmentation or distribution function multiplying the
coefficient function $d\hat\sigma$. We will consider those two
``representative" cases (only initial or final observed hadrons)
of Eq.\eq{fac} separately.

We start with:
\begin{equation}
d\sigma(x,Q^2,m^2) = \sum_{i}~ f_{i/N}(x,\mu^2)~\otimes~
d\hat\sigma_{i\to X}(x,Q^2,m^2,\mu^2), \label{fac f}
\end{equation}
which corresponds to the case of a single hadron (nucleon $N$) in
the initial state and no observed hadrons in the final state. A
prominent example is the case of inclusive Neutral Current (NC) or
Charged Current (CC) DIS:
\begin{equation}
\lepton~+~N~\to~\lepton'~+~X, \label{DIS reaction}
\end{equation}
with $\lepton$ and $\lepton'$ being leptons, $N$ a hadron (usually
a nucleon) and $X$ stands for any unobserved hadrons produced in
the reaction \eq{DIS reaction}. This case describes reactions with
so-called space-like evolution.

As a representative for a reaction with a single fragmentation
function we take:
\begin{equation}
d\sigma(x,Q^2,m^2) = \sum_{f}~d\hat\sigma_{e^+e^-\to
f}(x,Q^2,m^2,\mu^2) ~\otimes~ D_{h/f}(x,\mu^2), \label{fac D}
\end{equation}
which corresponds to the case of inclusive production of a single
hadron $h$ in a non-hadronic collision. These reactions are known
as having time-like evolution. An example is the case of a single
particle inclusive $e^+e^-$ annihilation:
\begin{equation}
e^+~+~e^-~\to~h~ +~ X \label{e+e- reaction}
\end{equation}
with $h$ being an observed hadron and, as usual, $X$ stands for
any unobserved hadrons produced in the reaction \eq{e+e-
reaction}. Also, for brevity, we have omitted the remainders in
Eqns.\eq{fac f} and \eq{fac D}.

\subsection{The Case of Space-like Evolution}

Let us concentrate on Eq.\eq{fac f}. Since the LHS is independent
of $\mu$ we can set that scale to any value we like. Among all the
possible values, the choice $\mu^2=Q^2$ is particularly appealing
as will become clear below.

From dimensional considerations the coefficient function can be
written as
\begin{equation}
d\hat\sigma_{i\to X}(z,Q^2,\mu^2) = \sigma_B C\left(z, {Q^2\over
\mu^2},\as(\mu^2) \right).\label{fac f born}
\end{equation}
where for the time being we consider the case when no heavy quarks
are present; we will generalize our considerations in the next
section. In Eq.\eq{fac f born} $\sigma_B$ is the Born
cross-section for the partonic subprocess and the function $C$ is
a dimensionless function that has a power series decomposition in
the strong coupling $\as$.

It is now obvious that upon setting $\mu^2=Q^2$ the coefficient
function takes the form $C(1,\as(Q^2))$ and depends only on the
strong coupling and on no other large (or small) parameters. Since
$\as$ is evaluated at the large scale $Q^2$ where the former is
small, the coefficient function can be easily and efficiently
calculated to some fixed order in perturbation theory. However, as
a result of the choice of scale we have made, the distribution
function has now become $Q$-dependent. That dependence is very
important. It indicates that the universality of a distribution
function may be reduced since the PDF is specific to the
experimental setup where it is extracted and therefore cannot be
applied to processes with different hard scale.

Fortunately, there exists a way to relate distribution functions
at different scales. The scale dependence of the fragmentation
functions $f_i,~ i=q,\bar{q},g$ is perturbatively controlled and
is given as a solution to a system of integro-differential
equations known as Dokshitzer-Gribov-Lipatov-Altarelli-Parisi
(DGLAP) equations \cite{AP,DGL}. That way the universality of the
distribution function is retained; we only need to extract from
experiment the distribution functions at one given scale $Q_0$.
Then that input can be used as the initial condition for the DGLAP
equations and the PDF at any other scale can be predicted. In
practice that procedure works in the following way: at some low
scale $Q_0\sim 1\GeV$ one writes down a function of $z$, that
contains small number of free parameters. Then one evolves that
initial condition via the DGLAP equations to different scales
where experimental data exist, and one tries to fit those data by
adjusting the parameters of the initial condition.

The DGLAP equations are:
\begin{equation}
{d\over{d\ln\mu^2}}f_i(z,\mu) = \sum_j\int_{z}^1{{{d\xi}\over \xi}
P_{ij}\left({z\over \xi},\alpha_S(\mu)\right) f_j(\xi,\mu)},
\label{dglap}
\end{equation}
and describe in general a system of $2n_f+1$ equations for the
distribution functions of all flavors of quarks, antiquarks and
the gluon. The kernels $P_{ij}$ have perturbative expansions in
powers of the strong coupling:
\begin{equation}
P_{ij}(z,\alpha_S (\mu))={{\alpha_S(\mu)}\over {2\pi}}
P_{ij}^{(0)}(z)+\left({{\alpha_S(\mu)}\over {2\pi}}\right)^2
P_{ij}^{(1)}(z)+{\cal O} (\alpha_S^3). \label{split}
\end{equation}
$P_{ij}^{(0)}(z)$ are the Altarelli--Parisi splitting functions
\cite{AP} that also appeared in Eq.\eq{parton pdf}. Because color
and flavor commute, those functions are independent of the quark
flavor i.e. $P_{q_iq_j}^{(0)}(z) = \delta_{ij}P_{qq}^{(0)}(z)$.
They also satisfy other relations as a result of probability
conservation:
\begin{equation}
\int_{0}^1 dz P_{qq}^{(0)}(z) = 0, \label{probab}
\end{equation}
and momentum conservation:
\begin{eqnarray}
&& \int_{0}^1 dz~z\left( P_{qq}^{(0)}(z) + P_{gq}^{(0)}(z)\right)
= 0, \nonumber\\
&& \int_{0}^1 dz~z\left( 2n_f P_{qg}^{(0)}(z) +
P_{gg}^{(0)}(z)\right) = 0. \label{momentum}
\end{eqnarray}
Note that because of the property \eq{probab}, the functions
$P_{qq}^{(0)}(z)$ are not positive definite. They are
distributions instead. The kernels $P_{ij}$ satisfy also the
following important relations as a result of charge invariance and
the $SU(n_f)$ flavor symmetry:
\begin{eqnarray}
&& P_{q_iq_j} =  P_{\bar{q}_i\bar{q}_j}~~;~~
P_{q_i\bar{q}_j} =  P_{\bar{q}_iq_j}  \nonumber\\
&& P_{q_ig} =  P_{\bar{q}_ig} = P_{qg}~~;~~ P_{gq_i} =
P_{g\bar{q}_i} = P_{gq}. \label{AP relation}
\end{eqnarray}
The explicit expressions for the four independent leading order
splitting functions are:
\begin{eqnarray}
P_{qq}^{(0)}(z) &=& C_F\left( {1+z^2\over (1-z)_+} +{3\over
2}\delta(1-z)\right),\nonumber\\
P_{qg}^{(0)}(z) &=& \half\left( z^2 + (1-z)^2\right),\nonumber\\
P_{gq}^{(0)}(z) &=& C_F\left({1 + (1-z)^2\over z}\right),\nonumber\\
P_{gg}^{(0)}(z) &=& 2C_A\left( {z\over (1-z)_+}+ {1-z\over z} +
z(1-z)\right)\nonumber\\
&+& {11C_A -2n_f\over 6}\delta(1-z).\label{AP functions}
\end{eqnarray}
The distribution $(F(z))_+$ is defined as $\int_0^1(F(z))_+H(z) dz
= \int_0^1F(z)(H(z)-H(1)) dz$, and its properties are discussed in
the appendix.

One can substantially simplify the study of the DGLAP equations if
one takes advantage of the flavor symmetry. As we mentioned
earlier, the DGLAP equations describe the evolution of $2n_f+1$
partons that are all massless (see also the discussion following
Eq.\eq{parton pdf reg}). From the form of the QCD lagrangian with
$n_f$ massless quarks \eq{L QCD}, it is evident that the theory
has an additional global $SU(n_f)$ flavor symmetry. Since the
gluons are flavor-neutral, they transform as singlets under the
flavor group. The quarks in general transform nontrivially under
that group. One can split the $2n_f$ (anti)quark fields into one
singlet:
\begin{eqnarray}
\Sigma(x,\mu) &=& \sum_i^{n_f}~\left(
f_{q_i}(x,\mu)+f_{\overline{q}_i}(x,\mu)\right),\nonumber
\end{eqnarray}
and $2n_f-1$ non-singlet (NS) combinations. $n_f$ of the NS fields
can be taken as the differences $ M^-_i = f_{q_i}(x,\mu) -
f_{\overline{q}_i}(x,\mu)$. The other $n_f-1$ combinations, which
we denote by $M^+_j$, depend on the value of $n_f$ and can be
found in \cite{purple}. Clearly, the NS combinations do not mix
with the singlet; in particular they do not mix with the gluon. To
LO, all NS fields also split from each other so we have a separate
equation for each NS field:
\begin{equation}
{d\over{d\ln\mu^2}}f^{NS}(z,\mu) = {{\alpha_S(\mu)}\over {2\pi}}
\int_{z}^1{{{d\xi}\over \xi} P_{qq}^{(0)}\left({z\over \xi}\right)
f^{NS}(\xi,\mu)}, \label{NS dglap}
\end{equation}
with $P_{qq}^{(0)}(z)$ given in Eq.\eq{AP functions}. Beyond the
leading order, however, the evolution kernels are no longer flavor
diagonal. One can still write the evolution equations in diagonal
form that is similar to the LO case Eq.\eq{NS dglap}, but the
kernels $P^{(1)+}_{NS}$ and $P^{(1)-}_{NS}$ corresponding to the
fields $M^+$ and $M^-$ are now different. Their explicit
expressions can be found in \cite{purple} as well.

In the singlet sector, there is non-trivial mixing between the
gluon density $g$ and the quark singlet state $\Sigma$. The
kernels of the evolution equations for the ``two-vector"
$\left(\Sigma(z,\mu), g(z,\mu)\right)$ form a $2\times 2$ matrix:
\begin{eqnarray}
\left(%
\begin{array}{cc}
  P_{qq}(z,\as(\mu^2)) & 2n_fP_{qg}(z,\as(\mu^2)) \\
  P_{gq}(z,\as(\mu^2)) & P_{gg}(z,\as(\mu^2)) \\
\end{array}%
\right). \nonumber
\end{eqnarray}
The NLO kernels in the singlet sector can be found in
\cite{purple}. Similarly to the NS case, we do not present them
here because of their length. The original derivations are
presented in \cite{pij,pij1}. For future reference we will only
present the large $z$ behavior of the $\MSbar$ splitting functions
at NLO:
\begin{eqnarray}
\lim_{z\to 1}~ P_{qq} &=& C_F{\as\over\pi}\left(1+K{\as\over
2\pi} +{\mathcal{O}}(\as^2)\right){1\over (1-z)_+} \nonumber\\
\lim_{z\to 1}~ P_{gg} &=& C_A{\as\over\pi}\left(1+K{\as\over 2\pi}
+{\mathcal{O}}(\as^2)\right){1\over (1-z)_+}\label{large x}
\end{eqnarray}
where:
\begin{equation}
K = C_A\left({67\over 18} - {\pi^2\over 6}\right) - {5\over 9}n_f.
\label{K}
\end{equation}

The origin of the DGLAP equations is in renormalization group
invariance. That invariance is manifested as independence of
physical quantities (for example a cross-section) of the
renormalization scale $\mu_r$ that is introduced as a result of
the renormalization procedure. Eq.\eq{fac f} is a typical example.
Before we exploit the fact that the LHS is independent of $\mu$,
let us take the Mellin transform of Eq.\eq{fac f} (with all masses
omitted for now):
\begin{equation}
d\sigma(N,Q^2) = \sum_{i}~ f_{i}(N,\mu^2) d\hat\sigma_{i\to
X}(N,Q^2,\mu^2), \label{fac f Mellin}
\end{equation}
where the Mellin transform $F(N)$ of a function $F(z)$ is defined
as
\begin{equation}
F(N) = \int_0^1 dz~ z^{N-1} F(z).\label{Mellin}
\end{equation}
To derive Eq.\eq{fac f Mellin} we used the following important
property: the Mellin transform of a convolution of two functions
is the ordinary product of the Mellin transforms of those
functions, i.e.
\begin{equation}
\int_0^1dz~ z^{N-1} (F~\otimes~ G)(z) = F(N) G(N). \label{Mellin
fac}
\end{equation}

Taking $\mu$-derivatives of both sides of Eq.\eq{fac f Mellin} and
omitting the summation over the partons (as in the NS case for
example), one has:
\begin{equation}
\mu{d \over d\mu} \ln f(N,\mu^2) = -\mu{d \over d\mu} \ln
C\left(N,{Q^2\over \mu^2}, \as(\mu^2)\right) =
2\gamma(N,\as(\mu^2)). \label{fac eq}
\end{equation}
In Eq.\eq{fac eq} we have also used Eq.\eq{fac f born}. From the
derivation of Eq.\eq{fac eq} it is clear that the function
$\gamma(N,\as(\mu^2))$ plays the role of a ``separation constant"
that depends on the common variables of the functions $f$ and $C$.
Therefore, that equation is of importance only if the function
$\gamma$ is known. Let us assume for the moment that this is the
case; then with simple integration one can obtain the scale
dependence of the moments of the distribution function $f$:
\begin{equation}
f(N,\mu^2) = f(N,\mu_0^2)~\exp\left(\int_{\mu_0^2}^{\mu^2}
{dk^2\over k^2} \gamma(N,\as(k^2) )\right), \label{solution f N}
\end{equation}
if the value of the fragmentation function at some initial scale
$\mu_0$ is known. Combining this result with Eq.\eq{fac f} and
after setting $\mu=Q$ we can finally complete the factorization
program discussed in the previous section. To calculate a physical
observable we only need the PDF at some initial scale and the
coefficient function for the hard process which is computable in
perturbation theory since it does not contain any large
parameters:
\begin{equation}
d\sigma(N,Q) = \sigma_B f(N,\mu_0^2) C(N,1,\as(Q^2))
~\exp\left(\int_{\mu_0^2}^{Q^2} {dk^2\over k^2} \gamma(N,\as(k^2)
)\right). \label{factorization}
\end{equation}
Of course, in practical applications one needs to carefully
account for the mixing between the various distribution functions
that would appear in the relation \eq{factorization}.

To make contact with the DGLAP equations, we need only notice that
by setting
\begin{equation}
\gamma_{ij}(N,\as(\mu^2)) = \int_0^1 dz~ z^{N-1}
P_{ij}(z,\as(\mu^2)), \label{gamma P}
\end{equation}
and applying a Mellin transformation to both sides of the DGLAP
equation we get exactly the equation \eq{fac eq} for the
fragmentation function $f$. That way we conclude that the
functions $\gamma_{ij}$ are the Mellin moments of the evolution
kernels $P_{ij}$.

The functions $\gamma_{ij}$ indeed have a deep physical meaning
which is not hard to understand. For that purpose let us rewrite
schematically Eq.\eq{fac eq} for $f$:
\begin{equation}
\left( \mu{\partial \over \partial\mu} +\beta(g_S){\partial \over
\partial g_S} - 2\gamma(g_S)\right)~\ln f ~=~0 \label{CalSimm}
\end{equation}
Eq.\eq{CalSimm} resembles the renormalization-group equation of a
matrix element of an operator in a massless theory with anomalous
dimension $\gamma$. The functions $\gamma_{ij}(N,\as(\mu^2))$ are
indeed anomalous dimensions but of local operators of spin-$N$ and
twist-two that appear in the light-cone operator product expansion
(OPE) of two currents. The derivation of this result is rather
lengthy and can be found in \cite{GP,GW}; for a more pedagogical
discussion see \cite{Muta}. Here we shall only mark the main steps
in the example of DIS. The DIS cross-section is of the form
$L^{\mu\nu}W_{\mu\nu}$ where $L$ is a tensor constructed from the
lepton's four-momentum and $W$ has the form:
\begin{equation}
W_{\mu\nu} \sim \int d^4y e^{iq.y}\langle P\vert
[J_\mu(y),J_\nu(0)]\vert P\rangle . \label{W}
\end{equation}
The non-local product of the two electromagnetic currents
$J_\mu(y)J_\nu(0)$ is sandwiched between the hadronic state $\vert
h(P)\rangle$ and its conjugate. This operator product can be
expanded in a series over local operators $O_n$:
\begin{equation}
J(y)J(0) = \sum_{n,i} C^i_n(y^2)O_{n,i}^{\nu_1\dots\nu_n}
y_{\nu_1}\dots y_{\nu_n}. \label{JJ}
\end{equation}
Sandwiching the current product in the hadronic states:
\begin{equation}
\langle P\vert O_{n,i}^{\nu_1\dots\nu_n}\vert P\rangle =
A_{n,i}(\mu^2)P^{\nu_1}\dots P^{\nu_n}, \label{A}
\end{equation}
results in the following expression:
\begin{equation}
F(n,Q^2) = A_{n,i}(\mu^2)C^i_n(Q^2/\mu^2,\as). \label{FF}
\end{equation}
Above, $F$ is the $n$-th Mellin moment of any of the structure
functions. Eq.\eq{JJ} implies that the coefficient function $C$
satisfies the renormalization group equation with anomalous
dimension $ 2\gamma_J -\gamma_O$. Since the current $J$ is
conserved its anomalous dimension vanishes. The anomalous
dimension corresponding to an operator $O$ can be computed in
perturbation theory from the relation:
\begin{equation}
\gamma_O = {\mu\over 2}{\partial \over \partial\mu} \ln Z_O,
\label{gamma O}
\end{equation}
where $Z_O$ is the renormalization constant for the operator $O$.

Collecting the results above, we see that after setting $\mu=Q$ we
obtain Eq.\eq{factorization} for the DIS structure functions, and
the anomalous dimension $\gamma(N)$ presented there corresponds to
the local operators $O_N$.

Let us make few more brief comments on the DGLAP equations. First,
the DGLAP equations in the form that we discussed were first
introduced by Altarelli and Parisi in \cite{AP}. There the authors
were following a more or less reversed approach: starting from the
OPE results we discussed above, they introduced the AP splitting
functions through the relations \eq{gamma P}. The important result
is that not only are the $z$-space evolution equations equivalent
to the renormalization group ones, but the former offer new way of
deriving the splitting functions based on their partonic
interpretation. Second, similarly to the beta-function, the
one-loop splitting functions $P_{ij}^{(0)}(z)$ are renormalization
scheme independent. However, that is not true for the higher order
functions which are renormalization scheme dependent. The most
common choice is to work in $\MSbar$ scheme.

\subsection{The Case of Time-like Evolution}

All the considerations that were made for the case of Eq.\eq{fac
f} can also be made for the fragmentation case \eq{fac D}. There
are, however, a few differences between those two cases and we
will discuss them now.

The fragmentation functions $D$ have an interpretation different
from that of the parton distribution functions. The function
$D_{i/h}(z,\mu^2)$ represents the probability density that a
parton $i$ produced at scale $\mu^2$ will fragment to an observed
(and therefore on-mass-shell) hadron $h$. Similarly to the
distribution functions, the evolution of the fragmentation
functions is also described by the DGLAP equations. The one-loop
splitting functions $P_{time-like,~ ij}^{(0)}(x)$ coincide with
those in the space-like case \eq{AP functions}. However the
time-like and the space-like evolution kernels differ beyond the
leading order. The NLO time-like functions can be found in
\cite{pij,pij1}. The large $z$ behavior of the NLO time-like
evolution kernels is the same as for the space-like kernels,
Eq.\eq{large x}.

\section{Infrared Effects}\label{IR Effects}

In our previous discussion we neglected the presence of masses of
the quarks. However for a theory which is sensitive to the IR such
as QCD, a detailed account for those effects is needed.

It is well known that in a gauge theory with massless matter
fields, in addition to the usual UV divergences, there is another
type of divergences that occurs in the evaluation of the Feynman
diagrams. These are known as infrared (IR) divergences and as
shown by Sterman for QCD in \cite{Sterman} can be divided into two
types: collinear and soft.

The collinear divergences are due to the vanishing mass of the
radiating particle (usually the quarks). When a quark radiates a
gluon that is almost collinear to it, the corresponding real or
virtual emission diagrams diverge. One can regulate such a
divergence by introducing a small quark mass or by working in
$D=(4-\epsilon)/2$ space-time dimensions. Then a collinear
divergence shows up as logarithmic singularity $\sim\ln(m^2)$ or
as $1/\epsilon$ pole respectively. In principle, since quarks have
non-vanishing masses, the quantities calculated should be finite
and free of IR collinear singularities. That is not the case,
however. In the perturbative regime it is not the absolute value
of the quark that is important but its value with respect to some
typical scale. If that typical scale - usually the hard scale $Q$
- is much larger than the quark mass, then in a perturbative
calculation there appear large logs $\ln(m^2/Q^2)$. Although these
logs are finite, they appear to any order in perturbation theory
and systematically multiply the strong coupling constant. Thus,
the effective perturbation parameter is not $\as$ anymore but
$\as\times ({\rm a}\ {\rm power}\ {\rm of} \ln m^2/Q^2$). The
latter can be quite large and can even invalidate the perturbation
series. In effect, small but non-zero quark mass leaves the result
finite but unphysical; one should sum up to all orders terms of
this type in order to be able to make definite perturbative
predictions. Such large logs are called quasi-collinear logs and
are classified in the following way: a term at order $\as^n$ has
the form
\begin{equation}
\as^n\sum_{k=0}^{n} c_k\ln^k\left( {m^2\over Q^2}\right).
\label{quasi logs}
\end{equation}
Terms with $k=n$ are known as leading logarithmic (LL) terms, the
ones with $k=n-1$ are the next-to-leading logarithms (NLL) etc. It
is a peculiar feature of QCD that due to its non-abelian gauge
group not only quarks but also gluons can radiate collinear
gluons. Unlike the quarks however the gluons are exactly massless
due to the gauge symmetry.

The origin of the soft divergences is in the vanishing mass of the
gauge fields (the gluon). These divergences manifest themselves as
singularities in the loop integrals over gauge boson lines in the
kinematical region where the energy $\omega_q$ of the gluon
vanishes or in the real emission diagrams where gluons with
vanishing energy are emitted. The most convenient way to
regularize those divergences is to work with dimensional
regularization since it preserves both Poincare and gauge
invariance (unlike gluon-mass regularization).

It was understood long ago \cite{IR QED} that the problem of IR
divergences is rooted in the way the physical observables are
defined. It is intuitively clear that a state containing a hard
parton cannot be distinguished from a state containing in addition
arbitrary number of soft (or collinear) gluons. In those singular
limits the particle nature of the real soft (or collinear) gluon
is not well defined and as a result we need to deal with
degenerate states. The conditions for cancellation of the IR
divergences are stated in the following theorem:

{\bf Kinoshita-Lee-Nauenberg Theorem} \cite{KLN}: In a theory with
massless fields, transition rates are free of IR divergences if a
summation over the initial and final degenerate states is carried
out.

The proof of the theorem can be found in \cite{Muta}. Its content
is however clear: a physical state is one that contains arbitrary
number of soft (collinear) gluons. When applied to perturbative
calculations, the KLN theorem means that to some fixed order in
the coupling constant one should take into account the
contributions from virtual and real emission diagrams with
arbitrary numbers of radiated gluons. Only their sum will be IR
finite. That way we arrive at the idea of an inclusive observable:
a calculated cross-section will be IR finite if it does not
distinguish a state with one particle from a state with a number
of soft (collinear) gluons. We will see examples in the next
Chapter when we discuss the decay of the top quark.

In terms of Feynman diagrams (to all orders), the IR divergences
are generated only from real or virtual emission lines connected
exclusively to external (hard) lines in the diagram. The reason is
that the internal lines are typically highly off-shell and thus
regulate any possible divergence. That observation is important
for constructing an explicit proof of cancellation of the soft
divergences and was used first by Weinberg \cite{Weinberg IR} in
the context of QED. Let us also mention that in fact that property
leads in the context of QCD to the factorization of the IR
singularities in the hard diagrams. This is a very important
property that we will make use of in the next section. For an
excellent discussion see \cite{Catani Grazzini}.

\subsection{Heavy Quark Masses}\label{HQM}

Let us return to Eq.\eq{fac f} or \eq{fac D} and now take into
account the masses of the quarks that we neglected in the
discussion in the previous section. For example, Eq.\eq{fac f
born} now generalizes to:
\begin{equation}
d\hat\sigma_{i\to X}(x,Q^2,m^2,\mu^2) = \sigma_B C\left(x,
{Q^2\over \mu^2},{m^2\over \mu^2},\as(\mu^2) \right).\label{fac f
born mass}
\end{equation}
It is clear that whatever choice we make for the scale $\mu$, we
cannot set equal to one both mass ratios that appear in the RHS of
Eq.\eq{fac f born mass}. Just for the sake of being less abstract,
let us set as before $\mu^2=Q^2$. Then \eq{fac f born mass} takes
the form:
\begin{equation}
d\hat\sigma_{i\to X}(x,Q^2,m^2,\mu^2=Q^2) = \sigma_B C\left(x,
1,{m^2\over Q^2},\as(Q^2) \right).\label{fac f born mass 2}
\end{equation}
Since the ratio $m^2/Q^2$ can take any non-negative real value, we
anticipate a strong dependence on the value of the quark mass. The
case when $m^2 \gg Q^2$ was already discussed in \refsec{Fac Th}:
one can simply integrate out the heavy quark and work in an
effective theory where that flavor is omitted. As a result, the
definition of the coupling constant becomes dependent on the
number of flavors lighter than $Q$ and satisfies non-trivial
conditions at the switching points. To understand the dependence
on $m^2$ in the case when $m^2$ is not (much) larger than $Q^2$,
we first need to know if the cross-section $d\hat\sigma$ is IR
safe, i.e. if it is finite in the limit $m^2\to 0$. If it is
collinearly safe, we can represent it as:
\begin{equation}
C\left(x, 1,{m^2\over Q^2},\as(Q^2) \right) = C\left(x,
1,0,\as(Q^2) \right) + {\mathcal{O}}\left({m^2\over Q^2}
\right).\label{fac f born mass 3}
\end{equation}
Such cross-sections are well behaved and can be obtained by
explicit calculation to any order in perturbation theory. A
typical example is the case when $d\hat\sigma$ is a total partonic
cross-section, e.g. for the process $e^+e^-\to {\rm hadrons}$.

The case when $d\hat\sigma$ is not IR safe is more complicated and
at the same time perhaps more common. In this case the limit $m\to
0$ is singular, i.e. $d\hat\sigma$ diverges. Examples are the
cases where $d\hat\sigma$ is an inclusive differential
cross-section for production of a parton, or a process which is
initiated by a single parton (the generalization to multiple
partons is straightforward). According to \eq{fac}, the hadron
level result is a convolution of the partonic cross-section with
distribution/fragmentation function. However in the presence of IR
divergences we should first understand how to make sense of such
divergent results.

Let us first consider the case when the parton in consideration is
light i.e. we take $m^2\approx 0$. The first thing to note is that
since it is experimentally measurable, the physical process of
creation of a hadron or the process that is initiated by a single
hadron is not IR divergent. Such a process undergoes complicated
stages that are controlled by long-distance physics. However, our
present ignorance about confinement makes us simplify the
calculation by assuming that the produced hadron has been created
from a single parton, which later non-perturbatively hadronizes
(the latter process being described by the fragmentation function
$D$). Such a description is typical for production of light
hadrons (i.e. hadrons that are constructed out of light quarks).
Thus in simplifying the process we necessarily introduce mass
singularities. The understanding of their origin suggests the
method to cure them: one calculates the partonic cross-section in
perturbation theory that is (usually) regulated in dimensional
regularization. As we discussed in \refsec{Fac Th} those
divergences factorize and can be subtracted in a particular scheme
(usually $\MSbar$). Then the subtracted partonic cross-section is
convoluted with a non-perturbative fragmentation function which is
process independent but subtraction scheme dependent. The same
considerations apply for a process with space-like evolution. The
justification for such a procedure is that the subtraction is
physically equivalent to the absorbtion of the effects sensitive
to long-distance physics into the distribution/fragmentation
functions.

After the subtraction of the mass singularities and defining the
corresponding distribution/fragmentation functions we are in the
situation described by Eq.\eq{factorization}:
\begin{equation}
d\sigma(N,Q) = \sigma_B C(N,1,\as(Q^2))
~\exp\left(\int_{\mu_0^2}^{Q^2} {dk^2\over k^2} \gamma(N,\as(k^2)
)\right)~D(N,\mu_0^2)  \label{factorization 2}
\end{equation}
where we again consider the non-singlet case for simplicity. The
function $C$ here describes the ($\MSbar$) subtracted partonic
cross-section while $D$ is an initial condition for the
fragmentation function that is extracted from experiment where
light hadrons are measured at some low scale $\mu_0$.

\subsection{Perturbative Fragmentation Function Formalism}

Let us turn our attention to the case of collinearly-sensitive
processes that involve heavy quarks and typical hard scale $Q$
somewhat larger than the quark mass. As usual, by heavy we mean
$c,b$ or $t$ quarks. Although the partonic cross-section for such
processes is divergent in the zero-mass limit, we cannot really
set the masses of the heavy quarks to zero since they are not that
small. We are then in a situation that we previously described:
the results are formally finite but in practice perturbation
theory cannot be applied in a straightforward way since, as shown
in Eq.\eq{quasi logs}, potentially large logs $\ln (m^2/Q^2)$
appear to all orders in $\as$. There are two physically different
cases where such a situation can occur.

The first is the case of partonic processes that are initiated by
quarks with non-zero mass. Although this is an interesting
problem, it is outside the scope of this Thesis since it mostly
concerns the non-trivial treatment of heavy flavors in the
nucleons. We will only indicate that in such situations one
typically would subtract the divergent part of the coefficient
function and then convolute it with the usual massless parton
density. An example is the case of \cite{strange mass} where the
strange quark was treated as having non-zero mass and after the
subtraction of the quasi-collinear logs the effect of the (finite)
power corrections of the mass of the strange quark was studied.

The second case is the one where a heavy quark (usually $c$ or
$b$) is created in a hard collision with a typical scale $Q$. Such
processes contribute to the creation of $c$- or $b$-flavored
hadrons and can be described within the formalism of the
perturbative fragmentation function (PFF) \cite{Mele Nason}. This
approach has been extensively used for $e^+e^-$ annihilation
\cite{CC,colnas,cagre1,nasole,canas}, hadron collisions
\cite{cagre2,cagre3} and photoproduction \cite{cagre4,cagre1}. In
the next Chapter we describe its application for bottom quark
production in top quark decay $t\to bW$ \cite{cormit}.

According to the factorization theorem \eq{fac}, we can write the
cross-section for creating a heavy-flavored hadron $H$ in the
following way:
\begin{equation}
{1\over {\sigma_0}} {{d\sigma^H}\over{dz}}
(z,Q,m)={1\over{\sigma_0}} \int_{z}^1 {{{d\xi}\over
\xi}{{d\sigma^q}\over {d\xi}}(\xi,Q,m) D_{np}^H\left({z\over
\xi}\right)}, \label{npff gen}
\end{equation}
where $\sigma^H$ ($\sigma^q$) is the cross-section for production
of hadron $H$ (heavy quark $q$) and $D_{np}$ is a non-perturbative
function that describes the transition $q\to H$ at the scale set
by the mass $m$ of the heavy quark $q$. That function is to be
obtained from a comparison with experiment. The kinematical
variable $z$ describes the parameter(s) of the observed final
state and typically is an appropriately normalized energy
fraction. The normalization of the cross-sections is chosen in
such a way that the Born term equals exactly $\delta(1-z)$.

The partonic cross-section $\sigma^q$ can be calculated in
perturbation theory. It is finite because of the finite mass $m$
but is not IR safe since the process is not completely inclusive.
It contains large quasi-collinear logs to all orders in $\as$ that
must be resummed as we previously discussed. The resummation of
those logs is done in complete analogy with Eq.\eq{factorization
2}. One writes:
\begin{equation}
{1\over {\sigma_0}} {{d\sigma^q}\over{dz}} (z,Q,m) = {1\over
\sigma_0} \sum_i\int_{z}^1 {{d\xi}\over \xi}{d\hat\sigma_i\over
{d\xi}}\left({z\over \xi},Q,\mu\right) D_i(\xi,\mu,m), \label{npff
part}
\end{equation}
where $\hat\sigma$ is the cross-section for production of massless
parton $i$ with the collinear singularity subtracted in the
$\MSbar$ scheme. The function $D_i(z,\mu,m)$ is the PFF and it
describes the transition of a massless parton $i$ to a massive
quark $q$.

The ansatz \eq{npff part} has the following physical
interpretation: in a hard collision set by the large scale $Q$ a
massive parton is produced at large transverse momentum. For that
reason it behaves like a massless parton. The replacement of a
massive parton with a massless one after the collinear divergence
is subtracted is justified up to powers of $m^2/Q^2$. The
formalism of the PFF is applicable when such power corrections are
small and can be neglected. For that reason, the cross-section
$\hat\sigma$ is insensitive to low energy (i.e. at the order of
$m$) physics and depends only on $Q$ but not on $m$. The scale
$\mu$ is the factorization scale that, as usual, separates the low
from the high energy regimes. Similarly, the function
$D_i(z,\mu,m)$ depends only on the mass and the factorization
scale, but is insensitive to the high energy part of the process.
In particular, the perturbative fragmentation function is
universal, {\it i.e.} independent of the process.

The PFF satisfy the DGLAP equations \eq{dglap}. The latter can be
solved with NLL accuracy. To completely specify the solution
however, one needs to specify at some scale $\mu_0$ an initial
condition $D^{ini}_i(z,\mu_0,m)$ that is also valid to NLL
accuracy. Clearly, such a condition is also universal and can be
obtained from a perturbative calculation \cite{Mele Nason}. To
that end one first needs to observe that if the scale $\mu_0$ is
chosen of the order of the mass $m$ then no large logs will be
present in the initial condition. Therefore, the initial condition
can be simply calculated in perturbation theory:
\begin{equation}
D^{ini}_i(z,\mu_0,m)=d_i^{(0)}(z)+
{{\alpha_S(\mu_0)}\over{2\pi}}d_i^{(1)}(z,\mu_0,m) +{\cal
O}(\alpha_S^2). \label{d01 gen}
\end{equation}
To obtain the above functions one needs to independently compute
to order $\as$ both $\sigma^q$ and $\hat\sigma$ for some process
and then plug the results into Eq.\eq{npff part}. Comparing the
terms order by order in $\as$ one gets in the $\MSbar$ scheme
\cite{Mele Nason}:
\begin{eqnarray}
d_i^{(0)}(z) &=& \delta_{iq} \delta(1-z),\nonumber\\
d_{i=q}^{(1)}(z,\mu_0,m) &=& C_F \left[{{1+z^2}\over{1-z}}
\left(\ln {{\mu_{0}^2}\over{m^2}}- 2\ln
(1-z)-1\right)\right]_+,\nonumber\\
d_{i=g}^{(1)}(z,\mu_0,m) &=& \half \left( z^2 + (1-z)^2 \right)
\ln\left( {\mu_0^2\over m^2}\right),\nonumber\\
d_{i\neq q,g}^{(1)}(z,\mu_0,m) &=& 0. \label{D ini}
\end{eqnarray}
Process independent derivations of these initial conditions also
exist \cite{CC, ini cond}.

The most convenient way to solve the DGLAP equations is to work
with the Mellin moments of the fragmentation functions. Then the
evolution equations take the following form:
\begin{equation}
{d\over{d\ln\mu^2}}D_i(N,\mu,m)= \sum_j \gamma_{ij}
(N,\alpha_S(\mu)) D_j(N,\mu,m), \label{dglap N}
\end{equation}
where we have introduced $D_i(N)$ through Eq.\eq{Mellin} and the
anomalous dimensions are defined in Eq.\eq{gamma P}. Also, one can
conveniently factorize the dependence on the initial conditions:
\begin{equation}
D_i(N,\mu,m)= \sum_j
E_{ij}(N,\mu,\mu_0)~D^{ini}_j(N,\mu_0,m),\label{solution fac}
\end{equation}
where the functions $E_{ij}(N,\mu,\mu_0)$ are universal solutions
to the DGLAP equation \eq{dglap N} subject to the initial
condition $E_{ij}(N,\mu_0,\mu_0)=\delta_{ij}$. From the
factorization property \eq{Mellin fac} it follows that the
$z$-space counterpart of this condition is
$E_{ij}(z,\mu_0,\mu_0)=\delta_{ij}\delta(1-z)$.

The DGLAP equations \eq{dglap N} couple all flavors that
potentially participate in the process thus making the system hard
to solve in general. An often made approximation is to neglect the
mixing of the heavy flavor $q$ with the light flavors and the
gluon. Such an approximation is justified in situations where the
probability to create a heavy quark from a secondary gluon is
small. The approximation amounts to considering only the
non-singlet contributions to the evolution kernels. We will
consider only this case hereinafter. In this approximation it is
very easy to solve Eq.\eq{dglap N} with NLL accuracy:
\begin{equation}
E_q(N,\mu,\mu_0) = \exp\Bigg\{P_N^{(0)}t + {1\over{4\pi^2b_0}}
\left[ \alpha_S(\mu_{0})-\alpha_S(\mu) \right]
\left[P_N^{(1)}-{{2\pi b_1}\over {b_0}}P_N^{(0)}\right]\Bigg\},
\label{dresum E}
\end{equation}
where we have introduced the variable $t$ as
\begin{equation}
t={1\over{2\pi b_0}}\ln {{\alpha_S(\mu_{0})}\over{\alpha_S(\mu)}}.
\end{equation}
The coefficients $b_{0,1}$ are defined in \eq{beta NLO} and the
functions $P_N^{(0,1)}$ are the Mellin transforms of the LO and
NLO coefficients in \eq{split}. Their explicit form as well as
$N$-moments of the initial conditions \eq{D ini} are given in the
appendix. Collecting the results in Eqns.\eq{npff gen}, \eq{npff
part}, \eq{solution fac}, \eq{dresum E} and  \eq{D ini N}, we can
write the differential cross-section for production of a
heavy-flavored hadron as:
\begin{equation}
{1\over {\sigma_0}} {{d\sigma^H}\over{dz}} (z,Q,m)=
{\mathcal{M}}^{-1}\left[ C(N,Q,\mu)~
E_q(N,\mu,\mu_0)~D^{ini}_q(N,\mu_0,m) D_{np}^H(N)\right](z).
\label{sigmaH}
\end{equation}
With ${\mathcal{M}}^{-1}$ we denote the inverse Mellin transform:
\begin{equation}
f(z) = {\mathcal{M}}^{-1}[f(N)](z) = {1\over 2\pi i}
\int_{C-i\infty}^{C+i\infty} dN z^{-N} f(N), \label{Mellin inv}
\end{equation}
which is typically performed numerically. We will give more
details in the next Chapters where we discuss our applications. In
Eq.\eq{sigmaH} we have also introduced the $N$-space coefficient
function:
\begin{equation}
C(N,Q,\mu) ={1\over{\sigma_0}} \int_{0}^1 dz~ z^{N-1}
{{d\sigma^q}\over {dz}}(z,Q,\mu), \label{coeff N}
\end{equation}
and the non-perturbative fragmentation function $D_{np}(N)$. The
latter is extracted from fits to experimental data. We will
present details for how that procedure works in the next Chapter
where we consider the case of top decay.

Eq.\eq{sigmaH} is the improved cross-section for production of a
hadron $H$ carrying longitudinal momentum fraction $z$, and with
all quasi-collinear logs $\ln (m^2/Q^2)$ resummed with NLL
accuracy. As we will also see in the next Chapter in the example
of top quark decay, the application of the resummation via the
perturbative fragmentation function represents a serious
improvement over the fixed order perturbative calculation.

\subsection{Soft-gluon Threshold Resummation}\label{secSGR}

Before we move to the discussion of our processes of interest, let
us describe another, more subtle source of large logarithmic
corrections that arise to all orders in perturbation theory and
also need to be resummed. Those logs appear in processes
containing massless and/or massive particles and are also of IR
origin.

Let us go back to Eq.\eq{factorization}. Since no masses were
present there, we concluded that such a renormalization group
improved cross-section does not contain any large parameters and
therefore can be unambiguously computed. We did not consider
however its behavior as a function of the dimensionless parameter
$N$, the Mellin conjugate of $z$. This parameter can be large
since the limit of large $z$ (i.e. $z\to 1$) corresponds through
the Mellin transform \eq{Mellin} to $N\to \infty$. Therefore in
the kinematical limit of large $z$ one may expect terms
$\sim\ln(N)$ to appear in the conjugate space and to spoil the
perturbative expansion in $\as$. Indeed, a typical perturbative
calculation of a hard cross-section shows terms like
\begin{equation}
\as^n\sum_{m=0}^{2n} c_m\ln^m(N). \label{soft logs}
\end{equation}
The terms with $n<m\leq 2n$ are called leading logs (LL), the ones
with $m=n$ are next-to-leading (NLL) logs and so on
\footnote{ We will give more convenient characterization of these
logs below; see Eq.\eq{eik sigma exp}.}.
An example is the large $N$ behavior of both space- and time-like
anomalous dimensions shown in Eq.\eq{large x} where the latter are
as singular as a single power of $\ln(N)$. Relations between the
large $z$ and large $N$ behavior of various relevant functions are
given in the appendix.

The physical origin of the above mentioned (soft) logs is in the
soft gluon radiation near the boundary of phase space. As we
discussed in \refsec{IR Effects}, the soft divergences cancel when
all the contributions from real and virtual radiation are taken
into account. However, this cancellation is incomplete near the
boundary of the phase space $z\to 1$ for semi-inclusive
cross-sections, and as a result large terms $\sim
\ln^k(1-z)/(1-z)$ are present in the perturbative calculation.

The all order resummation of classes of such soft logs is
performed in terms of the Mellin moments because in Mellin space
the factorization properties of the amplitudes are explicit. We
will consider cross-sections of the type \eq{sigmaH}, where $D$
can be a general fragmentation or distribution function. The form
of Eq.\eq{sigmaH} suggests the following approach for the
resummation procedure: one can obtain separate resummed
expressions for each of the factors appearing in the RHS of
\eq{sigmaH} that require such resummation. This way, one
explicitly separates the process-dependent from the universal
factors which makes the resummation procedure more universal. Of
course, this is not the only possibility; one can solve directly
for the whole cross-section. An example is the large-$z$
resummation performed in \cite{Mele Nason}.

In what follows, we will describe the method of Catani and
Trentadue \cite{CT} originally applied to Drell-Yan and DIS. We
will apply their method in the next two chapters for the
calculation of the resummed expressions for the coefficient
functions for top-quark decay and inclusive CC DIS. The
equivalence of their result with the Sterman's \cite{Sterman soft}
was shown in \cite{CT 2}.

The coefficient function represents the normalized cross-section
for production of a massless quark after the corresponding
collinear singularities are subtracted. Up to non-singular terms
in the limit $z\to 1$, it is approximated by the eikonal
cross-section \cite{CT}:
\begin{equation}
d\sigma = \sigma_B\overline{\sum_{n=0}} \int d\phi^{(n)}\langle
n\vert\hat{S}\vert0\rangle|^2, \label{eikonal crosssec}
\end{equation}
where $\int d\phi^{(n)}$ is the (partially integrated) phase space
for $n$-soft gluons, $\vert n\rangle$ is a state with n real soft
gluons and $\hat{S}$ is the (nonlocal) eikonal operator:
\begin{equation}
\hat{S} = TP \exp\left[ -ig_S\int_C dx^\mu
A_\mu\right].\label{eikonal op}
\end{equation}
The contour $C$ is the classical path of the radiating hard
partons and $A$ is the gluon field in the Heisenberg picture. $T$
is a time-ordering operation for the operators $A$ and $P$ denotes
path-ordering for the color matrices along to contour $C$. From
the unitarity property
\footnote{For the applications we consider in this Thesis it is
satisfied, since the unitarity of the eikonal operators is
explicit in the case when every two points of the contour $C$ are
separated by a time-like interval, i.e. the contour is
time-ordered; see \cite{eikonal}.}
of the eikonal operators \cite{eikonal}, it follows that the total
eikonal cross-section equals exactly the leading order
cross-section, {\it i.e.} the higher order contributions vanish
(in particular that also implies cancellation of the soft-gluon
divergences). For example at order $\as$:
\begin{equation}
\omega^{(0)} + \omega^{(1)} = 0\label{unit as1},
\end{equation}
where $\omega^{(i)}$, $i=0,1$ is the total probability for virtual
($i=0$) and real ($i=1$) soft gluon emission.

One can evaluate the eikonal cross-section to some fixed order in
$\as$ by working in the eikonal approximation. In this
approximation, the emission of a soft gluon with momentum $q$ from
a hard quark with momentum $p$ is described by vertex $2p_\mu$ and
with eikonal propagator $1/2p.q$. An important property of the
eikonal approximation is that real gluon emission factorizes
\cite{Catani Grazzini}. To order $\as$ the amplitude for a real
soft-gluon emission in a process involving $k$ hard particles is
proportional to the $k$-particle hard amplitude. In general
however, there are color correlations involved. In the case
$k=2,3$ one can show that the correlations are absent since the
product of color matrices can be expressed completely in terms of
the Casimir \eq{CFCA} \cite{Catani Grazzini}. For the purposes of
our applications we will only need the result for the case of two
hard partons (quarks):
\begin{equation}
d\omega^{(1)} = -g_S^2{ d^3q\over (2\pi)^3 2q_0} C_F\vert
J(q)\vert^2, \label{one loop fac}
\end{equation}
where $J^\mu$ is the eikonal current:
\begin{equation}
J^\mu = {p_1^\mu\over p_1.q} - {p_2^\mu\over p_2.q}, \label{J}
\end{equation}
and $p_{1,2}$ are the momenta of the two hard quarks that
constitute the path $C$ in Eq.\eq{eikonal op}. In case the hard
partons are gluons the color factor in Eq.\eq{one loop fac} should
be replaced with $C_A$.

Collecting the results above, one can easily obtain the eikonal
coefficient function to order $\as$:
\begin{equation}
C^{soft}(z) = {1\over \sigma_B}{d\sigma\over dz} =
\left(1+\omega^{(0)}\right)\delta(1-z) + \int
d\omega^{(1)}\delta\left(1-z-{q_0\over p_0}\right) +
{\mathcal{O}}(\as^2), \label{C eik 1}
\end{equation}
where $q_0$ is the energy of the soft gluon and $p_0$ is the c.m.
energy of the hard parton. To obtain the all-order result valid up
to NLL accuracy one has to take the Mellin moment of Eq.\eq{C eik
1} and then to exponentiate the $\as$ term. The resulting integral
typically takes the following form:
\begin{eqnarray}
\ln C^{soft}(N) &=& {{C_F}\over\pi}\int_0^1 {dz
{{z^{N-1}-1}\over{1-z}}} \left\{ \int_{\mu_F^2}^{(1-z)^aQ^2}
{{dq^2}\over {q^2}} \alpha_S(q^2)\right. \nonumber\\
&+& J\left[\alpha_S((1-z)^aQ^2)\right]\Bigg\}. \label{typical}
\end{eqnarray}
To put Eq.\eq{typical} in its present form, we have introduced the
variable $z=1-q_0/p_0$ (as follows from the argument of the
$\delta$-function in \eq{C eik 1}) and have worked in spherical
co-ordinates conveniently determined by the hard momenta. Two of
the integrations over the phase-space of the radiated gluon are
trivial. The one remaining integration is over the azimuthal angle
which in turn can be related to the transverse momentum of the
radiated gluon. The latter is known to set the scale of the
running coupling in the soft region \cite{ABCMV}, \cite{BCM} and
is an important leading order effect. More details on these
relations will be presented when we consider our applications. The
origin of the $-1$ term in Eq.\eq{typical} is in the virtual
emission term $\omega^{(0)}$ after \eq{unit as1} is used. The
number $a$ is process dependent; typically $a=1,2$. The
kinematical constraints imply that the lower limit of the $q^2$
integration is zero. It is convenient to regularize the low-$q^2$
collinear divergence by introducing a cut-off. The (singular)
dependence on the cut-off is later absorbed into the partonic
distribution/fragmentation function as was discussed in
\refsec{HQM}. As a result, in the $\MSbar$ factorization scheme
the lower limit is set by the factorization scale $\mu_F^2$
\cite{CC}, \cite{cmw}.

The function $J$ is process dependent and has a perturbative
expansion in $\as$. To NLL accuracy, only the ${\mathcal{O}}(\as)$
term participates. That function receives two contributions: the
first one is from the collinearly non-enhanced term of the square
of the eikonal current, after the $q^2$ integration has been
explicitly performed. Up to NLL accuracy one can use the following
result:
\begin{equation}
{1\over{Q^2(1-z)^a}} \int_0^{Q^2(1-z)^a}{dq^2\alpha_S(q^2)}=
\alpha_S\left(Q^2(1-z)^a\right),~~~a=1,2. \label{asint}
\end{equation}
To show Eq.\eq{asint}, one needs to use the running of the strong
coupling (see Eq.\eq{alpha s ii}) and then to carefully identify
and collect the NLL contributions. Both in \eq{asint} and
\eq{typical}, $Q$ stands for some process-dependent hard scale.

To specify the second contribution to the function $J$, we need to
first explain the physical origin of the various enhanced
contributions in the soft limit. From kinematical considerations
it is clear that in the limit $z\to 1$, typically, the initial
partons can emit only soft gluons. The hard partons in the final
state however may or may not be subject to that constraint. In the
case of inclusive DIS for example, the hard parton in the final
state is completely free to fragment with the only restriction
being on the virtuality of the final jet. If however the final
parton is observed (and considered as massless as is appropriate
for a coefficient function) it can only radiate soft gluons. Since
the eikonal approximation handles only the contribution from soft
gluon radiation (that itself can be collinear), one needs to
specify in addition to the eikonal cross-section the contribution
from the fragmentation region. Such a contribution was evaluated
in \cite{CT} after the soft-gluon radiation was completely
decoupled from the final massless quark (with appropriate choice
of the vector specifying the physical gauge). The jet function
that incorporates the effect of the collinear radiation in the
fragmenting jet was obtained from a modified evolution equation.
We will return to that point later when we consider our
applications.

Finally, the results to NLL order can be completed by introducing
an additional ${\mathcal{O}}(\as^2)$ contribution to the RHS of
Eq.\eq{typical}. One needs to make the replacement:
\begin{equation}
{{C_F}\over\pi}{{\alpha_S(q^2)}\over{q^2}} \to
{{A\left[\alpha_S(q^2)\right]}\over {q^2}}, \label{afun}
\end{equation}
where the function $A(\alpha_S)$ can be expanded as follows:
\begin{equation}
A(\alpha_S)=\sum_{n=1}^{\infty}\left({{\alpha_S}\over
{\pi}}\right)^n A^{(n)}. \label{function A}
\end{equation}
The first two coefficients are needed at NLL level and are given
by~\cite{CT,kt}:
\begin{equation}
A^{(1)}=C_F,
\end{equation}
\begin{equation}
A^{(2)}= {1\over 2} C_F \left[ C_A\left(
{{67}\over{18}}-{{\pi^2}\over 6}\right) -{5\over 9}n_f\right].
\end{equation}
A numerical estimate for $A^{(3)}$ is known too \cite{nv} (see
also \cite{vogt}). The function $A$ represents the coefficient of
the singular term of the anomalous dimension \eq{large x}.

After the replacement \eq{afun}, Eq.\eq{typical} contains all NLL
contributions. Its explicit evaluation would, however, contain
subleading terms as well. One can evaluate explicitly the
integrals and extract only the LL and NLL terms. After that is
performed, the eikonal cross-section takes the form:
\begin{equation}
C^{soft}(N,Q,\alpha_S(\mu^2),\mu,\mu_F)=\exp\left[\ln N
g^{(1)}(\lambda)+ g^{(2)}(\lambda,Q,\mu,\mu_F)\right]\, \label{eik
sigma exp}
\end{equation}
where we have specified all arguments, restored the separate
dependence on the renormalization and factorization scales, and
introduced:
\begin{equation}
\lambda=b_0\alpha_S(\mu^2)\ln N. \label{lambda}
\end{equation}
The functions $g^{(1)}$ and $g^{(2)}$ contain all LL
($\ln(N)(\as\ln(N))^n$) and NLL ($(\as\ln(N))^n$) contributions
respectively. The results at NNLL order and beyond can be also
written in terms of functions
$\as^k(N)g^{(k+2)}(\lambda),~~k=1,\dots$ (see e.g. \cite{aglric}).
To evaluate the integrals in \eq{typical} with NLL accuracy, one
makes use of the running coupling Eq.\eq{alpha s ii} and the
following identity, valid to NLL \cite{CT}:
\begin{equation}
z^{N-1}-1\to -\Theta\left( 1-{{e^{-\gamma_E}}\over N}-z\right).
\label{NLL eval}
\end{equation}
We would like to emphasize the fact that the use of the
replacement \eq{NLL eval} makes it possible to evaluate the
integral in \eq{typical}. It is clear that the integrals in
\eq{typical} are divergent for any value of $N$ simply because of
the presence of the Landau pole in the running coupling $\as$; see
Eq.\eq{alpha s}. The role of the replacement \eq{NLL eval} is that
it extracts finite values from the integrals in \eq{typical} up to
NLL accuracy, while neglecting subleading terms that are formally
divergent. Consistency is however, preserved because of the
neglecting of similar subleading terms in the derivation of
\eq{typical}. A detailed description can be found in \cite{cmnt}
and \cite{BB}.

So far we discussed only the evaluation with NLL accuracy of the
coefficient function. However, the ${\mathcal{O}}(\as)$ expression
for the initial condition of the quark initiated perturbative
fragmentation function \eq{D ini} also shows enhancement in the
limit $z\to 1$. The resummation of the soft-gluon contribution to
the (universal) initial condition for the PFF was performed in
\cite{CC} with NLL accuracy. We shall only present the final
result here:
\begin{eqnarray}
\ln D^{ini}_q(N,\mu_0^2,\mu_{0F}^2,m^2,\as(\mu_0^2) &=& \int_0^1
{dz {{z^{N-1}-1}\over{1-z}}} \left\{\int_{(1-z)^2m^2}^{\mu_{0F}^2}
{{dq^2}\over {q^2}} A\left[\alpha_S(q^2)\right]\right. \nonumber\\
&+& H\left[\alpha_S\left((1-z)^2m^2\right)\right]\Bigg\}.
\label{deltares}
\end{eqnarray}
Here $m$ is the mass of the heavy quark and the function $H$ also
has a perturbative expansion in $\as$:
\begin{equation}
H(\alpha_S)=\sum_{n=1}^{\infty}\left({{\alpha_S}\over
{\pi}}\right)^n H^{(n)}.
\end{equation}
At NLL level, we are just interested in the first term of the
above expansion:
\begin{equation}
H^{(1)}=-C_F.
\end{equation}
The explicit evaluation of the initial condition \eq{deltares} to
order NLL gives:
\begin{equation}
D_q^{ini} = \exp\left[\ln N g^{(1)}_{ini}(\lambda_0)+
g^{(2)}_{ini}(\lambda_0)\right]\, \label{D ini exp}
\end{equation}
where we have suppressed the arguments for brevity, and
$\lambda_0$ is given in \eq{lambda} with the replacement
$\mu^2\to\mu_0^2$. The functions $g_{ini}^{(1,2)}$ are:
\begin{eqnarray}
g^{(1)}_{ini}(\lambda_0)&=& -\frac{A^{(1)}}{2\pi b_0 \lambda_0} \;
[ 2\lambda_0 + (1-2\lambda_0) \ln (1-2\lambda_0)] \;,\\
g^{(2)}_{ini}(\lambda_0) &=& \frac{A^{(1)}}{2 \pi b_0}\left[\ln
\frac{\mu_{0F}^2}{m^2}
+ 2\gamma_E\right] \ln(1-2\lambda_0)\nonumber\\
&-&\frac{A^{(1)}  b_1}{4 \pi b_0^3} \left[ 4\lambda_0 + 2 \ln
(1-2\lambda_0) +
\ln^2 (1-2\lambda_0) \right]\nonumber\\
&+& \frac{1}{2\pi b_0} \left[2\lambda_0 + \ln (1-2\lambda_0)
\right] \left(\frac{A^{(2)}}{\pi b_0} +
A^{(1)}\ln\frac{\mu_0^2}{\mu_{0F}^2}\right)
\nonumber\\
&+& \frac{H^{(1)}}{2\pi b_0} \ln (1-2\lambda_0).
\end{eqnarray}

So far we discussed how one can evaluate the approximate
expressions for the coefficient function and the initial condition
of the PFF in the limit $z\to 1$. However, to obtain a result
equally applicable for all $z$, one needs to give a matching
prescription for the fixed order result and its approximation in
the soft limit. In this Thesis we will use the following matching
procedure: one adds the fixed order result and the result valid in
the soft limit together, and then subtracts what they have in
common. Note that for consistency, the decomposition of the result
valid in the soft limit in powers of $\as$ should coincide with
the soft limit of the fixed order result. We will give details for
that procedure in the next Chapters.

%
%%%%%%%%%%%%%%%%%%%%%%%%%%%%%%%%%%%%%%%%%%%%%%%%%%%%%%%%%%%%%%%%%%
\chapter{$b$-quark Fragmentation in $t$-quark Decay}

In this Chapter we present our original results for
$b$-fragmentation in top decay. In their derivation we apply
several techniques within perturbative QCD such as the fixed order
calculation and resummation of large quasi-collinear and soft
logs. Those techniques were discussed in the previous Chapter. We
also extract the non-perturbative fragmentation function from fits
to $e^+e^-$ data and thus make a prediction for the spectrum of
$b$-flavored hadrons in top decay. This Chapter is based on the
papers \cite{cormit} and \cite{CCM}.

\section{Motivation: Top Physics}

The existence of the top quark is a prediction of the Standard
Model. It was discovered at the Fermilab Tevatron \cite{top} and
its mass was measured \cite{top mass} to be $174.3\pm 5.1\GeV$.
This large mass applies, among other things, a large coupling to
the Higgs boson (the top-Yukawa coupling $g_{ttH}\sim 1$). Unlike
the other five quarks, top's CKM matrix elements are dominated by
a single element -- $V_{tb}$ -- which is very close to one. As a
result top decays almost exclusively to a real $W$ boson and a
$b$-quark. Its electric charge (as predicted by the SM) is
$+(2/3)e$ but is not yet measured directly \cite{topcharge}. The
present status of these and other top-quark properties can be
found in the extensive recent review \cite{top review}.

At present, and for the foreseeable future, the study of processes
involving the top quark is of great interest. On the experimental
side, because of its very large mass, the top is still a challenge
to present experimental facilities. The CDF and D\O\
collaborations in Run I at the Fermilab Tevatron produced a small
but sufficient number of top events to directly prove its
existence in 1995 \cite{top} and to pinpoint its mass. Since then
the Tevatron has been upgraded and Run II is presently collecting
data. The expectations are that during Run II more detailed
measurements of the top quark properties will be possible. In the
near future (around 2007-8) the LHC proton-proton collider is
expected to start operation. Due to its much larger energy of
operation it is expected to produce vast numbers of top quarks
with a corresponding improvement in top studies \cite{lhc}. There
is another high-energy machine -- the electron-positron Linear
Collider \cite{lc} -- the prospects of building which are being
studied at present. Such a machine will not produce as many top
events as LHC, but will have the potential for precise
measurements of the top quark parameters.

On the theoretical side, the large mass of the top quark makes it
decay very fast to a $b$-quark and a $W^+$ boson. That implies a
very large top decay width, the NLO value being
$\Gamma_{top}\approx 1.4\GeV$. The corresponding life-time is
about an order of magnitude shorter than the typical hadronization
times set by the scale $\L$. An immediate consequence is that the
top quark cannot form $t$-flavored hadrons; it simply decays
before it is able to hadronize \cite{top hadron}. For that reason
the top behaves much like a free particle. That property of the
top as well as the very small ratio of its width to its mass
suggest that it is reasonable for most purposes to assume that top
production and top decay factorize to a good approximation.

It is not only the study of the SM that draws attention to the top
quark. As we mentioned, the top has very large Yukawa coupling,
i.e. its coupling to the Higgs sector is largest among all known
particles. For that reason it is reasonable to expect that new
physics that may be related to the mechanism of electroweak
symmetry breaking may first show up in the top sector. For a more
detailed account of such models see \cite{top review}.

Clearly, a precise knowledge of the properties of the top quark is
necessary not only in order to verify experimentally the SM
predictions but also to be able to discriminate non-SM effects
should they be discovered in forthcoming experiments. However,
because the top has such a small life-time, the only way one can
get any information about the top itself is through its decay
products. Clearly, this requires -- among other observables -- a
reliable description of the fragmentation of the $b$-quark in
top-quark decay.

As shown in \cite{tev}, $b$-fragmentation is indeed one of the
sources of uncertainty in the measurement of the top mass at the
Tevatron. At the LHC, recent studies \cite{avto} have suggested
that final states with leptons and a $J/\psi$, with the $J/\psi$
coming from the decay of a $b$-flavored hadron and the isolated
lepton from the $W$ decay, will be a promising channel to
reconstruct the top mass. In \cite{avto}, the expected
experimental error, including statistics and systematics, has been
estimated to be $\Delta m_t\simeq 1$~GeV and the $b$ fragmentation
is the largest source of uncertainty, accounting for about
0.6~GeV.

In this Chapter we will study the $b$-spectrum in top decay in the
following framework: neglecting the interference effects between
top production and decay, one has
\begin{equation}
{1\over{\sigma}}{{d\sigma}\over {dx_b}}={1\over {\Gamma}}
{{d\Gamma}\over{dx_b}}. \label{top factorization}
\end{equation}
In the above equation, $(1/\sigma) d\sigma/dx_b$ is the normalized
differential cross section for the process of production of a $b$
quark with energy fraction $x_b$ (defined in Eq.\eq{xb}) via top
quarks for any production mechanism. Similarly, $(1/\Gamma)
d\Gamma/dx_b$ is the normalized differential distribution of
$b$-quark in the decay of a real top quark. Our purpose in this
Chapter is to give a precise prediction for the distribution on
the RHS of Eq.\eq{top factorization}; through the relation \eq{top
factorization} our results will then be applicable to top quarks
produced in $p\bar p$ (Tevatron), $pp$ (LHC) or $e^+e^-$ (Linear
Collider) collisions.

We start by first calculating in \refsec{sec NLO QCD} the
differential top width with respect to the $b$-energy fraction in
NLO QCD taking full account of the mass of the $b$ quark. Then in
\refsec{sec top PFF} we resum with NLL accuracy the large
quasi-collinear logs $\ln(m_b^2/m_t^2)$. Applying soft-gluon
resummation techniques in \refsec{sec soft} we improve the
prediction for the $b$-energy spectrum at large $x_b\to 1$. Then
we discuss the improvement to the fixed order calculation due to
the various resummations. Finally, in \refsec{sec top np} we
extract the non-perturbative fragmentation function using $e^+e^-$
data to make predictions for the energy spectrum of $b$-flavored
hadrons in top decay.

\section{$t\to b W$ in NLO QCD}\label{sec NLO QCD}

In this section we present our results for the NLO QCD calculation
of the differential width:
\begin{equation}
{1\over{\Gamma_0}}{{d\Gamma_b}\over {dx_b}}, \label{width}
\end{equation}
for the Standard Model top-quark decay:
\begin{equation}
t(q,m_t)\to b(p_b,m_b)W(p_W,m_W)\left( g(p_g,0) \right),
\label{dec}
\end{equation}
of an on-shell top quark at next-to-leading order in $\alpha_S$,
with respect to the energy fraction of the produced $b$-quark
(defined below). We assume that the decay process \eq{dec} is the
only decay mode of the top quark. That assumption is consistent
with recent measurements of the CDF Collaboration of the ratio
$R=B(t\to Wb)/B(t\to Wq)$, where $q$ is a $d$, $s$ or $b$ quark,
and the subsequent extraction of the Cabibbo--Kobayashi--Maskawa
matrix element $V_{tb}$ \cite{vtb}. We present the results from
two independent calculations: when the exact dependence on the
bottom mass is kept, thus regulating the collinear divergences,
and when the bottom mass is set to zero from the outset. In the
latter case additional collinear divergences occur. In all our
calculations we use dimensional regularization to regulate the UV
and all IR divergences, that is, we perform the calculations in
$D=4-2\epsilon$ space-time dimensions (see appendix). We use the
on-shell renormalization scheme to remove the UV divergences. Let
us note that in principle one does not need to calculate
Eq.\eq{width} both in the $m_b=0$ and $m_b\neq 0$ cases; one can
infer the massless result for the normalized distribution
\eq{width} (with the collinear singularity subtracted in the
$\MSbar$ scheme) from the massive one, using the initial
conditions for the PFF, Eq.\eq{D ini}. In fact with our
independent calculation of those two cases we were able to
reproduce those initial conditions in the case of top-quark decay.

\subsection{ Calculation with $m_b\neq 0$}

In this section we will present in more detail the calculation in
the case when the full dependence on the $b$-mass is kept. Let us
first introduce our notation:
\begin{eqnarray}
b&=& {{m_b^2}\over {m_t^2}},\nonumber\\
w&=& {{m_W^2}\over {m_t^2}},\nonumber\\
s&=&{1\over 2} (1+b-w),\nonumber\\
\beta &=& {\sqrt{b}\over s},\nonumber\\
Q&=& s\sqrt{1-\beta^2},\nonumber\\
G_0&=&{1\over 2}\left[ 1+b-2w+{{(1-b)^2}\over w}\right],\nonumber\\
\Phi (x_b)&=&s\left[\sqrt{x_b^2-\beta^2}- {\mathrm{arcth}} \left(
{{\sqrt{x_b^2-\beta^2}}\over {x_b}}\right)\right],
\label{notations}
\end{eqnarray}
where:
\begin{equation}
x_b={{x_E}\over {x_{E,\mathrm{max}}}}={{{ p_b\cdot p_t}\over
{m_t^2s}}} \ \ \ , \ \ \ \beta\leq x_b \leq 1.\label{xb}
\end{equation}
It is obvious that $x_b$ is an invariant, which in the top rest
frame coincides with the energy of the $b$-quark, normalized with
respect to its maximum value. The LO width $\Gamma_0$ is:
\begin{equation}
\Gamma_0={{\alpha m_t|V_{tb}|^2}\over{16\sin^2\theta_W}}~4QG_0,
\label{LO width m}
\end{equation}
where $\alpha$ is the electromagnetic coupling constant and
$\theta_W$ is the Weinberg angle.

Next we proceed with the details of the calculation. At order
${\mathcal{O}}(\as)$ one has to calculate the contributions from
the diagrams in Figures \eq{born level}, \eq{virt level} and
\eq{real level}.
\begin{figure}[ht!]
\centerline{\epsfig{file=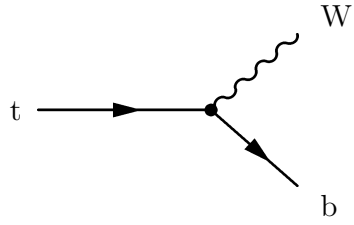,height=1.2in,width=1.9in}}
\caption{\footnotesize Tree level (Born) diagram.} \label{born
level}
\end{figure}
\begin{figure}[ht!]
\centerline{\epsfig{file=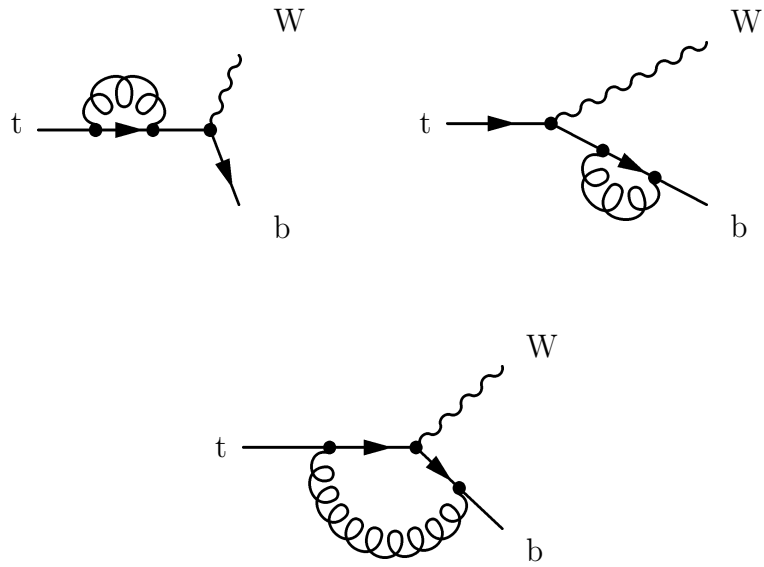,height=3in,width=4in}}
\caption{\footnotesize Contributing virtual emission diagrams.}
\label{virt level}
\end{figure}
\begin{figure}[ht!]
\centerline{\epsfig{file=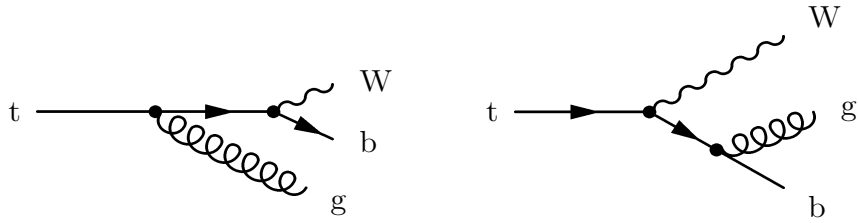,height=1.2in,width=4.5in}}
\caption{\footnotesize Contributing real emission diagrams.}
\label{real level}
\end{figure}
To obtain the contribution from the real emission diagrams --
Fig.\eq{real level} -- one has to add the two matrix elements and
square them. After averaging/summing over spins and colors (all
the needed details can be found in the appendix) one gets the
following expression for the square of the matrix element:
\begin{eqnarray}
\overline{\vert M_{real}\vert^2} &=& {2\over 3} g^2g_S^2\vert
V_{tb}
\vert^2 \left[ {m_t^4\over \pk\qk}(2sG_0-\epsilon 4s^2)\right.\nonumber\\
&+& {1\over \pk^2}\left( -{1+b\over w}\pk^2 +\left( 2 + {1+b\over
w}\right)\pk\qk \right. \nonumber\\
&+& \left. 2\pk m_t^2 G_0 +m_t^4( -G_0 + \epsilon
2s)\right) \nonumber\\
&+& {1\over \qk^2}\left( -{1+b\over w}\qk^2 +\left( 2 + {1+b\over
w}\right)\pk\qk \right.\nonumber\\
&-&  2\qk m_t^2 G_0 +m_t^4( -bG_0 + \epsilon 2bs)\Bigg)\Bigg],
\label{M real}
\end{eqnarray}
in terms of the scalar products of the four vectors. We have kept
those terms linear in $\epsilon$ that will give finite
contributions after the singular integration over the phase space
is performed. The next step is to integrate the matrix element
\eq{M real} over the three-body phase space. More details about
that procedure are relegated to the appendix. As a result, we
obtain the following contribution from the real emission diagrams
to the differential width \eq{width} at order $\as$:
\begin{eqnarray}
{1\over{\Gamma_0}}{{d\Gamma_{real}^{(\as)}}\over{dx_b}}&=&
{{C_F\alpha_S(\mu)}\over{\pi Q}}\left\{ \left[ \Phi(x_b=1)\left(
{1\over \epsilon} -2\gamma_E+2-\ln\left({Q^2m_t^2\over
4\pi^2\mu^2}\right)\right)\right.\right.\nonumber\\
&+& s\left(  \Sp{ {{2Q}\over{1-s+Q}} } -
\Sp{ {{2Q}\over{s-b+Q}} } \right. \nonumber\\
&-&\left. \Ln{ {s+Q\over \sqrt{b}} } \left(\Ln{ {{s+Q}\over
{\sqrt{b}}} } + 2\Ln{ {{1-s+Q}\over{2s(1-\beta)}} }
\right)\right) \nonumber\\
&+& \Ln{ {s+Q\over\sqrt{b}} }\left(s-b+2{s^2\over G_0}\right) +
\Ln{ {{1-s+Q}\over{\sqrt{w}}} }(1-b) \label{Gamma real}\\
&+& \left. 2Q \left( \Ln{ {\sqrt{w}\over 2s(1-\beta)} } -
{s\over G_0}\right)\right]\delta(1-x_b) \nonumber\\
&-&2\ {\Phi (x_b)}\left[ {1\over {(1-x_b)_+}}+
{s\over{G_0}}\left(1+{{1+b}\over{2w}}\right)(1-x_b) -
1\right]\nonumber\\
&+&\left. 2s\sqrt{x_b^2-\beta^2}\left[ 2{s^2\over G_0}\left(
{{1-x_b}\over{1-2sx_b+b}}\right) + {s\over{G_0}}\left(1+{{1+b}
\over{2w}}\right)(1-x_b) - 1\right]\right\}\nonumber
\end{eqnarray}

As expected, the real emission contribution to the differential
distribution contains an explicit soft divergence $\sim
1/\epsilon$. The latter will be cancelled after we add the
contribution from the virtual diagrams. Let us explain their
evaluation.

The diagrams with virtual gluon emission that contribute to order
$\as$ are shown in Fig.\eq{virt level}. They represent the same
physical process as the leading order (Born) diagram in
Fig.\eq{born level}. Because the virtual diagrams contain loop
integrals, they are UV divergent. We regularize these divergences
by working in $D$ space-time dimensions. The cancellation of the
UV divergences requires an appropriate introduction of
counterterms. The counterterms are determined from the on-shell
condition:
\begin{equation}
\Sigma({\not\! p}=m)=0~,~~{\partial\Sigma({\not\! p}=m)\over
\partial {\not\! p}}=0,\label{self energy}
\end{equation}
where $\Sigma$ is the quark self-energy for quark $q$ with mass
$m$. From the second condition in \eq{self energy}, one can derive
the quark wave function renormalization constant (it coincides
with the one in \cite{wid2}):
\begin{equation}
Z_2^{(q)} = 1-{\as\over 3\pi} \left( {1\over
\epsilon_{UV}}+{2\over\epsilon_{IR}} +4 -3\gamma_E -3
\Ln{{m^2\over 4\pi\mu^2}}\right).\label{Z quark}
\end{equation}
This then determines the vertex renormalization constant:
\begin{equation}
Z^{(\Lambda)} = 1-{\as\over 3\pi} \left( {1\over
\epsilon_{UV}}+{2\over\epsilon_{IR}} +4 -3\gamma_E -{3\over 2}
\Ln{{m_t^2\over 4\pi\mu^2}} -{3\over 2} \Ln{{m_b^2\over
4\pi\mu^2}}\right).\label{Z vertex}
\end{equation}
The renormalized vertex thus becomes:
\begin{equation}
\Lambda_\mu^{reg} = \Lambda_\mu -i\left(
Z^{(\Lambda)}-1\right)\gamma_\mu. \label{ren Lambda}
\end{equation}
The function $\Lambda_\mu$ contains the higher order contributions
to the vertex amplitude, and is defined through the relation:
$M=g/\sqrt{8}V_{tb} \bar{u}\Lambda_\mu(1-\gamma^5) u\epsilon^\mu$.

Let us make a remark on the evaluation of the second term in
Eq.\eq{self energy}. The function $\Sigma(p)$ is UV divergent and
its finite part is a regular function of the momentum $p$ for
$p^2\leq m^2$. The point $p=m$ is a (finite) logarithmic branching
point, i.e. there $\Sigma$ contains terms behaving as:
$$\lim_{y\to 0}\Sigma \sim y\Ln{y} + {\rm regular\ terms}.$$
For that reason, although $\Sigma$ is finite for $p=m$, its
derivative with respect to $p$ diverges at that point. This is an
IR divergence which we have regulated and distinctly parameterized
in Eq.\eq{Z quark}.

The evaluation of the vertex correction and of the self energy
diagrams can be done by applying the Passarino-Veltman method
\cite{VeltPass} for reduction of tensor integrals. As a result,
after some extensive algebra, one can reduce the calculation to
the evaluation of the three scalar integrals $A_0,~B_0$ and $C_0$.
The latter are defined as the integration in $D$ space-time
dimensions of a product of one, two and three scalar propagators
with, generally, different masses. For their evaluation we use
standard techniques and Feynman parametrization. In the present
case of one-to-two particle decay, the phase space is very simple
(see the appendix). To obtain the differential cross-section one
does not need to perform any nontrivial integrations. The
resulting expression is:
\begin{eqnarray}
{1\over{\Gamma_0}}{{d\Gamma_{virt}^{(\as)}}\over{dx_b}}&=&
{{C_F\alpha_S(\mu)}\over{\pi Q}}\left\{ -\Phi(x_b=1)\left( {1\over
\epsilon} -2\gamma_E+2-\ln\left({Q^2m_t^2\over
4\pi^2\mu^2}\right)\right)\right.\nonumber\\
&+& s\left[  \Sp{ {2Q\over 1-s+Q} } -
\Sp{ {{2Q}\over{s-b+Q}} } \right. \nonumber\\
&-&\left. 2\Ln{ s+Q }\Ln{ {{1-s+Q}\over \sqrt{w}} } -\Ln{
{s+Q\over \sqrt{b}} } \Ln{ (s+Q)\sqrt{b} }\right] \nonumber\\
&+& {3Q^2-2s^2\over G_0}\Ln{ {s+Q\over\sqrt{b}} } +2Q\left(
{s\over G_0}-1 \right) \nonumber\\
&-& {Q\over 4wG_0}(1 +4b +bw-5b^2-5w+4w^2)\Ln{ \sqrt{b} }
\Bigg\}\delta(1-x_b) \label{Gamma virt}
\end{eqnarray}

To get the needed distribution we need to only add the
contributions \eq{Gamma virt} and \eq{Gamma real}. After also
adding the LO contribution, the result reads:
\begin{eqnarray}
{1\over{\Gamma_0}}{{d\Gamma}\over{dx_b}}&=&\delta(1-x_b)+
{{C_F\alpha_S(\mu)}\over{\pi Q}}\left\{\left\{2s\left[
{\mathrm{Li}}_2\left( {{2Q}\over{1-s+Q}}\right)-
{\mathrm{Li}}_2\left(
{{2Q}\over{s-b+Q}}\right)\right.\right.\right.
\nonumber\\
&-&\ln(s+Q)\left(\ln\left({{1-s+Q}\over {\sqrt{w}}}\right)+
\ln\left( {{s-b+Q}\over{2s(1-\beta)}}\right)\right)\nonumber\\
&+&\left.{1\over 2} \ln (b)\ln
\left({{s-b+Q}\over{2s(1-\beta)}}\right) \right]+ \left(
3{{Q^2}\over{G_0}}+s-b\right)\ln\left({{s+Q}\over \sqrt{b}}\right)
\nonumber\\
&+&(1-b)\ln\left({{1-s+Q}\over {\sqrt{w}}}\right)+ Q
\left[\left(6{{(w-b)(s-b)}\over{wG_0}}-1\right){{\ln (b)}\over 4}
\right.\label{fullmass}\\
&-&\left.\left.2\ \ln\left({{2s(1-\beta)}\over{\sqrt{w}}}\right)-2
\right]\right\}\delta(1-x_b)\nonumber\\
&-&2\ {\Phi (x_b)}\left[ {1\over {(1-x_b)_+}}+
{s\over{G_0}}\left(1+{{1+b}\over{2w}}\right)(1-x_b) - 1\right]\nonumber\\
&+&\left. 2s\sqrt{x_b^2-\beta^2}\left[ 2{s^2\over G_0}\left(
{{1-x_b}\over{1-2sx_b+b}}\right) + {s\over{G_0}}\left(1+{{1+b}
\over{2w}}\right)(1-x_b) - 1\right]\right\}\nonumber.
\end{eqnarray}
Separately, the contributions from the virtual and from the real
emission are IR divergent but those divergences cancel in the sum.
In the derivation of the above result, we have extensively made
use of various relations between the dilogarithms (Spence
functions); see the appendix for details. Finally, one can verify
that the integral of the differential distribution \eq{fullmass}
coincides with the total top-decay width to order $\as$ evaluated
in \cite{wid2}. As expected from the KLN theorem (see Section
(1.4)), all IR sensitive contributions disappear. Let us also note
that the top width is known at present to order $\as^2$
\cite{wid3}.

Let us also present here the result \eq{fullmass} in the limit
$m_b\to 0$, where all power corrections of the type $m_b/m_t$ can
be neglected. The result reads:
\begin{equation}
{1\over{\Gamma_0}}{{d\Gamma}\over {dx_b}}=\delta(1-x_b)
+{{\alpha_S(\mu)}\over{2\pi}}A_1(x_b), \label{gammass}
\end{equation}
with
\begin{eqnarray}
A_1(x_b)&=&C_F\ \left\{ \left[ {{1+x_b^2}\over{(1-x_b)_+}}
+{3\over 2}\delta(1-x_b)\right]\ln{{m_t^2}\over{m_b^2}}\right.\nonumber\\
&+& 2{{1+x_b^2}\over
{(1-x_b)_+}}\ln[(1-w)x_b]-{{4x_b}\over{(1-x_b)_+}}
+{{4w(1-w)}\over{1+2w}}{{x_b(1-x_b)}\over {1-(1-w)x_b}}\nonumber\\
&+& \delta(1-x_b)\left[ 4{\mathrm{Li}}_2 (1-w)+2\ln w\ln (1-w)
-{{2\pi^2}\over 3}\right.\nonumber\\
&-& {{2(1-w)}\over{1+2w}}\ln(1-w) -{{2w}\over {1-w}}\ln w
-4\Bigg]\Bigg\}. \label{mass}
\end{eqnarray}
The origin of the plus-prescription appearing in the above results
as well as some its properties are discussed in the appendix.

\subsection{ Calculation with $m_b$=0}

In a similar fashion to the above derivation one can also obtain
the massless results with the bottom mass set to zero from the
outset. In this calculation we proceeded in the same way as in the
$m_b\neq 0$ case. The integrals to be evaluated simplify
significantly in this case. We use the same renormalization scheme
as in the case $m_b\neq 0$. The essential difference is that in
the massless case, the contributions to the differential width
from real and virtual gluon emission, after the UV renormalization
is performed, contain IR divergent terms up to $1/\epsilon^2$. An
extra power of $1/\epsilon$ appears because the $b$-mass no longer
prevents collinear divergences. For that reason, in general, one
need to keep terms to order ${\mathcal{O}}(\epsilon^2)$. Similarly
to the case $m_b\neq 0$, in the sum of the real and virtual
contributions to the differential width the leading IR divergences
$\sim 1/\epsilon^2$ cancel. The resulting expression contains a
singularity $\sim 1/\epsilon$ which is in one-to-one
correspondence with the quasi-collinear log $\ln(m_b/m_t)$ present
in the massive result \eq{mass}. Finally, the massless result
reads:
\begin{eqnarray}
{1\over{\Gamma_0}}  {{d\hat\Gamma_b}\over {dx_b}}&=&
\delta(1-x_b)+{{\alpha_S(\mu)}\over{2\pi}}\left\{C_F \left[
{{1+x_b^2}\over{(1-x_b)_+}}+{3\over 2}\delta (1-x_b)\right]\right.
\nonumber\\
&\times&\left(-{1\over\epsilon}+\gamma_E-\ln 4\pi\right) +\hat
A_1(x_b)\Bigg\}, \label{cmseps}
\end{eqnarray}
with $\hat A_1(x_b)$ defined as:
\begin{eqnarray}
\hat A_1 (z) &=& C_F\left\{ \left[ {{1+z^2}\over{(1-z)_+}}+{3\over
2}\delta (1-z)\right] \left[\ln{{m_t^2}\over {\mu_F^2}}+2
{{1+w}\over {1+2w}}-2\ln (1-w)\right] \right.
\nonumber\\
&+&{{1+z^2}\over {(1-z)_+}} \left[ 4\ln \left[(1-w)z\right]-
{1\over{1+2w}}\right]\nonumber\\
&-&{{4z}\over{(1-z)_+}}\left[1-{{w(1-w)(1-z)^2}\over
{(1+2w)(1-(1-w)z)}}\right]\nonumber\\
&+&2(1+z^2)\left[\left({1\over {1-z}}\ln(1-z) \right)_+-{1\over
{1-z}}
\ln z\right]\nonumber\\
&+& \delta(1-z) \left[4{\mathrm {Li}}_2 (1-w)+ 2\ln(1-w)\ln w
-{{2\pi^2}\over 3}
+{{1+8w}\over {1+2w}}\ln (1-w)\right.\nonumber\\
&-&{2w\over{1-w}}\ln w+ {{3w}\over {1+2w}}-9\Bigg]\Bigg\}.
\label{cms}
\end{eqnarray}
In order to get the correct finite term in the normalized
differential width, the Born width $\Gamma_0$ (see Eq.\eq{LO width
m}) will have to be evaluated in dimensional regularization to
order ${\cal O} (\epsilon)$. We find:
\begin{eqnarray}
\Gamma_0 &=& {{\alpha m_t|V_{tb}|^2}\over{16\sin^2\theta_W}}
{{(1-w)^2(1+2w)}\over w}\left\{1+\epsilon \right. \nonumber\\
&\times& \left[-\gamma_E+\ln 4\pi- 2\ln (1-w) +2 {{1+w}\over
{1+2w}}\right]\Bigg\}.
\end{eqnarray}

One can easily verify that the difference
$A_1(x_b)-\hat{A}_1(x_b)$ exactly reproduces the (heavy quark
initiated) initial condition for the perturbative fragmentation
function, Eq.\eq{D ini}. We also checked that the integral of
Eq.~(\ref{cmseps}) agrees with the result of \cite{wid1}, where
the ${\cal O}(\alpha_S)$ corrections to the top width have been
evaluated in the approximation of a massless $b$ quark. Of course,
it also coincides with the massless limit of the total width
obtained from our calculation with non-zero bottom mass. It is
worth noting that in the total width the IR divergences disappear
together with the dependence on the unphysical factorization scale
$\mu_F$. Technically that is obvious since the terms that depend
on $\mu_F$ are multiplied in \eq{cmseps} by the LO
Altarelli-Parisi splitting function \eq{AP functions} whose
integral vanishes (see Eq.\eq{probab}).

\section{NLL Resummation of logs $\ln(m_b^2/m_t^2)$}\label{sec top PFF}

In principle the perturbative result Eq.\eq{fullmass} gives a
finite prediction for the $b$-spectrum in top decay. However, that
quantity is not IR safe (in the sense of Section (2.4)) since it
diverges in the limit $m_b\to 0$. As follows from our general
discussions in the previous Chapter, to obtain a reliable
perturbative prediction one needs to resum the quasi-collinear
logs $\ln(m_b^2/m_t^2)$ to all orders
\footnote{There is another mass parameter in our problem - the $W$
mass $m_W$. However, we will treat it as not small with respect to
$m_t$, and therefore will not resum terms like
$\ln(m_W^2/m_t^2)$.}
in the strong coupling. We will show later in this section that
the numerical difference between the two expressions for the
$b$-spectrum -- Eqns.\eq{fullmass} and \eq{gammass} -- is totally
negligible. Since the two expressions differ by power corrections
$\sim (m_b/m_t)^n$ which are small since
$m_b/m_t={\mathcal{O}}(10^{-2})$, one can conclude that the power
corrections can indeed be safely neglected. Let us recall that
this is precisely the condition for validity of the method of the
perturbative fragmentation function described in detail in Section
(2.4.2). Therefore, in the following we will present the
application of this method for resummation of the above mentioned
quasi-collinear logs in top-quark decay with NLL accuracy.

Following the general discussion in Section (2.4.2), the
differential width for the production of a massive $b$ quark in
top decay can be expressed via the following convolution:
\begin{equation}
{1\over {\Gamma_0}} {{d\Gamma}\over{dx_b}} (x_b,m_t,m_W,m_b)=
{1\over{\Gamma_0}} \int_{x_b}^1 {{{dz}\over z}{{d\hat\Gamma}\over
{dz}}(z,m_t,m_W,\mu,\mu_F) D\left({x_b\over z},\mu_F,m_b \right)},
\label{pff}
\end{equation}
where we have restricted the above result to the flavor
non-singlet case, i.e. to the perturbative fragmentation of a
massless $b$ into a massive $b$. In doing so, we have ignored the
strongly suppressed mixing of the $b$-quark with the gluon and the
other flavors through DGLAP evolution. In Eq.~(\ref{pff}) $\mu$
and $\mu_F$ are the renormalization and factorization scales
respectively, the former associated with the renormalization of
the strong coupling constant. In principle, one can use two
different values for the factorization and renormalization scales;
however, a choice often made consists of setting $\mu=\mu_F$ and
we shall adopt this convention for most of the results which we
shall show.

The NLO, $\MSbar$-subtracted coefficient function is:
\begin{equation}
{1\over{\Gamma_0}}{{d\hat\Gamma_b}\over
{dz}}^{\overline{\mathrm{MS}}}=
\delta(1-z)+{{\alpha_S(\mu)}\over{2\pi}}\hat A_1 (z), \label{diff}
\end{equation}
with $\hat{A}_1$ given in Eq.\eq{cms}. The consistent NLO plus NLL
(as defined in Eq.\eq{quasi logs}) evaluation of the
$b$-production rate requires that $\as$ is the NLO strong coupling
constant \eq{alpha s} with $n_f=5$ for the number of active
flavors. The initial condition for the perturbative fragmentation
function must be evaluated to order $\as$ as well. The initial
condition is presented in \eq{D ini} with $m$ being the bottom
mass $m_b$ and, in order to avoid large logs there, we need to
take the initial scale $\mu_{0F}$ of the order of $m_b$.

As follows from our discussion of Eq.\eq{sigmaH}, we will work in
analytical form in terms of the Mellin moments $N$; the Mellin
transformation is defined in \eq{Mellin}. The Mellin transform of
Eq.~(\ref{pff}) is:
\begin{equation}
\Gamma_N(m_t,m_W,m_b)=\hat\Gamma_N(m_t,m_W,\mu,\mu_F)
D_{b,N}(\mu_F,m_b), \label{gamman}
\end{equation}
with
\begin{equation}
\Gamma_N (m_t,m_W,m_b)= {1\over{\Gamma_0}}\int_0^1 {dx_b \
x_b^{N-1} {{d\Gamma}\over{dx_b}}(x_b,m_t,m_W,m_b) }.
\end{equation}
The fragmentation function $D_{b,N}$ is a product of the solution
of the DGLAP equations Eq.\eq{dresum E} with the initial condition
Eq.\eq{D ini N}. The explicit $N$-space expression for the
coefficient function $\hat\Gamma_N$ can also be found in the
appendix.

For our numerical study, we shall assume $m_t=175$~GeV,
$m_W=80$~GeV, $m_b=5$~GeV and $\Lambda=200$~MeV.

The $b$-quark energy distribution in $x_b$-space will finally be
obtained by inverting the $N$-space result (\ref{gamman})
numerically, by a contour integral in the complex $N$-plane. We
will use the so-called minimal prescription introduced in
\cite{cmnt}, where the integration contour is deformed with
respect to the one in Eq.\eq{Mellin inv}. We will discuss that
point in more detail in the next section.

We already have all the ingredients needed to obtain the
expression for the $b$-quark spectrum in top-quark decay with
collinear logs $\sim \ln(m_b/m_t)$ resummed with NLL accuracy. We
normalize our plots to the total NLO width $\Gamma$, obtained by
neglecting powers $\sim (m_b/m_t)^p$. Its expression can be found
in \cite{wid1}. In fact, that normalization is not affected by the
factorization on the right-hand side of Eq.~(\ref{pff}), the DGLAP
evolution of the relevant factors, or the resummation of soft
logarithms in the initial condition of the perturbative
fragmentation function and the coefficient function.

\begin{figure}[t]
\centerline{\resizebox{0.65\textwidth}{!}{\includegraphics{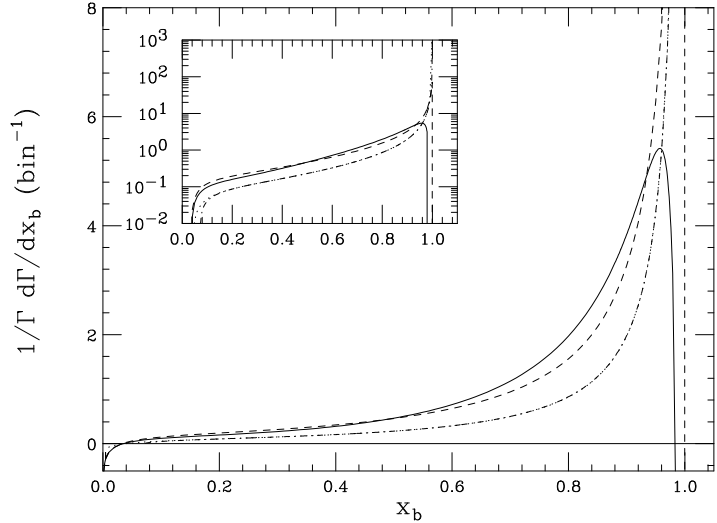}}}
\caption{\footnotesize $b$-quark energy distribution in top decay
according to the perturbative fragmentation approach, with (solid
line) and without (dashes) NLL soft-gluon resummation in the
initial condition of the perturbative fragmentation function, and
according to the exact NLO calculation, with (dot-dashes) and
without (dots) inclusion of powers of $m_b/m_t$. In the inset
figure, we show the same curves on a logarithmic scale.}
\label{figpart}
\end{figure}

We first present our results for the $x_b$ spectrum in
Fig.~\ref{figpart}. We plot the $x_b$ distribution according to
the perturbative fragmentation approach. For the sake of
comparison, we also show the exact ${\cal O}(\alpha_S)$ result for
a massive $b$ quark, Eq.\eq{fullmass}. We set  $\mu=\mu_F=m_t$ and
$\mu_0=\mu_{0F}=m_b$. As an illustration, the effect of NLL
soft-gluon resummation only in the initial condition of the
perturbative fragmentation function is also presented; it will be
systematically discussed in the next section.
Clearly, the use of perturbative fragmentation functions has a
strong impact on the $x_b$ distribution. The fixed-order result
lies well below the perturbative fragmentation results for about
$0.1\lsim x_b\lsim 0.9$ and diverges once $x_b\to 1$, due to a
behavior $\sim 1/(1-x_b)_+$. Moreover, as we previously noted, the
full inclusion of powers of $m_b/m_t$ has a negligible effect on
the $x_b$ spectrum; the dot-dashed and dotted lines in
Fig.~\ref{figpart} are in fact almost indistinguishable. As for
the perturbative fragmentation results, the distribution shows a
very sharp peak, though finite, once $x_b$ approaches unity. This
behavior will be smoothed out after we resum the soft NLL
logarithms appearing in the initial condition of the perturbative
fragmentation function and the coefficient function; an indication
is the behavior of the distribution after the soft-resummation
(Eq.\eq{D ini exp}) in the initial condition \eq{D ini} is
performed. Both perturbative fragmentation distributions become
negative for $x_b\to 0$ and $x_b\to 1$, which is a known result,
already found for heavy-quark production in $e^+e^-$ annihilation
\cite{CC},\cite{nasole}. For $x_b\to 0$, the coefficient function
(\ref{diff}) contains large logarithms $\sim\alpha_S(\mu) \ln x_b$
which have not been resummed yet. Likewise, in the soft limit
$x_b\to 1$, Eq.~(\ref{cms}) contains contributions
$\sim\alpha_S(\mu)/(1-x_b)_+$ and
$\sim\alpha_S(\mu)[\ln(1-x_b)/(1-x_b)]_+$ whose resummation is
discussed in the next section. As stated in \cite{CC}, once $x_b$
gets closer to unity, non-perturbative contributions also become
important and should be taken into account. The region of
reliability of the perturbative calculation at large $x_b$ may be
related to the Landau pole in the expression for the strong
coupling constant \eq{alpha s}, and is estimated to be $x_b\lsim
1-\Lambda/m_b$.
\begin{figure}
\centerline{\resizebox{0.49\textwidth}{!}{\includegraphics{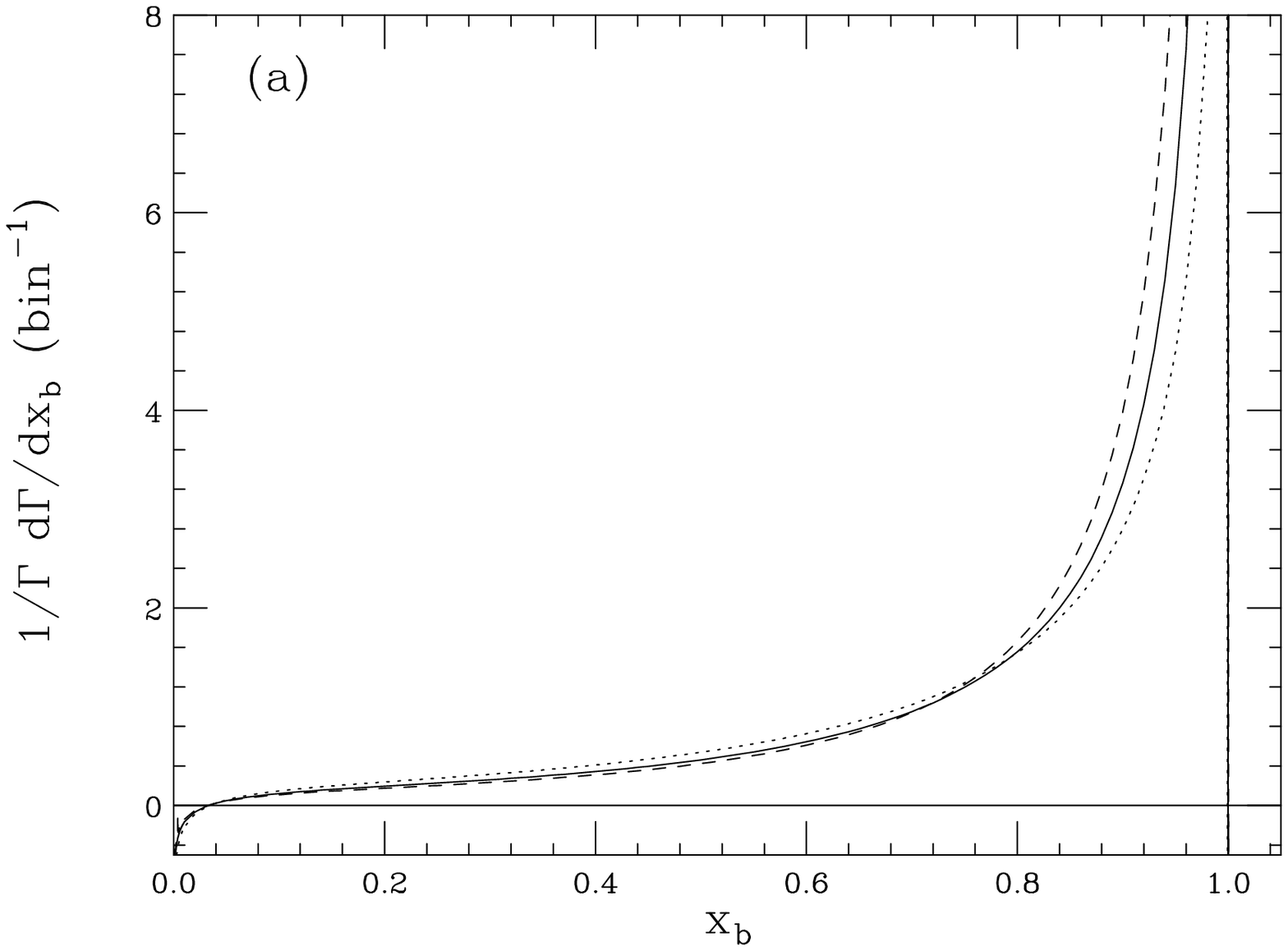}}%
\hfill%
\resizebox{0.49\textwidth}{!}{\includegraphics{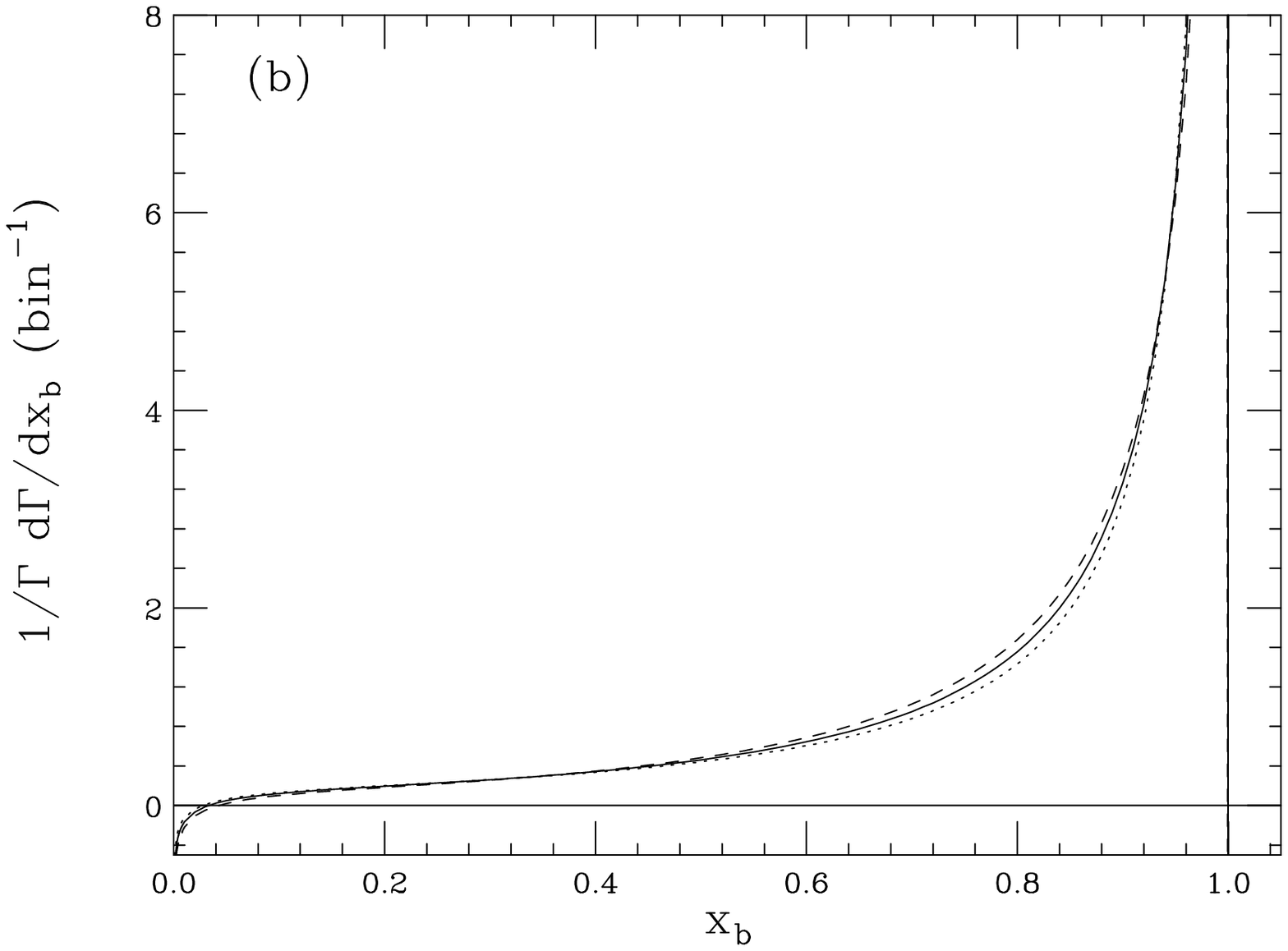}}}
\caption{\footnotesize (a): $x_b$ spectrum for $\mu_F = m_t$ and
$\mu_{0F}=m_b/2$ (dots), $\mu_{0F}=m_b$ (solid) and
$\mu_{0F}=2m_b$ (dashes); (b): $\mu_{0F}=m_b$ and $\mu_F=m_t/2$
(dots), $\mu_F=m_t$ (solid) and $\mu_F=2m_t$ (dashes). The
renormalization scales are kept at $\mu=m_t$ and $\mu_0=m_b$.
Though not visible, all distributions show a finite, sharp peak
once $x_b$ is close to 1.} \label{figmunos}
\end{figure}
Fig. \ref{figmunos} shows the dependence of the perturbative
fragmentation $x_b$ distributions on the factorization scales
$\mu_{0F}$ (a) and $\mu_F$ (b). It is obvious that the results
exhibit non-trivial dependence on the choice of those non-physical
scales. Still, this dependence is concentrated close to the region
of the Sudakov peak and is relatively mild far from it. Since the
presence of scale dependence indicates the importance of higher
order terms that are unaccounted for, we expect that after we
resum large soft-logs the result will be less sensitive to the
choice of those scales. That will be discussed in the next
section. As a check of our results we have verified that for
$\mu_F$ approaching $\mu_{0F}$, the distribution gets closer to
the fixed-order, unevolved one shown in Fig.~\ref{figpart}.

\section{NLL Threshold Resummation}\label{sec soft}

The $\overline{\mathrm{MS}}$ coefficient function \eq{diff} and
the initial condition of the perturbative fragmentation function
\eq{D ini} contain terms behaving like $1/(1-x_b)_+$ or
$[\ln(1-x_b)/(1-x_b)]_+$, which become arbitrarily large when
$x_b$ approaches one. This is equivalent to contributions
proportional to $\ln N$ and $\ln^2 N$ in moment space, as can be
seen by writing the $\overline{\mathrm{MS}}$ coefficient function
in the large-$N$ limit
\footnote{Following \cite{aglric}, we note that, by defining $n =
N \exp(\gamma_E)$, we could rewrite this expression in terms of
$\ln(n)$ rather than $\ln N$, with no $\gamma_E$ terms explicitly
appearing.}:
\begin{eqnarray}
\hat\Gamma_N (m_t,m_W,\mu_F) &= & 1+{{\alpha_S C_F}\over{2\pi}}
\left\{2\ln^2 N +\left[4\gamma_E +2-4\ln(1-w)
-2\ln{{m_t^2}\over{\mu_F^2}}\right]\ln N\right.\nonumber\\
&+&K(m_t,m_W,\mu_F)+{\cal O}\left( {1\over N}\right)\Bigg\}
\label{largen}
\end{eqnarray}
In Eq.\eq{largen} we have introduced the function
$K(m_t,m_W,\mu_F)$, which contains terms which are constant with
respect to $N$; more details for the extraction of the terms which
$\sim const$ when $N\to\infty$ can be found in the appendix. The
function $K$ reads:
\begin{eqnarray}
K(m_t,m_W,\mu_F)&=& \left({3\over
2}-2\gamma_E\right)\ln{{m_t^2}\over{\mu_F^2}}+2\gamma_E^2
+2\gamma_E\left[1-2\ln(1-w)\right]\nonumber\\
&+& 2\ln w\ln(1-w)-2{{1-w}\over{1+2w}}\ln(1-w)-
{{2w}\over{1-w}}\ln w\nonumber\\
&+&4 {\mathrm{Li}}_2(1-w) -6-{{\pi^2}\over 3}\, . \label{kappa}
\end{eqnarray}
The $x_b\to 1$ ($N\to\infty $) limit corresponds to soft-gluon
radiation in top decay. These soft logarithms need to be resummed
to all orders in $\alpha_S$ \cite{CT}, \cite{Sterman soft} to
improve our prediction.

We can perform the soft-gluon (threshold) resummation for the
coefficient function following the general procedure described in
Section (1.4.3). We treat the $b$-quark as exactly massless and
therefore the result is collinearly divergent. That divergence is
subtracted in the $\MSbar$ scheme. The eikonal current reads:
\begin{equation}
|J(p_t,p_b,p_g)|^2=\left| {{m_t^2}\over{(p_t\cdot p_g)^2}}-
2{{(p_t\cdot p_b)}\over{(p_t\cdot p_g)(p_b\cdot
p_g)}}\right|.\label{J top}
\end{equation}
We express the ${\cal O}(\alpha_S)$ width in the soft
approximation as an integral over the variables
$q^2=(p_b+p_g)^2x_g$ and $z=1-x_g$, with $0\leq q^2\leq
m_t^2(1-w)^2(1-z)^2$ and $0\leq z\leq 1$. The limits $z\to 1$ and
$q^2\to 0$ correspond to soft and collinear emission respectively.
In the soft approximation, $z\simeq x_b$; $x_g$ is the gluon
energy fraction that is defined similarly to $x_b$ \eq{xb}. In the
$m_b=0$ case which we consider here:
\begin{equation}
x_b={1\over{1-w}}{{2 p_b\cdot p_t}\over {m_t^2}},\ \ \
x_g={1\over{1-w}}{{2 p_g\cdot p_t}\over {m_t^2}},\ \ \ 0\leq
x_{b,g} \leq 1. \label{xbpart}
\end{equation}
We point out that our definition of the integration variable $q^2$
is analogous to the quantity $(1-z)k^2$ to which the authors of
Ref.~\cite{CT} set the scale for $\alpha_S$ for soft-gluon
resummation in Drell--Yan and Deep-Inelastic-Scattering processes.
For small-angle radiation, $q^2\simeq q_T^2$, the gluon transverse
momentum with respect to the $b$-quark line. The variable $z$ is
analogous to $z=1-E_g/E_q$ of Ref.~\cite{CT}.

Performing the operations discussed in Section (1.4.3) we obtain:
\begin{eqnarray}
\hat\Gamma_N(m_t,m_W,\mu_F)&=&{{C_F}\over{\pi}} \int_0^1{dz
{{z^{N-1}-1}\over {1-z}}}
\left[\int_{\mu_F^2}^{m_t^2(1-w)^2(1-z)^2}{{dq^2}\over{q^2}}
\alpha_S\right.\nonumber\\
&-&\left.{1\over{m_t^2(1-w)^2(1-z)^2}}
\int_0^{m_t^2(1-w)^2(1-z)^2}{dq^2 \alpha_S}\right]. \label{eik}
\end{eqnarray}
We note that it corresponds to Eq.\eq{typical} with $a=2$ and
$Q^2=m_t^2(1-w)^2$. The origin of the second term in \eq{eik} is
in the collinearly non-enhanced term in the squared eikonal
current \eq{J top}.

In order to include all NLL contributions we only need to perform
the replacement \eq{afun} and use the property \eq{asint}. No
additional contribution from collinear radiation from the final
state is presented here because of the character of the
cross-section that our coefficient function represents; the final
$b$-quark is ``observed" and has its virtuality fixed by the
on-shell condition $p_b^2=0$. Exponentiating the result we obtain
the following form-factor that describes the soft limit of the
corresponding production rate:
\begin{eqnarray}
\ln \Delta_N &=& \int_0^1 {dz {{z^{N-1}-1}\over{1-z}}}
\left\{\int_{\mu_F^2}^{m_t^2(1-w)^2(1-z)^2}
{{dq^2}\over {q^2}} A\left[\alpha_S(q^2)\right]\right. \nonumber\\
&+& S\left[\alpha_S\left(m_t^2(1-w)^2(1-z)^2\right)\right]\Bigg\}.
\label{resum}
\end{eqnarray}
The function $S(\alpha_S)$ can be expanded according to:
\begin{equation}
S(\alpha_S)=\sum_{n=1}^{\infty}\left({{\alpha_S}\over
{\pi}}\right)^n S^{(n)}. \label{function S}
\end{equation}
At NLL level, we are just interested in the first term of the
above expansion:
\begin{equation}
S^{(1)}=-C_F.
\end{equation}
The integral in Eq.\eq{resum} can be performed, up to NLL
accuracy, according to the methods described in Section (1.4.3).
This leads to the following result:
\begin{equation}
\Delta_N(m_t,m_W,\alpha_S(\mu^2),\mu,\mu_F)=\exp\left[\ln N
g^{(1)}(\lambda)+ g^{(2)}(\lambda,\mu,\mu_F)\right]\, ,
\label{deltaint}
\end{equation}
with $\lambda$ defined in \eq{lambda} and the functions $g^{(1)}$
and $g^{(2)}$ given by:
\begin{eqnarray}
g^{(1)}(\lambda)&=& \frac{A^{(1)}}{2\pi b_0 \lambda} \;
[ 2\lambda + (1-2\lambda) \ln (1-2\lambda)] \;,\\
g^{(2)}(\lambda,\mu,\mu_F) &=& \frac{A^{(1)}}{2 \pi b_0}\left[\ln
\frac{m_t^2(1-w)^2}{\mu_F^2}
- 2\gamma_E\right] \ln(1-2\lambda)\nonumber\\
&+&\frac{A^{(1)}  b_1}{4 \pi b_0^3} \left[ 4\lambda + 2 \ln
(1-2\lambda) +
\ln^2 (1-2\lambda) \right]\nonumber\\
&-& \frac{1}{2\pi b_0} \left[2\lambda + \ln (1-2\lambda) \right]
\left(\frac{A^{(2)}}{\pi b_0} +
A^{(1)}\ln\frac{\mu^2}{\mu_{F}^2}\right)
\nonumber\\
&+& \frac{S^{(1)}}{2\pi b_0} \ln (1-2\lambda).
\end{eqnarray}
In Eq.\eq{deltaint} the term $\ln N g^{(1)}(\lambda)$ accounts for
the resummation of leading logarithms $\alpha_S^n\ln^{n+1}N$ in
the Sudakov exponent, while the function
$g^{(2)}(\lambda,\mu,\mu_F)$ resums NLL terms $\alpha_S^n\ln^nN$.

Furthermore, we follow Ref.~\cite{CC} and in our final
Sudakov-resummed coefficient function we also include the constant
terms of Eq.\eq{kappa}:
\begin{eqnarray}
\hat\Gamma_N^S(m_t,m_W,\alpha_S(\mu^2),\mu,\mu_F)&=&
\left[1+{{\alpha_S(\mu^2)\ C_F}\over{2\pi}}
K(m_t,m_W,\mu_F)\right]\nonumber\\
&\times & \exp\left[\ln N
g^{(1)}(\lambda)+g^{(2)}(\lambda,\mu,\mu_F)\right]. \label{delta}
\end{eqnarray}
One can check that the ${\cal O} (\alpha_S)$ expansion of
Eq.\eq{delta} yields Eq.\eq{largen} as it must for consistency.
Next we match the resummed coefficient function to the exact
first-order result, so that also $1/N$ suppressed terms, which are
important in the region $x_b < 1$, are taken into account. We
adopt the same matching prescription as in \cite{CC}: we add the
resummed result to the exact coefficient function and, in order to
avoid double counting, we subtract what they have in common, i.e.
the up-to-${\cal O}(\alpha_S)$ terms in the expansion of
Eq.\eq{delta}. Our final result for the resummed coefficient
function reads:
\begin{eqnarray}
\hat\Gamma_N^{\mathrm{res}}(m_t,m_W,\alpha_S(\mu^2),\mu,\mu_F)
&=&\hat\Gamma_N^S(m_t,m_W,\alpha_S(\mu^2),\mu,\mu_F)\nonumber\\
&-& \left[\hat\Gamma_N^S(m_t,m_W,\alpha_S(\mu^2),\mu,\mu_F)
\right]_{\alpha_S}\nonumber\\
&+&\left[\hat\Gamma_N(m_t,m_W,\alpha_S(\mu^2),\mu,\mu_F)\right]_{\alpha_S},
\label{resum Gamma}
\end{eqnarray}
where $[\hat\Gamma_N^S]_{\alpha_S}$ and
$[\hat\Gamma_N]_{\alpha_S}$ are respectively the expansion of
Eq.\eq{delta} up to ${\cal O}(\alpha_S)$ and the  full fixed-order
top-decay coefficient function at ${\cal O} (\alpha_S)$, presented
in the appendix.

We would like to compare our resummed expression with other
similar results obtained in heavy quark decay processes
\cite{korster,roth,aglietti,aglric}. Besides the obvious
replacement of a bottom quark with a top in the initial state, our
work presents other essential differences. We have resummed large
collinear logarithms $\alpha_S\ln(m_t^2/m_b^2)$, while
Refs.~\cite{korster,roth,aglietti,aglric} just address the decay
of heavy quarks into massless quarks. Moreover, this work still
differs in a critical issue. Those papers are concerned with
observing the lepton produced by the $W$ decay or the photon in
the $b\to X_s \gamma$ process, while we wish instead to observe
the outgoing $b$ quark. This is immediately clear from the choice
of the $z$ variable whose $z\to 1$ endpoint leads to the Sudakov
logarithms. In our case it is the normalized energy fraction of
the outgoing bottom quark; in \cite{korster,roth,aglietti,aglric}
it is instead related to the energy of either the lepton or the
radiated photon.

The most evident effect of this different perspective is that an
additional scale, namely the invariant mass of the recoiling
hadronic jet, enters the results
\cite{korster,roth,aglietti,aglric}, but as we noted above, is
absent in our case. An additional function (called
$\gamma(\alpha_S)$ in \cite{korster,roth}, $C(\alpha_S)$ in
\cite{aglietti}, $B(\alpha_S)$ in \cite{aglric}) appears in those
papers.  The argument of $\alpha_S$ in this function is related to
the invariant mass of the unobserved final state jet constituted
by the outgoing quark and the gluon(s). It is worth noting that an
identical function, called
$B\left[\alpha_S\left(Q^2(1-z)\right)\right]$, also appears in the
$e^+e^-$~\cite{CC} and DIS~\cite{CT} massless coefficient
functions, where it is again associated with the invariant mass of
the unobserved jet. We also observe that to order $\alpha_s$ the
coefficient $S^{(1)}$ coincides with the corresponding $H^{(1)}$
of the function $H\left[\alpha_S\left(m_b^2(1-z)^2\right)\right]$
(see Eq.\eq{deltares}), which resums soft terms in the initial
condition of the perturbative fragmentation function. It will be
very interesting to compare the functions $S(\alpha_S)$ and
$H(\alpha_S)$ at higher orders as well.

One final comment we wish to make is that, as expected, in our
final result, Eq.\eq{gamman}, which accounts for NLL soft
resummation in both the coefficient function and the initial
condition of the perturbative fragmentation function,
$\alpha_S^n\ln^{n+1} N$ terms do not appear, since they are due to
soft {\sl and} collinear radiation. Both the quarks being heavy,
only the former leads to a logarithmic enhancement. Double
logarithms are generated by a mismatch in the lower and upper
$q^2$ integration limits over the $A[\alpha_S(q^2)]$ function in
the exponent of the resummation expression. In our case both of
them have the same functional dependence with respect to $z$, i.e.
$(1-z)^2$ in Eq.\eq{resum} and Eq.\eq{deltares} (corresponding to
$a=2$ in \eq{typical}). The cancellation of the
$\alpha_S^n\ln^{n+1} N$ term can be explicitly seen at order
$\alpha_S$ by comparing the large-$N$ limit for the coefficient
function, Eq.~(\ref{largen}), and the initial condition (Eq.\eq{D
ini N}): the $\ln^2 N$ terms have identical coefficients and
opposite signs.

As follows from Eq.\eq{gamman}, to obtain the resummed
differential distribution in $x_b$-space, one needs to perform the
inverse Mellin transformation \eq{Mellin inv}. As we previously
emphasized, we perform that inversion numerically, using the
so-called Minimal Prescription \cite{cmnt}: the integration
contour in the complex $N$ plane crosses the real line at a value
$C_{MP}$ such that $2< C_{MP}< N_L$. That way all singularities of
the integrand are to the left of the contour with the exception of
the Landau pole $N_L=\exp(1/(2\as(\mu)b_0))$. The origin of $N_L$
is in the Landau pole associated with the integration over the
running strong coupling in the Sudakov exponent \eq{resum}, when
combined with the relations \eq{NLL eval}, \eq{deltaint} and
\eq{lambda}.

Next we present the results for the $b$-quark energy distribution
in top decay including the soft-gluon resummation in the initial
condition of the perturbative fragmentation function and the
coefficient function. The rest of the settings are the same as the
ones we discussed in the end of the previous section.

\begin{figure}[t]
\centerline{\resizebox{0.65\textwidth}{!}{\includegraphics{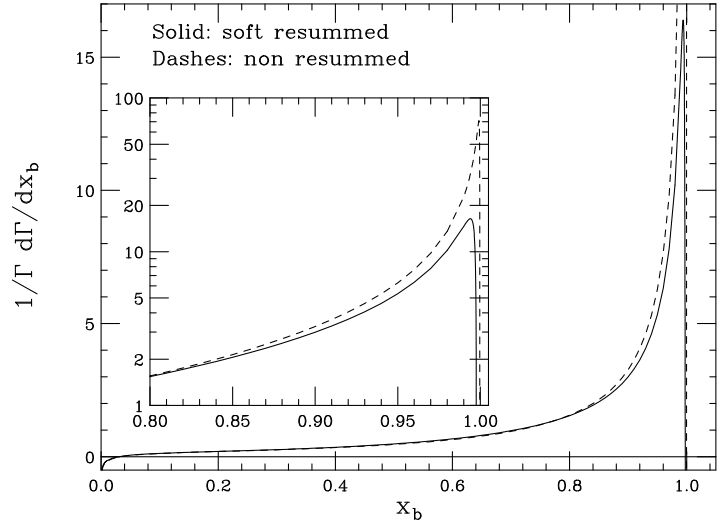}}}
\caption{\footnotesize $b$-quark energy distribution in top decay
according to the perturbative fragmentation approach, with (solid
line) and without (dashes) NLL soft-gluon resummation. In the
inset figure, we show the same curves on a logarithmic scale, for
$x_b>0.8$. We have set $\mu_F=\mu=m_t$ and $\mu_{0F}=\mu_0=m_b$.}
\label{fig1}
\end{figure}

In Fig.~\ref{fig1} we present the $x_b$ distribution according to
the approach of perturbative fragmentation, with and without NLL
soft-gluon resummation. For the scales we have set $\mu_F=\mu=m_t$
and $\mu_0=\mu_{0F}=m_b$. We note that the two distributions agree
for $x_b\lsim 0.8$, while for larger $x_b$ values the resummation
of large terms $x_b\to 1$ smoothes out the distribution, which
exhibits the Sudakov peak. Both distributions become negative for
$x_b\to 0$ and $x_b\to 1$. The negative behavior at small $x_b$
can be related to the presence of unresummed $\alpha_S\ln x_b$
terms in the coefficient function. At large $x_b$, we approach
instead the non-perturbative region, and resumming leading and
next-to-leading logarithms is still not sufficient to correctly
describe the spectrum for $x_b$ close to $1$. In fact, the range
of reliability of the perturbative calculation has been estimated
to be $x_b\lsim 1-\Lambda/m_b\simeq 0.95$ \cite{CC}.
\begin{figure}[p]
\centerline{\resizebox{0.65\textwidth}{!}{\includegraphics{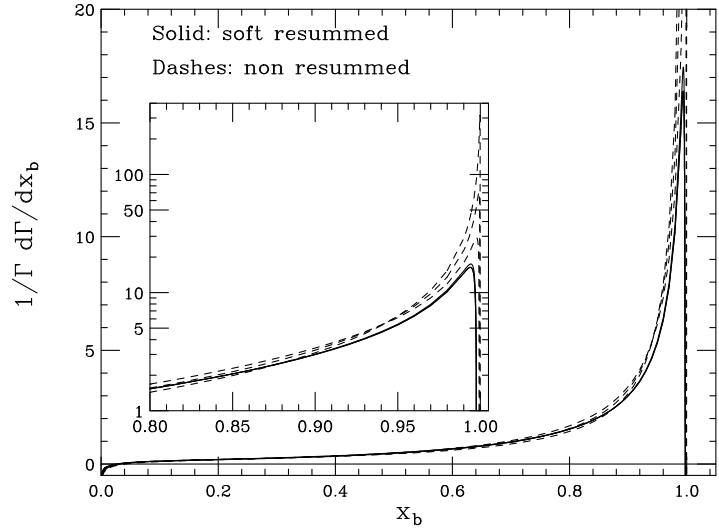}}}
\caption{\footnotesize $b$-quark energy spectrum for different
values of the factorization scale $\mu_F$, with (solid) and
without (dashes) NLL soft-gluon resummation. The other scales are
fixed at $\mu=m_t$, $\mu_0=\mu_{0F}=m_b$. As in Fig.~1, in the
inset figure, we present the same plots for large values of $x_b$,
on a logarithmic scale.} \label{figfac}
\end{figure}
\begin{figure}
\centerline{\resizebox{0.65\textwidth}{!}{\includegraphics{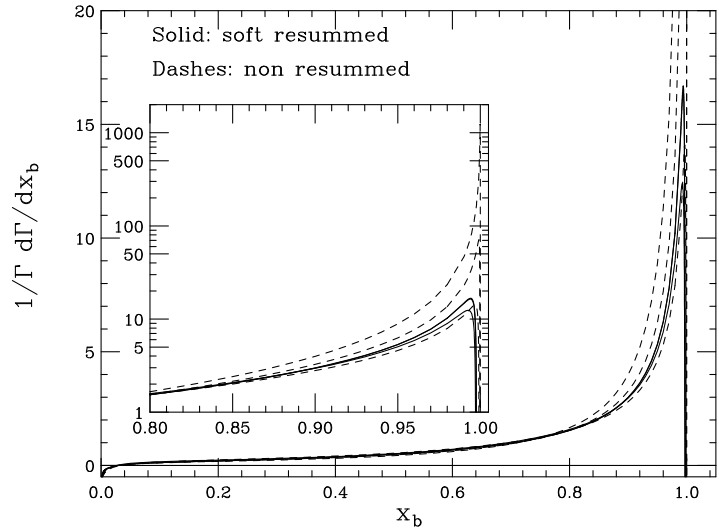}}}
\caption{\footnotesize As in Fig.~\ref{figfac}, but for different
values of $\mu_{0F}$. The other scales are fixed at
$\mu=\mu_F=m_t$, $\mu_0=m_b$.} \label{figfac0}
\end{figure}
\begin{figure}
\centerline{\resizebox{0.65\textwidth}{!}{\includegraphics{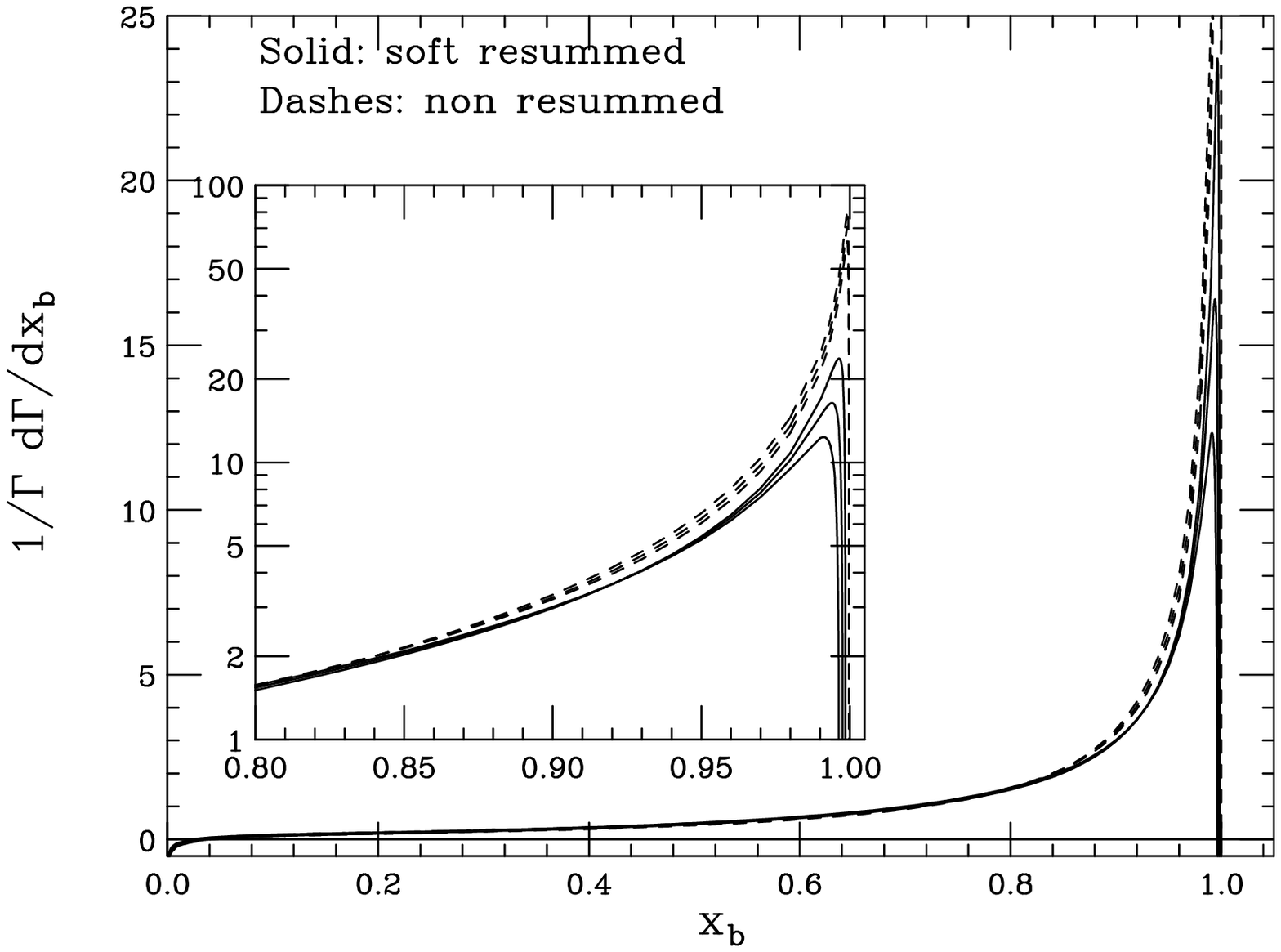}}}
\caption{\footnotesize As in Fig.~\ref{figfac}, but for different
values of the renormalization scale $\mu$. The other scales are
fixed at $\mu_F=m_t$, $\mu_0=\mu_{0F}=m_b$.} \label{figren}
\end{figure}
\begin{figure}
\centerline{\resizebox{0.65\textwidth}{!}{\includegraphics{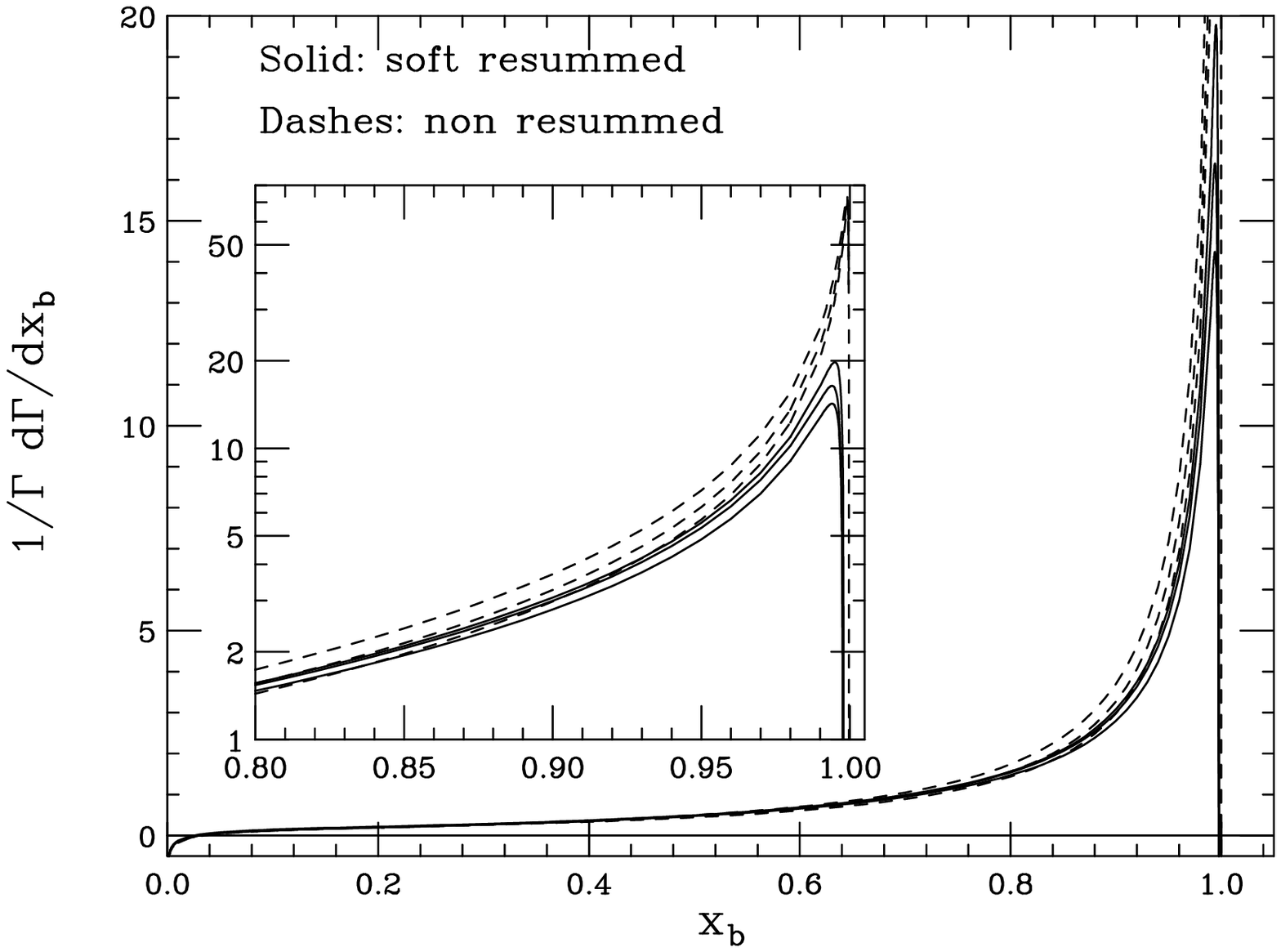}}}
\caption{\footnotesize As in Fig.~\ref{figfac}, but for different
values of the renormalization scale $\mu_0$. The other scales are
fixed at $\mu=\mu_F=m_t$, $\mu_{0F}=m_b$.} \label{figren0}
\end{figure}
It is interesting to investigate the dependence of
phenomenological distributions on the renormalization and
factorization scales which enter the coefficient function ($\mu$
and $\mu_F$) and the initial condition of the perturbative
fragmentation function ($\mu_0$ and $\mu_{0F}$). In particular, it
is worth comparing the $b$-energy spectra with and without soft
resummation. For the scales $\mu$ and $\mu_F$ we consider the
values $m_t/2$, $m_t$ and $2m_t$; for $\mu_{0F}$ and $\mu_0$ the
choices are $m_b/2$, $m_b$ and $2m_b$. Figs.~\ref{figfac} and
\ref{figfac0} show the dependence of the $x_b$ spectrum on the
factorization scales $\mu_F$ and $\mu_{0F}$; the dependence on the
renormalization scales $\mu$ and $\mu_0$ is exhibited in
Figs.~\ref{figren} and \ref{figren0}.

We note that all distributions which include soft-gluon
resummation exhibit a reduced dependence on the factorization  and
renormalization scales.

Fig.~\ref{figfac} shows that curves obtained using different
values of $\mu_F$ are almost indistinguishable once soft
resummation is included; the unresummed plots exhibit a stronger
effect of the chosen value for $\mu_F$. A similar result also
holds for the scale $\mu_{0F}$: the dependence of the plots on its
actual value for $x_b>0.8$ is small if soft logarithms are
resummed and quite strong if the prediction is unresummed
(Fig.~\ref{figfac0}).

The choice of the value for the renormalization scale $\mu$
appearing in Eq.~(\ref{largen}) affects only the neighborhood of
the Sudakov peak of the resummed predictions, at $x_b$-values very
close to one (Fig.~\ref{figren}), where, as we have pointed out,
our perturbative approach is anyway unreliable. The effect of the
choice of the renormalization scale $\mu_0$ on the soft-resummed
spectra is slightly larger than the effect of $\mu$ and visible at
$x_b<1$ as well (Fig.~\ref{figren0}). As for the non-soft-resummed
predictions, although all dashed curves in Figs.~\ref{figren} and
\ref{figren0} seem to converge to the same point for $x_b\to 1$,
the overall dependence on $\mu$ and $\mu_0$ for $x_b<1$ is
stronger than for the resummed predictions.

As a whole, one can say that the implementation of NLL soft-gluon
resummation, along with the NLL DGLAP evolution for the
perturbative fragmentation function, yields a remarkable
improvement of our phenomenological results, since the reduced
dependence on the choice of factorization and renormalization
scales in the region where the perturbative approach is reliable
corresponds to a reduction of the theoretical uncertainty.

\section{Energy Spectrum of $b$-flavored Hadrons in Top Decay}
\label{sec top np}

In this section we consider the inclusion of a non-perturbative
component in addition to the perturbative result, so as to make
predictions for observable $b$-flavored hadrons (like $B$ mesons)
in top decay. At the same time, we also account for the inclusion
of NLL soft and collinear  resummation.

We write the normalized rate for the production of $b$-hadrons $B$
as in Eq.\eq{sigmaH}, i.e. as a convolution of the rate for the
production of $b$ quarks in top decay, given by Eq.\eq{gamman} and
a non perturbative fragmentation function $D^{np}(x)$:
\begin{equation}
{1\over {\Gamma}} {{d\Gamma^B}\over{dx_B}}
(x_B,m_t,m_W,m_b)={1\over{\Gamma}} \int_{x_B}^1 {{{dz}\over
z}{{d\Gamma^b}\over {dz}}(z,m_t,m_W,m_b) D^{np}\left({x_B\over
z}\right)}, \label{npff}
\end{equation}
where $x_B$ is the $B$ normalized energy fraction:
\begin{equation}
x_B={1\over{1-w}}{{2p_B\cdot p_t}\over {m_t^2}},
\end{equation}
$p_B$ being the $B$ four-momentum. Since $D^{np}(x)$ contains
non-perturbative information, it cannot - for the time being - be
calculated from first principles in QCD, but can only be extracted
from data.

We shall assume a universality property for such a function, i.e.
that it does not depend on the process but only on the
non-perturbative transition it describes, and extract it from fits
to $B$-production data collected at LEP in $e^+e^-$ collisions. In
particular, we can choose different functional forms for
$D^{np}(x)$, and  tune these hadronization models to the available
data. We shall consider three models: a power law with two tunable
parameters:
\begin{equation}
D^{np}(x;\alpha,\beta)={1\over{B(\beta +1,\alpha +1)}}(1-x)^\alpha
x^\beta, \label{ab}
\end{equation}
the model of Kartvelishvili et al. \cite{kart}:
\begin{equation}
D^{np}(x;\delta)=(1+\delta)(2+\delta) (1-x) x^\delta\; ,
\label{kk}
\end{equation}
and the Peterson et al. model \cite{peterson}:
\begin{equation}
D^{np}(x;\epsilon)={A\over {x[1-1/x-\epsilon/(1-x)]^2}}.
\label{peter}
\end{equation}
In Eq.\eq{ab}, $B(x,y)$ is the Euler Beta function; in \eq{peter}
$A$ is a normalization constant for which an explicit expression
can be found in \cite{cagre1}.

In order for our extraction procedure to be self-consistent, we
employ the same underlying perturbative description in both the
$e^+e^-\to b\bar b$ process (where the non-perturbative
contribution is fitted) and $t\to bW$ (where it is used). This
will be ensured by using in both cases NLO,
$\overline{\mathrm{MS}}$ coefficient functions, along with a fully
NLL soft-gluon resummed description, with the large collinear
logarithms resummed to NLL accuracy by DGLAP evolution. For the
coefficient functions in $e^+e^-$ annihilation we shall refer to
\cite{msbar}.

Fits to data points can be performed either in $x_B$-space, or, as
recently advocated \cite{canas}, in the conjugated moment space.
When fitting in $x_B$ space we discard data points close to
$x_B=0$ and $x_B=1$ and consider ALEPH~\cite{aleph} data in the
range $0.18\lsim x_B\lsim 0.94$. Good-quality data are also
available from SLD \cite{sld} on $b$-flavored mesons and baryons.
Recently, the OPAL~\cite{opal} and DELPHI~\cite{delphi}
Collaborations also published new results, which are fully
compatible with the ones from ALEPH. The results of our fits are
shown in Table~\ref{table1}.
\begin{table}
\begin{center}
\begin{tabular}{||c|c||}
\hline
$\alpha$&$0.51\pm 0.15$  \\
\hline
$\beta$&$13.35\pm 1.46$  \\
 \hline
$\chi^2(\alpha,\beta)$/dof&2.56/14 \\
\hline
$\delta$&$17.76\pm 0.62$ \\
\hline
$\chi^2(\delta)$/dof&10.54/15  \\
\hline
$\epsilon$&$(1.77\pm 0.16)\times 10^{-3}$\\
\hline
$\chi^2(\epsilon)$/dof&29.83/15\\
\hline
\end{tabular}
\end{center}
\caption{\footnotesize Results of fits to $e^+e^-\to b\bar b$
ALEPH data, using matched coefficient function and initial
condition, with NLL DGLAP evolution and NLL soft-gluon
resummation. We set $\Lambda^{(5)}=200$~MeV,
$\mu_{0F}=\mu_0=m_b=5$~GeV and $\mu_F=\mu=\sqrt{s}=91.2$~GeV.
$\alpha$ and $\beta$ are the parameters in the power law
(\ref{ab}), $\delta$ refers to (\ref{kk}), $\epsilon$ to
(\ref{peter}). The fits have been performed neglecting the
correlations between the data points.} \label{table1}
\end{table}

\begin{figure}[t]
\centerline{\resizebox{0.65\textwidth}{!}{\includegraphics{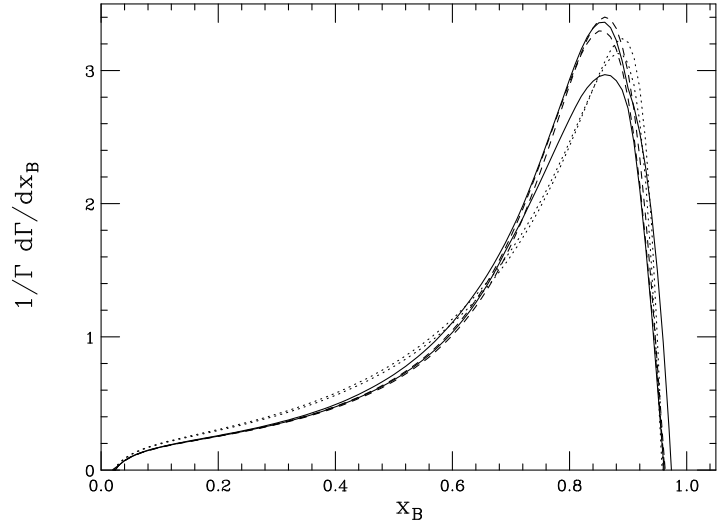}}}
\caption{\footnotesize $x_B$ spectrum in top decay, with the
hadronization modelled according to a power law (solid lines), the
Kartvelishvili et al. (dashes) and the Peterson (dots) model, with
the relevant parameters fitted to the ALEPH data. The plotted
curves are the edges of bands at one-standard-deviation confidence
level. NLL soft-gluon resummation is included. We set
$\mu_F=\mu=m_t$ and $\mu_{0F}=\mu_0=m_b$.} \label{had1}
\end{figure}

The models in Eqs.~(\ref{ab}) and (\ref{kk}) yield very good fits
to the data, while the model (\ref{peter}) is marginally
consistent. We have also fitted the SLD data and found
qualitatively similar results. However, the values for the
parameters which best fit the SLD data are different from the ones
obtained for ALEPH and quoted in Table~\ref{table1}. Using the
results in Table~\ref{table1} we can give predictions for the
spectrum of $b$-flavored hadrons in top decay. To account for the
errors on the best-fit parameters, we shall plot bands which
correspond to predictions at one-standard-deviation confidence
level. In Fig.~\ref{had1} we show our predictions for the $x_B$
distribution using the three models fitted to the ALEPH data. At
one-standard-deviation confidence level, the three predictions are
different, with the Peterson model yielding a distribution which
lies quite far from the other two and peaked at larger
$x_B$-values. Within two standard deviations, the predictions
obtained using the models (\ref{ab}) and (\ref{kk}) are
nonetheless in agreement.

We would like to emphasize that the differences between the
various models mainly originate from the varying quality of the
fits to $e^+e^-$ data where, as the $\chi^2$ values in
Table~\ref{table1} seem to suggest, a given model is sometimes not
really able to describe the data properly, due to its too
restrictive functional form.
\begin{table}[t]
\begin{center}
\begin{small}
\begin{tabular}{| c | c c c c |}
\hline
  & $\langle x\rangle$ & $\langle x^2\rangle$ & $\langle
x^3\rangle$ & $\langle x^4\rangle$ \\
\hline $e^+e^-$ data $\sigma_N^B$&0.7153$\pm$0.0052
&0.5401$\pm$0.0064 & 0.4236$\pm$0.0065 &0.3406$\pm$0.0064  \\
\hline \hline
$e^+e^-$ NLL $\sigma_N^b$ [A]   & 0.7666 & 0.6239 & 0.5246 & 0.4502  \\
$e^+e^-$ NLL $\sigma_N^b$ [B]   & 0.7801 & 0.6436 & 0.5479 & 0.4755  \\
\hline
$D^{np}_N$ [A]          & 0.9331 & 0.8657 & 0.8075 & 0.7566 \\
$D^{np}_N$ [B]          & 0.9169 & 0.8392 & 0.7731 & 0.7163 \\
\hline \hline
$t$-decay NLL $\Gamma^b_N$ [A]& 0.7750 & 0.6417 & 0.5498 & 0.4807 \\
$t$-decay NLL $\Gamma^b_N$ [B]& 0.7884 & 0.6617 & 0.5737 & 0.5072 \\
\hline
$t$-decay $\Gamma^B_N$ [A]        & 0.7231 & 0.5555 & 0.4440 & 0.3637 \\
$t$-decay $\Gamma^B_N$ [B]        & 0.7228 & 0.5553 & 0.4435 & 0.3633 \\
\hline
\end{tabular}
\end{small}
\end{center}
\caption{\footnotesize Experimental data for the moments
$\sigma^B_N$ from DELPHI~\protect\cite{delphi}, the resummed
$e^+e^-$ perturbative calculations for
$\sigma^b_N$~\protect\cite{CC}, the extracted non-perturbative
contribution $D^{np}_N$. Using the perturbative results
$\Gamma^b_N$, a prediction for the physical observable moments
$\Gamma^B_N$ is given. Set [A]: $\Lambda^{(5)} = 0.226$~GeV and
$m_b = 4.75$~GeV, set [B]: $\Lambda^{(5)} = 0.2$~GeV and $m_b =
5$~GeV. The experimental error should of course be propagated to
the final prediction.} \label{table2}
\end{table}

An alternative, and probably better, way of determining and
including non-perturbative information makes use of moment space
perturbative predictions and data \cite{canas}. The full
hadron-level result can be written in $N$-space as the product of
a perturbative and a non-perturbative contribution, $\Gamma^B_N =
\Gamma^b_N D^{np}_N$. For each value of $N$ one can then extract
the corresponding $D^{np}_N$ value from $e^+e^-$ data, with no
reference whatsoever to a specific hadronization model, and use it
to predict the same moment in top decay. The DELPHI Collaboration
\cite{delphi} has recently published preliminary results for the
moments of $B$-meson fragmentation in $e^+e^-$ collisions up to
$N=5$. From these data, and using the moments of the $e^+e^-$
perturbative contributions \cite{CC}, one can extract $D^{np}_N$.
The corresponding $\Gamma^B_N$ values can then be calculated
making use of the results for $\Gamma^b_N$ obtained in \cite{CCM}
and given by Eqns.\eq{gamman} and \eq{resum Gamma}. Calling
$\sigma^B_N$ and $\sigma^b_N$ the moments for the production rate
of $B$ mesons (measured) and $b$ quarks (calculated in
perturbative QCD) in $e^+e^-$ annihilation, we have
$\sigma_N^B=\sigma_n^b D^{np}_N$ and hence
\begin{equation}
\Gamma^B_N = \Gamma^b_N D^{np}_N = \Gamma^b_N
\frac{\sigma^B_N}{\sigma^b_N} \;.
\end{equation}

Table \ref{table2} shows a practical implementation of this
procedure. Predictions for the moments $\Gamma^B_N$ of $B$-meson
spectra in top decay are given, making use of the DELPHI
experimental data. Two sets of perturbative results ([A] and [B])
are shown, the first using $\Lambda^{(5)} = 0.226$~GeV and $m_b =
4.75$~GeV, the second one using our default parameters introduced
in \refsec{sec top PFF}. As expected, the perturbative
calculations and the corresponding non-perturbative components
differ, but the final predictions for the physical results
$\Gamma^B_N$ are to a large extent identical.

%
%%%%%%%%%%%%%%%%%%%%%%%%%%%%%%%%%%%%%%%%%%%%%%%%%%%%%%%%%%%%%%%%%%
\chapter{Charged--Current Deep Inelastic Scattering}

In this Chapter we present our result for the threshold
resummation of the coefficient function for heavy quark production
in Charged-Current Deep Inelastic Scattering (CC DIS) at low
transferred momentum $Q$ (i.e. when $Q$ is of the order of the
mass of the heavy quark) and with next-to-leading logarithmic
(NLL) accuracy. We then apply our result to the case of
charm-quark production in CC DIS processes with the masses of the
charm quark and the target taken fully into account. The
presentation in this Chapter is based on the paper
\cite{cormitDIS}.

\section{CC DIS: Notation and Overview}

DIS is a process of scattering of a lepton and a hadron (usually
nucleon) that produces another lepton and final hadronic state
$X$:
\begin{equation}
\ell_{in}(p_{\ell_{in}})+N(P) \to \ell_{out}(p_{\ell_{out}}) + X.
\label{DIS process}
\end{equation}
\begin{figure}[ht!]
\centerline{\epsfig{file=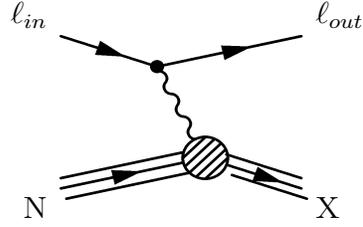,height=1.2in,width=1.9in}}
\caption{\footnotesize Deep Inelastic Scattering.} \label{fig DIS}
\end{figure}
In general, the leptons $\ell_{in}$ and $\ell_{out}$ can be of the
same type (neutral-current scattering) or be different (charged
current scattering). To lowest order in the electroweak coupling,
the DIS interaction has the structure shown in Fig.\eq{fig DIS}.
It represents the interaction of a leptonic current $j^\mu$ with a
hadronic current $J^\mu$ through the exchange of a virtual vector
boson. The vector boson is a combination of a virtual photon
$\gamma$ and $Z^0$ in the NC case and a virtual charged boson
$W^\pm$ in the CC case.

Let us introduce the following notation: as usual, we will denote
by $q$ the momentum of the transferred boson (of any type). The
momentum is space-like, and $Q^2\equiv -q^2>0$. The momentum of
the incoming nucleon is $P$ with $P^2=M^2$, $M$ being the mass of
the nucleon. We will consider two types of targets: one, where $N$
is a proton and the other an isoscalar target, where $N$ is a
``half-sum" of a proton and a neutron. In the latter case, the
partonic content of $N$ is determined by assuming isospin
invariance \eq{isospin}. We will not distinguish between the
masses of the two target types. Numerically, in our applications
we set $M=1\GeV$. A very important parameter is the Bjorken
variable:
\begin{equation}
x={{Q^2}\over{2P\cdot q}}, \label{Bj x}
\end{equation}
which is directly measured in the experiment. That variable has
the nice property that in the parton model it coincides with the
momentum fraction carried by the initial state parton, as was
discussed in \refsec{sec partmod}. Another common variable is:
\begin{equation}
\nu = {{P\cdot q}\over M}, \label{DIS nu}
\end{equation}
as well as the inelasticity $y$:
\begin{equation}
y={{P\cdot q}\over{P\cdot p_{\ell_{in}}}}. \label{DIS y}
\end{equation}

From the discussion following Eq.\eq{DIS process} it becomes clear
that the cross section for DIS process factorizes into leptonic
and hadronic parts \cite{AOT}:
\begin{equation}
d\sigma \sim L^{\mu\nu}W_{\mu\nu}, \label{LW}
\end{equation}
where:
\begin{equation}
L_{\mu\nu}={1\over Q^2}\overline{\sum}\langle \ell_{in}\vert
j_\nu^\dagger\vert \ell_{out}\rangle \langle \ell_{out}\vert
j_\mu\vert \ell_{in}\rangle, \label{L lepton}
\end{equation}
is a tensor containing the leptonic current and
\begin{equation}
W_{\mu\nu}={1\over 4\pi}\overline{\sum} (2\pi)^4\delta^4(P+q-P_X)
\langle P\vert J_\mu\vert P_X\rangle \langle P_X \vert
J_\nu^\dagger\vert P\rangle, \label{W hadron}
\end{equation}
is the tensor containing all the hadronic contributions.

It is obvious that unlike $L_{\mu\nu}$, its hadronic counterpart
$W_{\mu\nu}$ cannot be simply calculated, since the latter
contains non-perturbative information incorporated in the hadronic
states present in Eq.\eq{W hadron}. It is customary to
parameterize the hadronic tensor in the following way:
\begin{equation}
W_{\mu\nu}=- g_{\mu\nu} W_1 +{P_\mu P_\nu\over M^2} W_2
-i{\epsilon_{\mu\nu\sigma\rho}P^\sigma q^\rho \over 2M^2} W_3
+\dots\ , \label{W param}
\end{equation}
where the functions $W_{1,2,3}$ are scalar form-factors, and
``$\dots$" stands for other contributions that do not contribute
in the limit when the lepton masses
\footnote{For more details on the treatment of lepton masses see
\cite{AOT} and especially \cite{kr}.}
are neglected.  It is however more common to work in terms of the
re-scaled form-factors:
\begin{eqnarray}
F_1 &=& W_1\nonumber\\
F_2 &=& {\nu\over M} W_2\nonumber\\
F_3 &=& {\nu\over M} W_3. \label{W to F}
\end{eqnarray}
One can then write the cross-section in Eq.\eq{LW} for the case of
neutrino-nucleon scattering (which is what we need for our
applications) in the following form:
\begin{equation}
{d^2\sigma^{\nu(\bar\nu)}\over dxdy} = {G_F^2ME_\nu\over
\pi(1+Q^2/M_W^2)^2}\left[y^2xF_1 + \left(1-\left(1+{Mx\over
2E_\nu} \right)y\right)F_2 \pm y\left(1-{y\over 2}\right) x
F_3\right], \label{sigma}
\end{equation}
where $E_\nu$ is the neutrino energy in the nucleon rest frame and
$y$ is given by \eq{DIS y}. Similar results exist for the NC case.
The functions $F_{1,2,3}$ are known as structure functions. The
knowledge of these is equivalent to the knowledge of the DIS
cross-section. Since the structure functions are observables they
are free from ambiguity. In the following we turn our attention to
their evaluation.

Following \cite{AOT}, and in accordance with our discussion in
\refsec{Fac Th} of the factorization theorem, one can write the
DIS cross-section as a convolution of a parton distribution and a
partonic cross-section:
\begin{equation}
W_{\mu\nu}(q,P) = \sum_a \int{d\xi\over \xi}
f_{a/N}(\xi,\mu)w_{\mu\nu}^a(q,p,\as(\mu)). \label{W w}
\end{equation}
We prefer to work in terms of the tensor $W_{\mu\nu}$ rather than
$d\sigma$ since the former already contains all the relevant
information for the hadronic part of the reaction.

The partonic tensor $w_{\mu\nu}$ is defined in complete analogy
with $W_{\mu\nu}$ (cf. Eq.\eq{W hadron}), but with the hadron
target replaced by a partonic one. A parametrization similar to
Eq.\eq{W param} holds for $w_{\mu\nu}$ as well. One just needs to
replace in \eq{W param} the hadron momentum $P$ with the momentum
$p$ of the incoming parton $a$, and the factors containing
$W_{2,3}$ are normalized with respect to $Q^2$ instead of $M^2$.
The dependence of the partonic tensor $w_{\mu\nu}$ on the
convolution parameter $\xi$ (which is the energy fraction carried
by the parton $a$) is through the relation $p^+=\xi P^+$ between
the plus light-cone components of the partonic and hadron momenta
(see also the discussion following Eq.\eq{pdf q}). We will discuss
that point in more detail below.

As follows from our general discussion in \refsec{Fac Th} and
\refsec{HQM}, the partonic tensor $w_{\mu\nu}$ can be calculated
in perturbation theory to any order in the strong coupling. As it
was explained there, in the case of massless initial-state partons
\footnote{In our discussions we will only consider the case of
massless incoming partons. Results for the case $p^2>0$ can be
found in \cite{AOT}; see also \cite{strange mass}.}
the tensor $w_{\mu\nu}$ contains a collinear divergence. That
divergence must be subtracted (i.e. effectively absorbed into the
parton distribution functions), and one usually does that in the
$\MSbar$ scheme. This is also the subtraction scheme that we use
in our applications. The partonic tensor $w_{\mu\nu}$ depends on
the parton momentum $p$ through the partonic analogue $\hat{x}$ of
the Bjorken variable. $\hat{x}$ is defined by Eq.\eq{Bj x} but
with the momentum $P$ replaced by $p$.

In fact it is easy to see from the kinematics of the partonic
scattering process for production at NLO of quark with mass $m$
\begin{equation}
q_1(p)W^*(q)\to q_2(p_2) (+ X), \label{parton DIS}
\end{equation}
that the partonic cross-section can be evaluated in terms of the
following variable:
\begin{equation}
z = {\hat{x}\over \lambda} = {Q^2+m^2\over 2p\cdot q}~,~~ 0\leq
z\leq 1. \label{z variable}
\end{equation}
In Eq.\eq{z variable} we have also introduced the following
parameter:
\begin{equation}
\lambda = {Q^2\over Q^2+m^2} \leq 1.\label{lambda variable}
\end{equation}

Although the tensors $W_{\mu\nu}$ and $w_{\mu\nu}$ have the same
tensor structure, they are not simply proportional to each other
because of their dependence on different momenta ($P$ and $p$
respectively). Therefore the relation between the invariant
hadronic and partonic form-factors $W_i$ and $w_i$ depends on the
relation between the momenta $P$ and $p$. One can parameterize the
momentum $p$ of the parton $a$ in the following way \cite{AOT}:
\begin{equation}
p^\mu = c_P P^\mu + c_q q^\mu,  \label{parameter p}
\end{equation}
where the coefficients $c_P$ and $c_q$ read:
\begin{eqnarray}
c_P &=& {Q^2\over Q^2+M^2\eta^2}~\xi, \nonumber\\
c_q &=& -{M^2\eta \over Q^2+M^2\eta^2 }~\xi. \label{CPCq}
\end{eqnarray}
In Eq.\eq{CPCq} $\xi$ is the momentum fraction carried by the
incoming parton and $\eta$ is the Nachtmann variable
\cite{Nachtmann}, that corrects the Bjorken variable $x$ for
non-zero target mass $M$:
\begin{equation}
{1\over\eta} ={1\over 2x} +\sqrt{{1\over 4x^2}+{M^2\over
Q^2}}~,~~{\rm or:}~~~{1\over x} ={1\over \eta} -{M^2\over
Q^2}~\eta~. \label{eta variable}
\end{equation}
It is now obvious that the whole dependence of the partonic
cross-section $w_{\mu\nu}(z)$ on the momentum fraction $\xi$ (that
is also the convolution variable) enters implicitly through the
relation \eq{parameter p}. One can make this dependence explicit
by combining \eq{parameter p} with the other relations above.
After some algebra one can show that:
\begin{equation}
z = {\chi\over \xi}, \label{z conv}
\end{equation}
where, as is common in the DIS literature, we have introduced the
following parameter:
\begin{equation}
\chi = {\eta\over \lambda}. \label{chi}
\end{equation}
The implication of the relation \eq{z conv} is clear: one can
calculate the partonic tensor in perturbation theory as a function
of the variable $z$ (as defined in Eq.\eq{z variable}), and then
use the relation \eq{z conv} to show that indeed the product of
the two terms in the integrand of Eq.\eq{W w} is of the general
form \eq{convolution}. From simple kinematical considerations
(i.e. using that $s_{\rm partonic}\geq m^2$) one can also show
that $\chi\leq \xi \leq 1$. Clearly, that way the hadronic tensor
$W_{\mu\nu}$ in essence becomes a function of the variable $\chi$.
In the limit of vanishing quark and target masses, that variable
coincides with the usual Bjorken variable $x$: $\chi\to x$.

It will be, however, more convenient for us to work in terms of
the structure functions because the $w_{\mu\nu}$ depends on $\xi$
not only through the arguments of the scalar form-factors
$w_{1,2,3}$, but also through the partonic momenta defining the
tensor structure of $w_{\mu\nu}$ (cf. Eq.\eq{W param}). In the
following, in order to relate the hadron level structure functions
with their parton level counterparts, we follow the notation in
\cite{gkr} and \cite{kr}. First one defines the ``theoretical"
structure functions ${\cal F}_i$ which are related to the $F_i$'s
of Eq.\eq{W to F} and \eq{sigma} via the following relations:
\begin{eqnarray}
F_1 &=& {\cal F}_1\\
F_2 &=& {{2x}\over{\lambda\rho^2}} {\cal F}_2\\
F_3 &=& {2\over\rho} {\cal F}_3, \label{calf}
\end{eqnarray}
with:
\begin{equation}
\rho=\sqrt{1+\left( {{2Mx}\over Q} \right)^2}.
\end{equation}
The introduction of the structure functions ${\mathcal{F}}_i$ is
convenient since they can be straightforwardly expressed as a
convolution of parton distribution functions and
$\overline{\mathrm{MS}}$ coefficient functions:
\begin{eqnarray}
{\mathcal{F}}_i(x,Q^2) &=& \int_{\chi}^1 {d\xi\over \xi}\left[
C_i^q \left({\chi\over \xi },\mu^2,\mu_F^2,\lambda\right)~q_1
\left(\xi,\mu_F^2\right)\right. \nonumber\\
&+& \left. C_i^g\left({\chi\over \xi }, \mu^2,\mu_F^2,\lambda
\right)~g\left(\xi,\mu_F^2\right)\right]~,~~i=1,2,3. \label{conv}
\end{eqnarray}
In Eq.\eq{conv}, we have explicitly shown the relevant arguments
of the structure functions, although it is clear that the
dependence on $x$ comes through the variable $\chi$ in the
following way:
\begin{equation}
x={{\lambda \chi}\over{1-M^2\lambda^2\chi^2/Q^2}}.
\end{equation}
That relation follows from Eqns.\eq{eta variable} and \eq{chi}.
Bjorken $x$ is constrained to be in the range:
\begin{equation}
0< x\leq {{\lambda}\over{1-M^2\lambda^2/Q^2}}.
\end{equation}
As usual, in Eq.\eq{conv} $\mu$ and $\mu_F$ are the
renormalization and factorization scales; $C^q$ and $C^g$ are
correspondingly the coefficient functions for the quark-scattering
subprocess
\begin{equation}
q_1(p)W^*(q)\to q_2(p_2) \left( g(p_g) \right), \label{qs}
\end{equation}
and for the gluon-fusion sub-process
\begin{equation}
g(p_g)W^*(q)\to \overline{q}_1 (p)q_2(p_2). \label{gf}
\end{equation}
In Eq.\eq{conv}, by $q_1(x,\mu_F^2)$ and $g(x,\mu_F^2)$ we have
denoted the parton distribution functions of the initial-state
light quark and the gluon. For consistency, one has to use
$\overline{\mathrm{MS}}$ parton distribution functions.

\section{Behavior of the Coefficient Function in the Soft Limit}

In our applications, we would like to consider ${\cal
O}(\alpha_S)$ corrections to CC DIS. The coefficient functions can
be evaluated perturbatively and to $\mathcal{O}(\alpha_S)$ they
read:
\begin{eqnarray}
C_i^q(z,\mu^2,\lambda) &=& \delta(1-z) +{\alpha_S(\mu^2)\over
2\pi} H_i^q(z,\mu_F^2,\lambda) \label{cq}\\
C_i^g(z,\mu^2,\mu_F^2,\lambda) &=& {\alpha_S(\mu^2)\over 2\pi}
H_i^g(z,\mu_F^2,\lambda). \label{cg}
\end{eqnarray}
The explicit expressions for the functions $H^{q,g}_{1,2,3}$, with
the collinear divergences subtracted in the
$\overline{\mathrm{MS}}$ factorization scheme, have been obtained
in \cite{got} and \cite{gkr}. For our applications, we use the
results for $H^{q,g}_{1,2,3}$ as reported in \cite{kr}.

As we are inclusive with respect to the final-state heavy quark
$q_2$ the quark-scattering coefficient functions are free from
logarithms $\ln(m^2/Q^2)$. However, the functions $H_i^q$ contain
terms which behave like $\sim 1/(1-z)_+$ or $[\ln(1-z)/(1-z)]_+$
\cite{gkr} and are therefore enhanced once the quark energy
fraction approaches one. The $z\to 1$ limit corresponds to
soft-gluon emission. In Mellin moment space, such a behavior
corresponds to contributions $\sim\ln N$ or $\sim\ln N^2$ in the
limit $N\to\infty$; see the appendix for more details.

The gluon-initiated coefficient functions $H_i^g$ are not enhanced
in the limit $z\to 1$, since the splitting $g\to q\bar q$ is not
soft divergent. Instead, they contain terms $\sim \ln^k(1-z)$ of
collinear origin but such terms are suppressed in Mellin space by
an inverse power of $N$ and are therefore neglected at our NLL
level of accuracy. The gluon-fusion coefficients $H_i^g$ also
contain collinear logs $\ln(m^2/Q^2)$ originating in collinear
gluon splitting. The treatment of those terms is related to the
choice of the parton distribution. We will return to that point
later in this Chapter. To summarize, in the rest of our
discussions we will not be concerned with the soft limit of the
gluon-initiated sub-process.

We would like to make further comments on the light quark
initiated coefficient functions at large $z$. Omitting terms that
are not enhanced in the soft limit for {\it any} value of the mass
ratio $m^2/Q^2$, one has:
\begin{equation}
H^q_i(z\to 1,\mu_F^2,\lambda) = H^{soft}(z,\mu_F^2,\lambda)
\label{H-Hsoft}
\end{equation}
where:
\begin{eqnarray}
H^{soft}(z,\mu_F^2,\lambda) &=& 2C_F\left\{ 2\left({\ln(1-z) \over
1-z}\right)_+ - \left({\ln(1-\lambda z)\over 1-z}\right)_+
\right.\nonumber\\
&+& \left. {1\over 4} \left({1-z\over (1-\lambda z)^2} \right)_+ +
{1\over (1-z)_+}\left[\ln{Q^2+m^2\over \mu_F^2}-1\right]
\right\}.\label{Hsoft}
\end{eqnarray}

One immediately sees that the behavior of the coefficient function
(\ref{Hsoft}) at large $z$ strongly depends on the value of the
ratio $m/Q$. To illustrate that point, we consider the two extreme
regimes: the small-$m/Q$ limit, i.e. $\lambda\approx 1$, and the
large-$m/Q$ limit, i.e. $\lambda\ll 1$.

In the first case, setting $\lambda=1$ in (\ref{Hsoft}) one
recovers the large-$N$ limit of the $\overline{\mathrm{MS}}$
massless coefficient function reported in \cite{cmw}:
\begin{equation}
C^{soft}_N\vert_{\lambda=1} = 1 + {{\alpha_S(\mu^2)C_F}\over{\pi}}
\left\{{1\over 2}\ln^2 N + \left[\gamma_E + {3\over 4}
-\ln{Q^2\over\mu_F^2} \right]\ln N\right\}, \label{softnomass}
\end{equation}
where $\gamma_E=0.577\dots$ is the Euler constant.

In the second case, for $\lambda\to 0$ and $z\to 1$, one has
$\lambda z=\lambda$ in (\ref{Hsoft}). As a result, the term $\sim
[(1-z)/(1-\lambda z)^2]_+$ is regular in the soft limit. The
following expression holds:
\begin{equation}
C^{soft}_N\vert_{\lambda\ll 1} = 1+
{{\alpha_S(\mu^2)C_F}\over{\pi}} \left\{\ln^2 N+\left[2\gamma_E+1
-\Ln{{\mathcal{M}}^2\over\mu_F^2}\right]\ln N\right\}.
\label{dismass}
\end{equation}
For later convenience we have introduced the following scale:
\begin{equation}
{\mathcal{M}}^2= m^2\left(1+{Q^2\over m^2}\right)^2 \label{M}.
\end{equation}
In evaluating the Mellin transforms, we have made use of the
results presented in the appendix.

It is obvious that the coefficient functions $H_i^q$ are enhanced
in the limit $z\to 1$ and therefore must be resummed. Let us note,
however, the very different behavior of the NLO result in the soft
limit for small and large values of the ratio $m/Q$ (see
Eqns.\eq{softnomass} and \eq{dismass}). That suggests that the
resummed coefficient function may exhibit strong dependence on the
value of $m/Q$. In the next section we will present our results
for the resummation of the coefficient function with NLL accuracy
for low transferred momentum $Q^2\sim m^2$, and will discuss its
relation to the well known massless limit $m^2/Q^2\to 0$.

\section{Soft-gluon Resummation for the Quark-initiated
Coefficient Function}

One can perform the soft-gluon (threshold) resummation with NLL
accuracy for the coefficient function for the process \eq{qs} by
applying the general methods developed in \cite{CT} that we
presented in detail in \refsec{secSGR}. To that end, one first
needs to collect the contributions from soft-gluon emission. That
can be done using the eikonal techniques discussed also in
\refsec{secSGR}. Second, one needs to identify and -- if it is
present -- to collect the contribution from pure collinear
radiation.

Let us start with the calculation of the soft contributions in the
eikonal approximation. We consider the inclusive hard scattering
process
\begin{equation}
q_1(p)W^*(q)\to q_2(p_2) +\ {\rm soft\ gluons}, \label{qs soft}
\end{equation}
where the momenta $p,q$ and $p_2$ are hard and the momenta of the
additional gluons are all soft. In c.m. co-ordinates we have:
\begin{equation}
2p^0 \cong {Q^2+m^2\over \sqrt{(1-z)Q^2+m^2}}.\label{po}
\end{equation}
In the eikonal approximation the momentum of the outgoing hard
quark is taken \cite{CT} to be $\overline{p}=p+q$, with
$\overline{p}^2$ given by:
\begin{equation}
\overline{p}^2 \simeq (1-z)(Q^2+m^2) + m^2.\label{pbar}
\end{equation}
Next we calculate the contribution from real gluon emission to the
eikonal cross-section at order $\mathcal{O}(\alpha_S)$ and add the
virtual contribution using unitarity, as in \eq{C eik 1}. The
eikonal current \eq{J} is constructed out of the hard momenta $p$
and $\overline{p}$, and its square is given in \cite{CT}. We then
express the result as an integral over the variables $z$ and
$k^2=(p+p_g)^2(1-z)$ with $0\leq k^2\leq (2p^0)^2(1-z)^2$ and
$0\leq z\leq 1$. Finally, the exponentiated result for the
coefficient function for the process \eq{qs soft} in the soft
limit and in the $\MSbar$-subtraction scheme reads:
\begin{eqnarray}
\ln \Delta_N &=& \int_0^1 {dz {{z^{N-1}-1}\over{1-z}}}
\left\{\int_{\mu_F^2}^{ k_{\rm max}^2} {{dk^2}\over {k^2}}
A\left[\alpha_S(k^2)\right] + S\left[\alpha_S\left(k_{\rm
max}^2\right)\right]\right\}. \label{deltares any m}
\end{eqnarray}
The function $A$ is defined in \eq{function A} and the function
$S$ in \eq{function S}. To obtain the latter we used the relation
\footnote{After Eq.\eq{kmax} has been used as well; see below.}
\eq{asint}. The derivation of \eq{deltares any m} proceeds in
complete analogy with our discussions in \refsec{secSGR} and
\refsec{sec soft}, so we do not present each step in full detail
here. In the present case, however, where the kinematics of the
process we consider allows the ratio of the two mass parameters
$m^2/Q^2$ to vary widely, we encounter another complication that
we discuss next.

Ideally, one would like to be able to describe the soft limit
$z\to 1$ of the coefficient function \eq{cq} with a single
expression that is valid for any kinematically allowed value of
the ratio $m^2/Q^2$. It principle that seems possible because the
zero-mass limit $m^2\to 0$ of the coefficient function is regular
for any value of the kinematical variable $z$. This follows from
the fact that the inclusive quark-initiated coefficient function
is free from mass singularities from the final state. However, as
can be explicitly seen from the fixed order result we discussed in
the previous section, the limit $z\to 1$ of the coefficient
function is strongly affected by the value of $m^2/Q^2$. The
reason is obvious: if one wants to be able to cover all values of
$m^2/Q^2$, effectively one has to deal with a two variable
problem. As such one can consider $z$ and $\lambda z$ or, in the
conjugated Mellin space, $N$ and $(1-\lambda)N$ (recall the
definition of $\lambda$ in Eq.\eq{lambda variable}).

Let us demonstrate our point in a simple example. Consider a term
of the form $\ln[(1-\lambda)N]$. We are interested in its behavior
at large $N$ for any $\lambda\leq 1$. Although such a term is
formally divergent when $N\to\infty$ for any $\lambda<1$, it is
clear that for values of $\lambda$ sufficiently close to one, such
a term cannot be considered large; in particular it cannot be
considered on the same footing with pure $\ln(N)$ terms. Another
example is the scale $k_{\rm max}^2$. Its behavior strongly
depends on the relative magnitude of $m^2$ with respect to
$(1-z)Q^2$. For small enough mass, that scale is $\sim (1-z)$
while for sufficiently large mass it is $\sim (1-z)^2$:
\begin{eqnarray}
k_{\rm max}^2\ \cong\ (2p^0)^2(1-z)^2 &=& \left\lbrace
 \begin{array}{l l}
  Q^2(1-z) &{\rm when}~ m^2=0, \\
  {\mathcal{M}}^2(1-z)^2 &{\rm when}~ m^2\sim Q^2,\label{kmax}
 \end{array}
\right.
\end{eqnarray}
with ${\mathcal{M}}$ defined in Eq.\eq{M}. That difference in the
behavior of the scale $k_{\rm max}^2$ influences significantly the
soft limit of the coefficient function; we will see that the
coefficient of the leading logarithmic term in the perturbative
result \eq{softnomass} and \eq{dismass} is in fact proportional to
the value of the power of $(1-z)$ in $k_{\rm max}^2$. Yet another
illustration of the interplay between the large $N$ limit and the
value of the mass $m$ (relative to $Q$) is the result for the
factor $K$ \eq{Kfact}, \eq{Kfactor-zero}; the derivation of this
factor is presented in some detail in the appendix.

In order to avoid the above mentioned subtleties and to be able to
extract in a straightforward way the singular $z$ dependence of
the resummed coefficient function \eq{deltares any m}, in what
follows we will consider separately the following two cases (each
case effectively involves a single scale): the case of
sufficiently large value of the ratio $m^2/Q^2$ that we will refer
to as ``massive", and the ``massless" case $m^2/Q^2=0$. We will
present separately results for both cases. We will treat these two
results as complementing each other; we will apply the ``massive"
result for cases where $m^2/Q^2 \gsim 1/5$ while we can certainly
apply the massless result for $m^2/Q^2\lsim 1/10$. Supposedly in
the intermediate region neither of the results is correct although
the difference between the two results in that range is not large,
i.e. the correct results also should not differ much.

Finally, in order to collect all contributions one must also
account for possible collinear radiation from the final state. As
we will see in a moment, that contribution is also affected by the
value of $m^2/Q^2$.

\subsection{The case $m^2 \sim Q^2$}

If the final-state quark is massive, the virtuality of the final
state in the large-$z$ limit is given in Eq.\eq{pbar} and is never
small provided that $m/Q$ is sufficiently large. As a result,
there is no additional collinear enhancement for heavy quark
production. Therefore in the case $m^2/Q^2\sim 1$, by combining
Eqns.\eq{deltares any m} and \eq{kmax}, we get the following
result for the quark-initiated coefficient function in the limit
$z\to 1$:
\begin{eqnarray}
\ln \Delta_N &=& \int_0^1 {dz {{z^{N-1}-1}\over{1-z}}}
\left\{\int_{\mu_F^2}^{ {\mathcal{M}}^2 (1-z)^2} {{dk^2}\over
{k^2}} A\left[\alpha_S(k^2)\right]\right.\nonumber\\
&+& S\left[\alpha_S\left({{\cal M}}^2(1-z)^2\right)\right]\Bigg\}.
\label{deltares mass}
\end{eqnarray}
One can evaluate Eq.\eq{deltares mass} with NLL accuracy with the
help of Eq.\eq{NLL eval}. The result has the general form \eq{eik
sigma exp}:
\begin{equation}
\Delta_N(\alpha_S(\mu^2),\mu^2,\mu_F^2,m^2,Q^2)=\exp\left[\ln N
g^{(1)}(L)+ g^{(2)}(L,\mu^2,\mu_F^2)\right]\, ,
\label{deltaintDIS}
\end{equation}
where similarly to Eq.\eq{lambda} we have introduced:
\begin{equation}
L=b_0\alpha_S(\mu^2)\ln N \, .
\end{equation}
The functions $g^{(1)}$ and $g^{(2)}$ are given by
\begin{eqnarray}
g^{(1)}(L)&=& \frac{A^{(1)}}{2\pi b_0 L} \;
[ 2L + (1-2L) \ln (1-2L)] \;,\\
g^{(2)}(L,\mu,\mu_F) &=& \frac{A^{(1)}}{2 \pi b_0}\left[\ln
\frac{{\mathcal{M}}^2}{\mu_F^2}
- 2\gamma_E\right] \ln(1-2L)\nonumber\\
&+&\frac{A^{(1)}  b_1}{4 \pi b_0^3} \left[ 4L + 2 \ln (1-2L) +
\ln^2 (1-2L) \right]\nonumber\\
&-& \frac{1}{2\pi b_0} \left[2L + \ln (1-2L) \right]
\left(\frac{A^{(2)}}{\pi b_0} +
A^{(1)}\ln\frac{\mu^2}{\mu_{F}^2}\right)
\nonumber\\
&+& \frac{S^{(1)}}{2\pi b_0} \ln (1-2L).
\end{eqnarray}
In Eq.\eq{deltaintDIS} the term $\ln N g^{(1)}(L)$ accounts for
the resummation of leading logarithms (LL) $\alpha_S^n\ln^{n+1}N$
in the Sudakov exponent, while the function
$g^{(2)}(L,\mu^2,\mu_F^2)$ resums NLL terms $\alpha_S^n\ln^nN$.

Next, we present our final Sudakov-resummed coefficient function
with matched NLO and NLL contributions:
\begin{eqnarray} H_{N,i}^S(\mu^2,\mu_F^2,m^2,Q^2)&=&
\left[ 1+{{\alpha_S(\mu^2)C_F}\over{2\pi}}
K_i(\mu_F^2,m^2,Q^2)\right]
\nonumber\\
&\times& \exp\left[\ln N g^{(1)}(\lambda)+
g^{(2)}(\lambda,\mu,\mu_F)\right]~.
\end{eqnarray}
Furthermore, in analogy with the top decay case we considered in
the previous Chapter and following \cite{CC}, we include in our
final Sudakov-resummed coefficient function also the terms
$K(\mu_F^2,m^2,Q^2)$ of the NLO coefficient function \eq{cq} which
are constant in the limit $N\to\infty$:
\begin{eqnarray}
K_i(\mu_F^2,m^2,Q^2) &=& \left( {3\over 2}
-2\gamma_E\right)\ln\left({Q^2+m^2 \over \mu^2}\right)
+\ln(1-\lambda)\left(
2\gamma_E+{3\lambda-2\over 2\lambda} \right)\nonumber\\
&+& A_i -2{\rm Li}_2\left(-{Q^2\over m^2}\right)
+2(\gamma_E-1)(\gamma_E+2)\label{Kfact}.
\end{eqnarray}
The evaluation of the factor \eq{Kfact} is detailed in the
appendix. Following \cite{gkr} we have defined the quantities
$A_i$, which satisfy the following relations:
\begin{equation}
A_1=A_3=0,~ A_2={1\over\lambda} (1-\lambda)\ln(1-\lambda).
\end{equation}
We now match the resummed coefficient function to the exact
first-order result: we add the resummed result to the exact
coefficient function and, in order to avoid double counting, we
subtract what they have in common. The final expression for the
resummed coefficient function reads:
\begin{eqnarray}
\hat{H}_{N,i}^{\mathrm{res}}(\mu^2,\mu_F^2,m^2,Q^2)
&=&H_{N,i}^S(\mu^2,\mu_F^2,m^2,Q^2)\nonumber\\
&-& \left[H_{N,i}(\mu^2,\mu_F^2,m^2,Q^2)
\right]_{\alpha_S}\nonumber\\
&+& C_{N,i}^q(\mu^2,\mu_F^2,m^2,Q^2) \label{match}
\end{eqnarray}
where $C_{N,i}^q$ is the Mellin transform of the perturbative
result for the functions $C_i^q$ introduced in (\ref{cq}).

Before closing the discussion of the threshold resummation in the
massive case we would like to discuss a relation between the
result Eq.\eq{deltares mass} and the corresponding results for
other processes.

One can relate our result Eq.\eq{deltares mass} to the one for
heavy quark decay. An example is the top quark decay Eq.\eq{dec}
we discussed in the previous Chapter. One can easily see that the
expression Eq.\eq{deltares mass} can be obtained from the
corresponding result in top decay -- Eq.\eq{resum} -- after the
identification:
\begin{equation}
m_{top}\to m\ \ ; \ \ m_W^2\to -Q^2.\label{topDIS}
\end{equation}

The relation \eq{topDIS} is not accidental. It can be understood
in the following way: consider the coefficient functions for the
following two processes:
\begin{eqnarray}
&& {\mathcal{Q}} \to W + q, \label{top proc}\\
&& q + W^* \to {\mathcal{Q}}. \label{DIS proc}
\end{eqnarray}
By ${\mathcal{Q}}$ and $q$ we denote heavy and light quark,
respectively, and the states in Eqns.\eq{top proc}, \eq{DIS proc}
are the initial/final states in those processes
\footnote{Clearly, that is unphysical in view of the
Kinoshita-Lee-Nauenberg Theorem discussed in \refsec{IR Effects}.
However, one can think of Eqns.\eq{top proc} and \eq{DIS proc} as
the Born approximation to the corresponding physical processes.}.
Ignoring details like spins and charges, the processes \eq{top
proc} and \eq{DIS proc} are related
\footnote{Plus inversion of the sign of the mass of the $W$-boson
since the latter is space-like in \eq{DIS proc} and time-like (and
on-shell) in \eq{top proc}.}
by time-inversion which, in turn, implies \eq{topDIS}. Let us now
generalize those processes and consider more particles (denoted
generally by $X$) in the final states:
\begin{eqnarray}
&& {\mathcal{Q}} \to W + q + X, \label{top proc X}\\
&& q + W^* \to {\mathcal{Q}} +X. \label{DIS proc X}
\end{eqnarray}
Of interest to us are processes where $X$ stands for a collection
of soft gluons. In that case, up to contributions that are not
singular in the limit $N\to\infty$, the processes \eq{top proc X}
and \eq{DIS proc X} can also be related by Eq.\eq{topDIS}. To see
this, one needs to recall that the processes \eq{top proc X} and
\eq{DIS proc X} can be described with the help of the eikonal
cross-section discussed in \refsec{secSGR}. In this approximation
the ``back reaction" from soft gluon radiation on the hard
radiating particles is neglected. To derive the soft coefficient
function in the eikonal approximation, we integrate Eq.\eq{eikonal
crosssec} over all of the radiated gluons. The resulting
expression therefore depends only on the configuration of the hard
scattering particles that in turn is encoded in Eqns.\eq{top proc}
and \eq{DIS proc}.

An explicit example is the one-loop evaluation of the eikonal
amplitudes in our processes of interest: the results depend only
on the square of the eikonal current for which \eq{topDIS} is
manifest.

Of course the above considerations are complete only because there
is no additional contribution due to collinear radiation in any of
the processes \eq{top proc X} and \eq{DIS proc X}. We would also
like to note that the relation \eq{topDIS} is not restricted only
to the NLL approximation. For that reason we expect that the
function $S$ appearing in both Eq.\eq{deltares mass} and
Eq.\eq{resum} is the same even beyond the NLL level.

Let us point out that soft resummation for the coefficient
function in DIS has been first considered in \cite{CT,cmw} and
lately implemented in \cite{vogt}, but in the approximation in
which all participating quarks are massless. Soft-gluon
resummation for heavy quark production in DIS processes has been
investigated in \cite{NCfirst} and \cite{nkoy}, where the authors
have considered neutral current interactions.

\subsection{The case $m^2 \ll Q^2$}

In the limit $m^2/Q^2\to 0$ we recover the well known soft limit
of the massless coefficient function \eq{cq} derived in \cite{CT}.
As follows from Eq.\eq{pbar}, in this case the virtuality of the
final-state jet vanishes like $(1-z)Q^2$. Therefore the NLL
coefficient function receives an additional contribution due to
purely collinear radiation. As shown in \cite{CT}, one can account
for such a contribution by solving a modified evolution equation
for the jet distribution associated with the final state. As a
result, an additional term will contribute to the function
$S(\alpha_S)$ that enters the Sudakov exponent in Eq.\eq{deltares
any m}. Taking also into account the zero-mass limit in
Eq.\eq{kmax}, we get the well known massless result for the quark
initiated $\MSbar$ coefficient function \cite{cmw}:
\begin{equation}
\ln \Delta_N\vert_{m/Q\to 0} = \int_0^1 {dz
{{z^{N-1}-1}\over{1-z}}} \left\{\int_{\mu_F^2}^{Q^2 (1-z)}
{{dk^2}\over {k^2}} A\left[\alpha_S(k^2)\right] + {1\over 2}
B\left[\alpha_S\left(Q^2(1-z)\right)\right]\right\}.
\label{deltazero}
\end{equation}
The function $B(\alpha_S)$ generalizes the function $S$ from
Eq.\eq{deltares any m} and also includes the contribution from
collinear radiation from the final state. It also can be expanded
in powers of $\alpha_S$:
\begin{equation}
B(\alpha_S)=\sum_{n=1}^{\infty}\left({{\alpha_S}\over
{\pi}}\right)^n B^{(n)}
\end{equation}
and, to NLL level, one needs only the first term of the expansion:
\begin{equation}
B^{(1)}=-{3\over 2}C_F.
\end{equation}
The explicit evaluation of Eq.\eq{deltazero} to NLL level can be
found in \cite{vogt}, where functions analogous to our $g^{(1)}$
and $g^{(2)}$ are reported. Similarly to Eq.\eq{match} in the
massive case, we also match the resummed result \eq{deltazero} to
the exact coefficient function \eq{cq}, with functions $H_i^q$ now
evaluated in the limit $m/Q\to 0$, i.e. $\lambda\to 1$. The
constant factors $K_i$, analogous to the one in Eq.\eq{Kfact},
read in the massless approximation (see the appendix for details):
\begin{equation}
K_i(\mu_F^2,m^2,Q^2)\vert_{m/Q\to 0} = \left( {3\over 2}
-2\gamma_E\right)\ln\left({Q^2 \over \mu^2}\right) + \gamma_E^2 +
{3\over 2}\gamma_E - {\pi^2\over 6} - {9\over 2},
\label{Kfactor-zero}
\end{equation}
for $i=1,\ 2,\ 3$.

Finally, one can check that the ${\cal O}(\alpha_S)$ expansion of
the resummed results correctly reproduces the large-$N$ limits of
the coefficient function in both the massive and massless cases.

Before we move on to the phenomenological applications, let us
compare our results on soft-gluon resummation for heavy quark
decay Eq.\eq{resum} and for heavy-quark production in CC DIS
Eq.\eq{deltares mass} to the general results for soft-gluon
resummation discussed in \cite{cmn}. Both our results have the
general form:
\begin{equation}
\ln\Delta_N^{\rm res} = \ln\Delta_N^q  + \ln J_N^q,\label{gen res}
\end{equation}
where the factor:
\begin{eqnarray}
\ln \Delta_N^q &=& \int_0^1 {dz {{z^{N-1}-1}\over{1-z}}}
\int_{\mu_F^2}^{ {Q}^2 (1-z)^2} {{dk^2}\over {k^2}}
A\left[\alpha_S(k^2)\right]. \label{factor D}
\end{eqnarray}
is due to soft and collinear radiation from the final quark for
the case of top-decay (similarly to the $e^+e^-$ quark production
in \cite{CC}) and from the initial quark in CC DIS. The scale
$Q^2$ appearing there is the corresponding characteristic hard
scale. It equals $m_t^2(1-m_W^2/m_t^2)^2$ for the case of
top-decay and $m^2(1+Q^2/m^2)^2$ for the case of massive quark
production in CC DIS. The second term in Eq.\eq{gen res}
\begin{eqnarray}
\ln J_N^q &=& \int_0^1 {dz {{z^{N-1}-1}\over{1-z}}}
S\left[\alpha_S\left({Q}^2(1-z)^2\right)\right], \label{factor J}
\end{eqnarray}
describes soft-gluon radiation that is not collinearly enhanced,
and the scale $Q^2$ is the same as the one in \eq{factor D}. Our
result \eq{factor J} however differs in an obvious way from the
similar factors appearing in \cite{cmn,CC}: in both of our cases
(top decay and CC DIS) we have an observed massless quark and an
unobserved massive quark. For that reason in Eq.\eq{factor J} we
have the function $S$ instead of the function $1/2B$, where the
latter embodies collinear radiation from a massless quark. Also,
the integral over the (eikonal) anomalous dimension that appears
in those papers is also absent in our cases, since for us the two
limits of that integral coincide, i.e. in the massive case we
have:
\begin{eqnarray}
\ln J_N^q &=& \int_0^1 {dz {{z^{N-1}-1}\over{1-z}}}
\left\{\int_{Q^2(1-z)^2}^{ Q^2 (1-z)^a} {{dk^2}\over
{k^2}} A\left[\alpha_S(k^2)\right]\right.\nonumber\\
&+& J\left[\alpha_S\left({Q}^2(1-z)^a\right)\right]\Bigg\},
\label{General J}
\end{eqnarray}
where $a=(1,2)$ and $J=(1/2B, S)$ in the massless and massive
cases respectively. Again, $Q$ stands for the relevant hard scale
for the process under consideration.

\section{Phenomenological Results for Charm Quark Production}

In this section we apply our results for soft-gluon (threshold)
resummation to the study of charm quark production in charged
current (CC) events. Measurements of the charm quark structure
functions in the CC regime are in fact important in order to probe
the density of strange quarks and gluons in the proton.

Charged-current Deep Inelastic Scattering processes $\nu_\mu N\to
\mu X$ were recently studied at the NuTeV experiment at Fermilab,
where neutrino and antineutrino beams collide with an iron fixed
target (see, e.g., Ref.~\cite{nutev} for the updated results). In
particular, production of two oppositely-charged muons at NuTeV
\cite{nutev1} is mainly associated with a CC event with a charm
quark in the final state. The first generation of the H1 and ZEUS
experiments on electron-proton DIS at the HERA collider at DESY
(HERA I) did detect charged-current events $ep\to \nu_eX$
\cite{h1,zeus}, but it did not have sufficiently-high statistics
to reconstruct heavy quarks in the CC regime. However, such
measurements are foreseen in the current improved-luminosity run
(HERA II) and at the possible future generation of HERA III
experiments \cite{hera3}, which should be able to investigate the
proton structure functions also at low $Q^2$ values.

The structure functions are given as a convolution of
$\overline{\mathrm{MS}}$ coefficient functions and parton
distribution functions. Since the resummed coefficient function is
given in $N$-space, in principle one would like to use parton
distribution functions in Mellin space as well, in order to get
the resummed structure function in moment space and finally invert
it numerically from $N$- to $x$-space. However, all modern sets of
parton distribution functions \cite{CTEQ,mrst,GRV} are given
numerically in the form of a grid in the $(x,Q^2)$ space.

In order to overcome this problem, we follow the method proposed
in Ref.~\cite{csw} in the context of joint resummation, which
allows one to use $x$-space parton distributions, even when
performing resummed calculations in Mellin space.

This method consists of rewriting the integral of the inverse
Mellin transform of the resummed structure functions as follows:
\begin{eqnarray}
{\cal F}_i^{\mathrm{res}}(\chi,Q^2)&=& {1\over{2\pi
i}}\int_{\Gamma_N}{dN \chi^{-N} q_N(\mu_F^2)
H_{N,i}^{\mathrm{res}}(\mu^2,\mu_F^2,m^2,Q^2)}\nonumber\\
&=& {1\over{2\pi i}}\int_{\Gamma_N}{dN \chi^{-N}
[(N-1)^2q_N(\mu_F^2)]
{{H_{N,i}^{\mathrm{res}}(\mu^2,\mu_F^2,m^2,Q^2)}\over{(N-1)^2}}},
\end{eqnarray}
where $q_N(\mu_F^2)$ is the parton distribution of the
initial-state quark $q_1$, as if it were available in $N$ space,
and $\Gamma_N$ is the integration contour in the complex plane,
chosen according to the Minimal Prescription \cite{cmnt}.

One can integrate out the term with the parton distribution
function observing that the following relation holds:
\begin{equation}
{1\over{2\pi i}}\int_{\Gamma_N}{dN \xi^{-N} (N-1)^2q_N (\mu_F^2)}=
{d\over{d\xi}}\left\{\xi {d\over{d\xi}}\left[\xi
q(\xi,\mu_F^2)\right]\right\}= \Phi(\xi,\mu_F^2). \label{phi}
\end{equation}
Due to Eq.\eq{phi}, one no longer needs the $N$-space parton
distributions, but one should just differentiate the $x$-space
ones, which is numerically doable.

As one has the analytical expression for $H_{N,i}^{\mathrm{res}}$
in $N$-space \eq{match}, the following inverse Mellin transform is
straightforward:
\begin{equation}
{\cal H}(\xi,\mu_F^2)={1\over{2\pi i}}\int_{\Gamma_N}
{dN\xi^{-N}{{H_{N,i}^{\mathrm{res}}(\mu^2,\mu_F^2,m^2,Q^2)}\over{(N-1)^2}}}
\label{calh}
\end{equation}
which, thanks to the suppressing factor $1/(N-1)^2$ in the
integrand function, turns out to be smooth for $\xi\to 1$.

The resummed structure function will be finally expressed as the
following convolution
\begin{equation}
{\cal F}_i^{\mathrm{res}}(x,Q^2)=
\int_\chi^1{{d\xi}\over{\xi}}{\cal H}(\xi,\mu^2,\mu_F^2)\Phi
\left({\chi\over\xi},\mu_F^2\right).
\end{equation}

After having clarified how we are dealing with the parton
distribution functions, we are able to present results for
soft-resummed structure functions. We shall consider charm quark
production, i.e. $q_2=c$ in Eq.\eq{qs}, since processes with charm
quarks in the final state play a role for structure function
measurements at NuTeV and in HERA experiments.

As for the choice of the parton distribution set, in principle,
once data on heavy quark production CC DIS were to be available,
one should use the CC NLL coefficient function when performing the
parton distribution global fits and get NLL parton densities as
well. For the time being, we can just use one of the most-updated
NLO sets and convolute it with the fixed-order or resummed
coefficient function.

We shall present results based on the new generation of CTEQ NLO
$\overline{\mathrm{MS}}$ parton distribution functions
\cite{CTEQ}, the so-called CTEQ6M set, but similar results can be
obtained using, e.g., the MRST \cite{mrst} or the GRV \cite{GRV}
sets.

The elementary scattering processes which yield the production of
charm quarks in CC DIS are $dW^*\to c$ and $sW^*\to c$. For
$e^+(e^-)p$ scattering at HERA, our parton distribution function
$q_1(\xi,Q^2)$ in Eq.\eq{conv} will hence be:
\begin{equation}
q_1(\xi,Q^2)\vert_{\mathrm{HERA}}=|V_{cd}|^2d(\xi,Q^2)+|V_{cs}|^2s(\xi,Q^2),
\end{equation}
where $V_{cd}$ and $V_{cs}$ are the relevant
Cabibbo--Kobayashi--Maskawa matrix elements. For neutrino
scattering on an isoscalar target, that will have to be modified
in order to account for the possibility of an interaction with a
neutron as well:
\begin{equation}
q_1(\xi,Q^2)\vert_{\mathrm{NuTeV}}=|V_{cd}|^2
{{d(\xi,Q^2)+u(\xi,Q^2)}\over 2}+|V_{cs}|^2s(\xi,Q^2),
\end{equation}
where $d(\xi,Q^2)$, $u(\xi,Q^2)$ and $s(\xi,Q^2)$ are still the
proton parton distribution functions and we have applied the
isospin symmetry \eq{isospin}.

In order to be consistent with the use of the CTEQ parton
distribution functions, we shall use for the QCD scale in the
$\overline{\mathrm{MS}}$ renormalization scheme the values
$\Lambda_4=326$~MeV and $\Lambda_5=226$~MeV, for four and five
active flavors respectively. This corresponds to
$\alpha_S(m_Z)=0.118$. The charm and bottom quark masses have been
set to $m_c=1.3$~GeV and $m_b=4.5$~GeV, as done in \cite{CTEQ}.

Since mass effects are important for large values of the ratio
$m_c/Q$, we would like to have $Q$ as small as possible to be able
to apply our massive resummation. The NuTeV experiment is indeed
able to measure structure functions at small $Q^2$ values
\cite{nutev,nutev1}, as they are able to detect the final-state
muon produced in the $\nu_\mu N\to \mu X$ processes. In our
phenomenological analysis we shall consider charm production at
$Q^2=2$~GeV$^2$ and $Q^2=5$~GeV$^2$, which are values reachable at
NuTeV and such that $m_c^2/Q^2$ is relatively large.

The detection of CC events at HERA is more problematic, due to
backgrounds and to the presence of a neutrino, instead of a
charged lepton, in the final state. The current HERA II
experiments make it possible to detect heavy quarks in CC events,
but the $Q^2$ values for such events are still supposed to be much
larger than the charm quark squared mass, typically $Q^2\gsim
100$~GeV$^2$. We shall present nonetheless results for charm quark
production at HERA, but in this case we shall have to use the
massless result for the resummed $\overline{\mathrm{MS}}$
coefficient function, i.e. Eq.\eq{deltazero} and the formulas
reported in \cite{vogt}. We shall consider the typical HERA values
$Q^2=300$~GeV$^2$ and 1000 GeV$^2$.

We shall present results for the structure function
$F_2^c(x,Q^2)$, but we can already anticipate that the effect of
the resummation is approximately the same on all three structure
functions $F_1^c$, $F_2^c$ and $F_3^c$ and on the
single-differential cross section $d\sigma/dx$, obtained from
Eq.\eq{sigma}, after integrating over $y$.

For the sake of comparison with the experiments, we shall plot the
structure function $F_2^c$ in terms of Bjorken $x$, which is the
quantity which is measured. For most of our plots, we shall set
the renormalization and factorization scales equal to $Q$, i.e.
$\mu_F=\mu=Q$. Afterwards, we shall also investigate the
dependence of our results on the choice of these scales.
\begin{figure}
\centerline{\resizebox{0.65\textwidth}{!}{\includegraphics{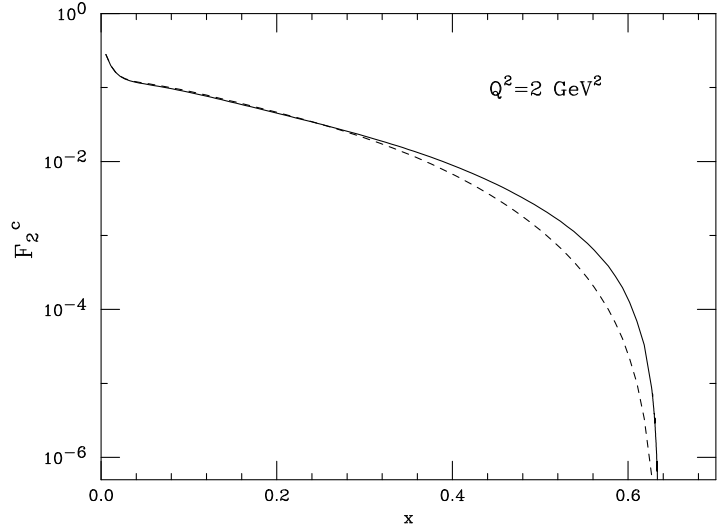}}}
\caption{\footnotesize Results for $F_2^c(x,Q^2)$ for charm quark
production in neutrino scattering at $Q^2=2$~GeV$^2$ with (solid)
and without (dashed) soft resummation in the coefficient function.
We have set $\mu_F=\mu=Q$.} \label{q2}
\end{figure}
\begin{figure}
\centerline{\resizebox{0.65\textwidth}{!}{\includegraphics{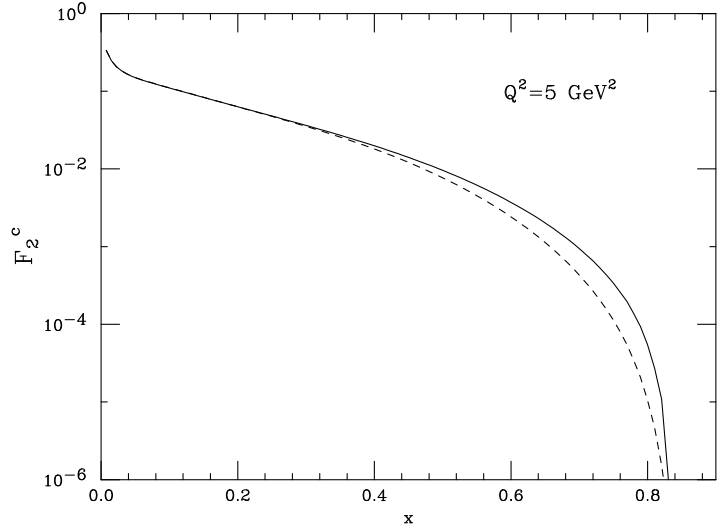}}}
\caption{\footnotesize As in Fig.~\ref{q2}, but for
$Q^2=5$~GeV$^2$.} \label{q5}
\end{figure}

In Figs.\eq{q2} and \eq{q5} we show $F_2^c(x,Q^2)$ in neutrino
scattering at $Q^2=2$~GeV$^2$ and 5 GeV$^2$ and compare the
results obtained using fixed-order and soft-resummed
$\overline{\mathrm{MS}}$ coefficient functions. We have set the
target mass to $M=1$~GeV, which is a characteristic nucleon mass.
We observe a relevant effect of the implementation of soft
resummation: the two predictions agree up to $x\simeq 0.2-0.3$;
afterwards one can see an enhancement of the structure function
due to the resummation. At very large $x$ resummed effects are
indeed remarkable: for $Q^2=2$~GeV$^2$ one has an enhancement of a
factor of 2 at $x=0.5$ and a factor of 5 at $x=0.6$. For
$Q^2=5$~GeV$^2$ one gets a factor of 2 at $x=0.7$ and 5 at
$x=0.8$.
\begin{figure}
\centerline{\resizebox{0.65\textwidth}{!}{\includegraphics{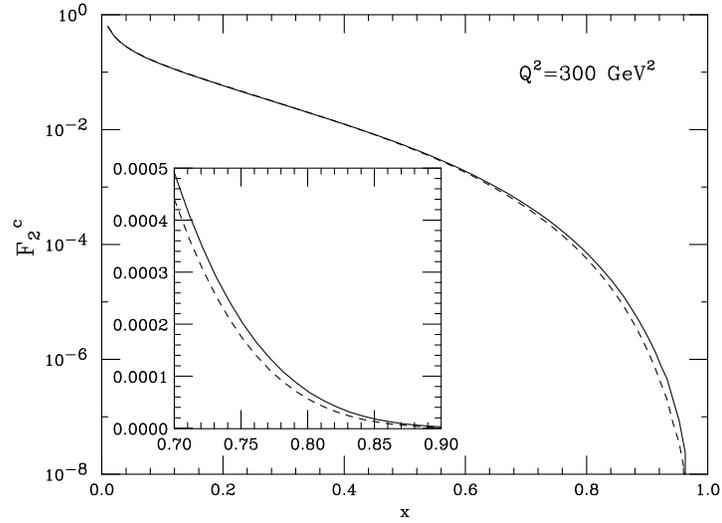}}}
\caption{\footnotesize Results for $F_2^c(x,Q^2)$ for charm quark
production in positron-proton scattering at HERA. for
$Q^2=2$~GeV$^2$ We have set $\mu_F=\mu=Q$. In the inset figure, we
show the same plots at large $x$ and on a linear scale.}
\label{q300}
\end{figure}
\begin{figure}
\centerline{\resizebox{0.65\textwidth}{!}{\includegraphics{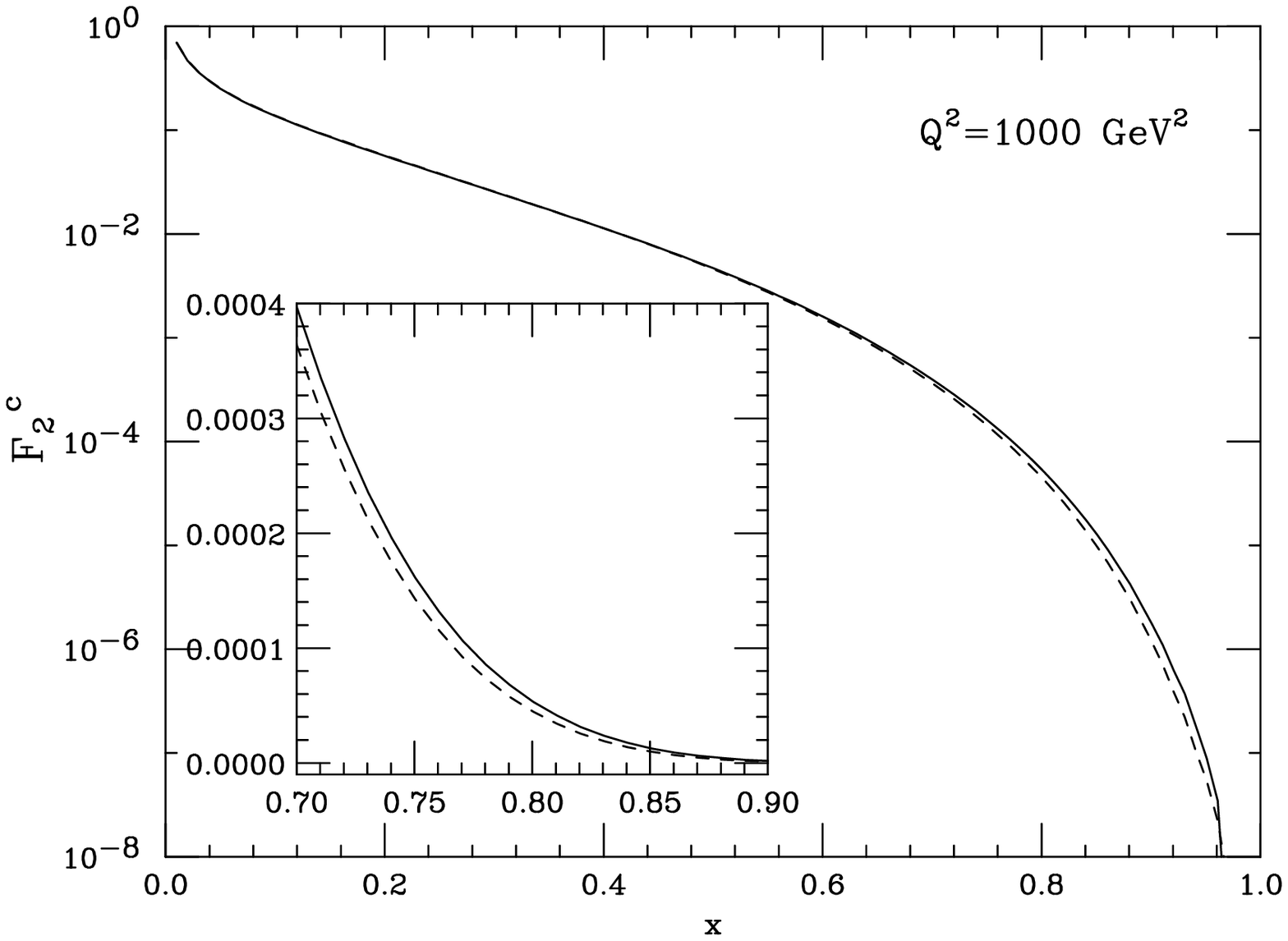}}}
\caption{\footnotesize As in Fig.~\ref{q300}, but for
$Q^2=1000$~GeV$^2$.} \label{q1000}
\end{figure}

In Figs.\eq{q300} and \eq{q1000} we present results for
$F_2^c(x,Q^2)$, but for charm production at HERA, in particular
for positron-proton scattering at $Q^2=300$ and 1000 GeV$^2$. In
this case, since $m_c/Q\ll 1$, we use the massless result
\eq{deltazero} for the resummed coefficient function. We observe
that the impact of the resummation is smaller than in the case of
low $Q^2$ values. This is a reasonable result: in fact, leading
and next-to-leading logarithms in the Sudakov exponent are
weighted by powers of $\alpha_S(\mu^2)$. As, e.g.,
$\alpha_S(2~\mathrm{GeV}^2)\simeq 3\
\alpha_S(300~\mathrm{GeV}^2)$, resummed effects are clearly more
important when $Q^2$ is small. Moreover, the larger $Q^2$ is, the
larger the values of $x$ are at which one is sensitive to Sudakov
effects.

We note in Figs.\eq{q300} and \eq{q1000} that, at about $x>0.6$,
the fixed-order and resummed predictions start to be
distinguishable. We estimate the overall impact of large-$x$
resummation on $F_2^c(x,Q^2)$ at $Q^2=300$~GeV$^2$ and
$Q^2=1000$~GeV$^2$ to be between $10\%$ and $20\%$.
\begin{figure}
\centerline{\resizebox{0.65\textwidth}{!}{\includegraphics{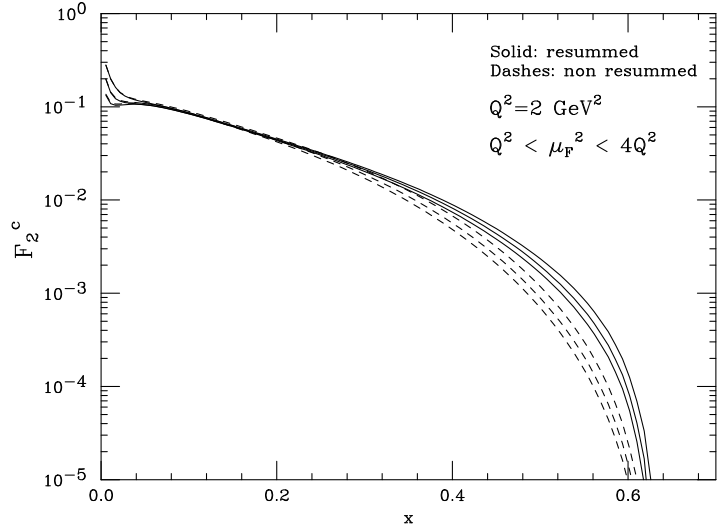}}}
\caption{\footnotesize Dependence of $F_2$ on the factorization
scale for neutrino scattering at NuTeV. Solid lines include soft
resummation in the coefficient function, dashed lines are
fixed-order predictions} \label{facnu}
\end{figure}
\begin{figure}
\centerline{\resizebox{0.65\textwidth}{!}{\includegraphics{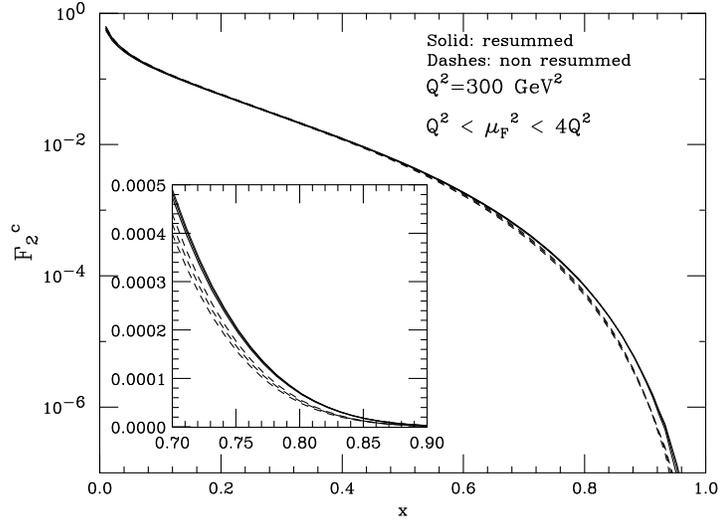}}}
\caption{\footnotesize As in Fig.~\ref{facnu}, but at HERA for
$Q^2=300$~GeV$^2$. In the inset figure, we plot the same curves at
large $x$ and on a linear scale.} \label{fach}
\end{figure}

In Figs.\eq{facnu} and \eq{fach} we investigate the dependence on
the factorization and renormalization scales. We still keep
$\mu=\mu_F$, which we allow to assume the values $Q^2$, $2Q^2$ and
$4Q^2$. We consider $Q^2$ values of 2 and 300 GeV$^2$ for the
experimental environments of NuTeV and HERA respectively. We see
that the curves which implement soft resummation in the
coefficient function show a weaker dependence on the chosen value
for the factorization/renormalization scales. In Fig.~\eq{facnu}
one can see that one still has a visible effect of the value of
such scales, but the overall dependence on $\mu_F$ and $\mu$ of
the resummed prediction is smaller than for the fixed-order. The
plots at large $Q^2$ exhibit in general a weak dependence on
$\mu_F$ and $\mu$ even at NLO, as shown in Fig.~\eq{fach}.
However, while the NLO structure function still presents a
residual dependence on the scales, the resummed predictions
obtained with three different values of $\mu_F$ are basically
indistinguishable. A smaller dependence on such scales implies a
reduction of the theoretical uncertainty of the prediction and is
therefore a remarkable effect of the implementation of soft gluon
resummation.

We finally would like to compare the impact that mass effects and
soft-gluon resummation have on the charm structure functions. To
achieve this goal, we plot in Fig.\eq{mpl} the theoretical
structure function ${\cal F}_2^c(\chi,Q^2)$ defined in
Eq.\eq{conv} as a function of the variable $\chi$ (see
Eq.\eq{chi}), for neutrino scattering at $Q^2=2$~GeV$^2$. We
compare fixed-order massive (dashed line), fixed-order massless
(dots) and massive resummed (solid) predictions. We observe that
the two fixed-order calculations yield different predictions
throughout the full $\chi$ range, which is a consequence of the
implementation of mass effects; however, at large $\chi$, where
one starts to be sensitive to the resummation, the impact of soft
resummation is competitive with that of mass contributions.
\begin{figure}
\centerline{\resizebox{0.65\textwidth}{!}{\includegraphics{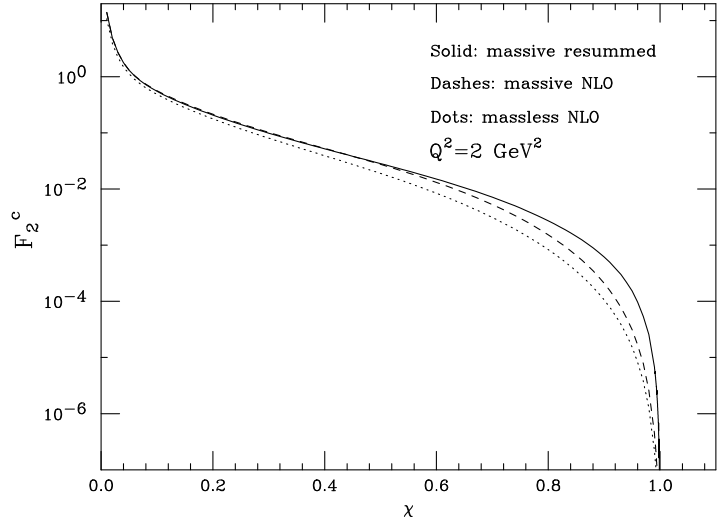}}}
\caption{\footnotesize Comparison of massive (dashes) and massless
(dots) fixed-order calculations with the resummed massive result
(solid) for $Q^2=2$~GeV$^2$ and $\mu=\mu_F=Q$. Plotted is the
theoretical structure function ${\cal F}_2^c(\chi,Q^2)$ defined in
Eq.(\ref{calf}).} \label{mpl}
\end{figure}

Before closing this section, we would like to point out that,
although we have improved our perturbative prediction by
implementing soft-gluon resummation in the coefficient function,
non-perturbative corrections are still missing. Non-perturbative
effects are important especially at small values of $Q^2$ and
large $x$. In fact, a more accurate investigation of the
very-large $x$ limit of the structure functions shows that they
exhibit an oscillatory behavior once $x$ gets closer to the
maximum value which is kinematically allowed.

%
%%%%%%%%%%%%%%%%%%%%%%%%%%%%%%%%%%%%%%%%%%%%%%%%%%%%%%%%%%%%%%%%%%
\chapter{Conclusions}

In this Thesis we have applied perturbative QCD to make precision
predictions for processes like top decay and heavy quark
production in charged-current DIS.

We have first discussed the general properties of QCD as a theory
of the strong interactions. The discussion is organized in a way
that emphasizes certain problems that arise in processes involving
heavy quarks, like the appearance of large quasi-collinear logs
and large soft-logs. Furthermore, in addition to the factorization
theorem and parton evolution, we presented a general description
of the resummation methods used to treat those problems.

Our first application was to $b$ quark fragmentation in top decay,
since this will be an important observable for the study of the
top quark once enough experimental data is available. We first
presented the results of our NLO QCD calculation for this process
and we commented on its shortcomings: the need for all-order
resummation of quasi-collinear and soft logs. Our NLO perturbative
result reproduces the known universal initial condition for the
perturbative fragmentation function as well as the NLO result for
the total top decay width. We performed the resummation of the
quasi-collinear logs $\ln(m_b^2/m_t^2)$ applying the formalism of
perturbative fragmentation function with NLL accuracy. The
resummation of those logs leads to very serious improvement of the
$b$-quark spectrum. To further improve the behavior of the
$b$-spectrum towards the region of large $b$-momentum fraction, we
have evaluated the soft-gluon resummed expression for the
top-decay coefficient function with NLL accuracy. The combined
result of the two types of all-order resummations that we have
performed is significant: the validity of our result is extended
towards higher values of the $b$-quark energy fraction $x_b$. The
resulting expression has very little dependence on the unphysical
factorization and renormalization scales. Finally, we used our
improved perturbative result for $b$-quark fragmentation to make a
prediction for the energy spectrum of certain $b$-flavored hadrons
in top decay. To that end we extracted information about the
non-perturbative transition $b$-quark to $b$-flavored hadron from
$e^+e^-$ collision data. We performed that extraction in two ways:
first we extracted the non-perturbative fragmentation function by
fitting $e^+e^-$ data with the fragmentation function of three
different functional forms having one or two fittable parameters.
We have observed that working this way, the largest uncertainty in
the prediction comes from the fits to the data. The second
approach we pursued in order to extract information for the
non-perturbative fragmentation, is to extract from data the four
lowest moments of the fragmentation function by working directly
in the space of Mellin moments. That way we do not have to extract
the non-perturbative function, i.e. no fit with its corresponding
uncertainty is involved.

The second application that we considered in this Thesis is heavy
quark production in inclusive CC DIS. We started by making the
observation that the NLO light quark initiated coefficient
function for heavy quark production in CC DIS contains terms that
become arbitrarily large towards high values of the Bjorken
variable. That fact signals the need for soft-gluon resummation
for that coefficient function. We carefully identified the sources
of such large contributions. Then we performed the soft-gluon
resummation separately in the kinematical regimes of large values
of the ratio of the mass of the heavy quark with respect to the
transferred momentum and small values of that ratio (i.e. the
massless regime). Our calculation reproduces the known results in
the limit of zero mass and of the NLO evaluation of the
coefficient function. We applied our result to charm production in
CC DIS. The most relevant application is to charm production in
neutrino-nucleon DIS at NuTeV, since that experiment can reach low
values of momentum transfer where our resummed result is most
relevant. We made a prediction for the charm-component of the
structure functions with full account for target mass effects. We
observed that at large Bjorken $x$, although the absolute value of
the structure function is small, the relative impact of the
resummation we performed is significant. The dependence on
factorization/renormalization scales is reduced as a result of the
resummation. We have checked that at low $Q$, the size of the
effect of the threshold resummation (at large $x$) is comparable
or even larger than the effect of the power corrections of the
charm mass. This is clearly another indication of the significance
of our result. It will be very interesting to compare our
prediction with the forthcoming data from NuTeV.

%
%%%%%%%%%%%%%%%%%%%%%%%%%%%%%%%%%%%%%%%%%%%%%%%%%%%%%%%%%%%%%%%%%%

%
%%%%%%%%%%%%%%%%%%%%%%%%%%%%%%%%%%%%%%%%%%%%%%%%%%%%%%%%%%%%%%%%%%
\appendix
\chapter{Some Supplementary Results}

\section{Relevant Feynman Rules}

We can specify the Feynman rules that we have used by presenting
the relevant amplitude for one of the diagrams with real gluon
emission.
\begin{eqnarray}
M &=& \overline{u}_{\sigma,c_1}(p_b) \left[-i{g_S\over
2}\gamma^\nu \lambda^a_{c_1,c_2} \right]
 G_\nu^a(p_g,\kappa)
\frac{i\left(\not{\!p}_b+\not{\!p}_g+m \right)}
{(p_b+p_g)^2-m_b^2+i\varepsilon}\nonumber\\
&\times& \left[ -i{g\over\sqrt{8}} V_{tb}
\delta_{c_2,c_3}\gamma^\mu\left(1-\gamma^5\right) \right]
\epsilon^*_\mu(p_W,\rho)u_{\lambda, c_3}(p_t).\label{Feynman}
\end{eqnarray}
The terms in square brackets correspond to the boson-quark
vertexes, $c_{1,2,3}$ are the quark color indexes, $\sigma,
\kappa, \rho$ and $\lambda$ are the polarizations (spins) of the
$b$-quark, the gluon, $W^+$-boson and the top-quark respectively.
Also, the propagator for a gluon with loop momentum $q$ is (we
choose to work in Feynman gauge):
\begin{equation}
-i\delta_{ab} {g_{\mu\nu}\over q^2+i\varepsilon}.\label{gluon
propagator}
\end{equation}
We work with cross-section that is summed over the final states
and averaged over the initial ones. The relevant results (for the
example of Eq.\eq{Feynman}) are:
\begin{equation}
{\overline{\sum_{in}}}\sum_{out} = {1\over 3}\sum_{c_1,c_2}\sum_a~
\half \sum_\lambda\sum_\sigma\sum_\rho\sum_\kappa,
\label{summation}
\end{equation}
\begin{equation}
\sum_\kappa G^a_\nu(p,\kappa)G^a_\mu(p,\kappa) = -g_{\nu\mu},
\label{gluon sum}
\end{equation}
\begin{equation}
\sum_\sigma u_{\sigma,\alpha}(p) \overline{u}_{\sigma,\beta} (p) =
(\not\! p +m)_{\alpha\beta}, \label{spinor sum}
\end{equation}
\begin{equation}
\sum_\rho \epsilon^*_\mu(p,\rho)\epsilon_\nu(p,\rho) = -g_{\nu\mu}
+{p_\mu p_\nu \over m_W^2}, \label{W sum}
\end{equation}
Our conventions for working in $D$ space-time dimensions are:
\begin{eqnarray}
\epsilon &=& {4-D\over 2}\nonumber\\
\{ \gamma_\mu,\gamma_\nu\} &=& 2g_{\mu\nu}\nonumber\\
\tr\left( \gamma_\mu\gamma_\nu\right) &=& 4g_{\mu\nu}\nonumber\\
\gamma_\mu\gamma^\mu &=& D \nonumber\\
\gamma_\mu\gamma_\nu\gamma^\mu &=& (2-D)\gamma_\nu \nonumber\\
g_\mu^\mu &=& D \nonumber\\
{d^Dk\over (2\pi)^D} &=& {\rm measure\ for\ the\ loop\ integrals}.
\end{eqnarray}
Also, only the strong coupling picks up the usual mass dimension:
$$g_S \to g_S\mu^\epsilon .$$

\section{Phase Space for Top Decay at NLO}

We consider the three particle top decay $t\to b W g$ in $D$
space-time dimensions, where each of the particles in that
reaction is on-mass-shell. The masses are $m_t, m_W,m_b$ and $0$
respectively. For simplicity, we work in the top rest frame. One
has:
\begin{eqnarray}
d\Gamma &=& N_D\; \overline{\vert
M\vert^2}(p_t;p_b,p_g,p_W)\; \delta^D(p_t-p_b-p_g-p_W)\nonumber\\
&\times & \delta(p_b^2-m_b^2) \delta(p_W^2-m_W^2)\delta(p_g^2) \;
d^Dp_bd^Dp_gd^Dp_W, \label{ps1}
\end{eqnarray}
with:
\begin{equation}
N_D\ = {1\over 2m_t}{1\over (2\pi)^{5-4\epsilon}}.\label{ND}
\end{equation}
With the help of the delta-functions, one can reduce \eq{ps1} to:
\begin{eqnarray}
d\Gamma &=& N_D\; \overline{\vert M\vert^2}\;
\delta\left(2\qk-2\pk-2\pq+m_t^2+m_b^2-m_W^2 \right)\nonumber\\
&\times& {d^{D-1}p_b\over 2E_b}{d^{D-1}p_g\over 2E_g}, \label{ps2}
\end{eqnarray}
The matrix element $\overline{\vert M\vert^2}$ is given in
Eq.\eq{M real}; we have used the relations implied by the delta
functions to simplify it.

Next, it will be convenient to work in spherical co-ordinates with
the $z$-axis along the $D-1$ dimensional vector
$\overrightarrow{p_g}$. Clearly, nothing depends on its
orientation; the only nontrivial integrations are over the
energies $E_b$ and $E_g$ and over $\cos\theta = \cos\left(
\overrightarrow{p_g},\overrightarrow{p_b}\right)$.

To carry out the trivial angular integrations, one may use the
following results: in $d$-dimensional Euclidean space (see also
\cite{Muta}),
\begin{equation}
d^dx = r^{d-1}dr d\Omega_d~,~~ d\Omega_d = \prod_{l=1}^{d-1}
\left(\sin\theta_l\right)^{d-l-1} d\theta_l.
\end{equation}
The trivial angular integrations (all for the $p_g$ case and all
but one for the $p_b$ case) amount to the factor $S_d\not\! S_d$,
where:
\begin{eqnarray}
S_d &=& \int d\Omega_d,\nonumber\\
\not\! S_d &:& d\Omega_d = \not\! S_d
\left(\sin\theta\right)^{d-2} d\theta,
\end{eqnarray}
i.e., the factor $\not\! S_d$ contains all the angular
integrations but the azimuthal one.

To evaluate those factors, it is convenient to introduce:
$$ J^{(n)} = \int_0^\pi \sin^n\theta d\theta = {\pi\over 2^n}
{\Gamma(n+1)\over \left(\Gamma(n/2+1)\right)^2}. $$
The latter satisfy the relation $J^{(n)} = (n-1)/n J^{(n-2)}$ with
$J^{(0)}=\pi$, $J^{(1)}=2$. Then one can easily show that
$S_d=2\prod_{n=0}^{d-2}J^{(n)}$ and $\not\!\! S_d
=2\prod_{n=0}^{d-3}J^{(n)}$. The relevant result for us is:
\begin{eqnarray}
S_{D-1} &=&  2{ \pi^{3/2-\epsilon}\over \Gamma(3/2-\epsilon) }
,\nonumber\\
\not\! S_{D-1} &=&  2{ \pi^{1-\epsilon}\over \Gamma(1-\epsilon)
}, \nonumber\\
S_{D-1}\not\! S_{D-1} &=& 8\pi^2\left[1 +2\epsilon\left(
1-\gamma_E-\ln(2\pi)\right)+{\mathcal{O}}(\epsilon^2)\right].
\end{eqnarray}

Finally, we can write the final result containing the two
non-trivial integrations over the two energies:
\begin{eqnarray}
d\Gamma &=&{ N_D\over 8} m_t^{2-4\epsilon}S_{D-1}\not\! S_{D-1} \;
{\overline{\vert M\vert^2} \over (Ay^2 +By+C)^\epsilon}dx_E dy,
\label{ps3}
\end{eqnarray}
where $x_E=E_b/m_t$, $y=E_g/m_t$, $A=x_E^2-b-(1-x_E)^2,\
B=2(1-x_E)(s-x_E),\ C=-(x_E-s)^2$. The squared matrix element
\eq{M real} should be evaluated using:
$$ \cos(\overrightarrow{p_b},\overrightarrow{p_g})= {m_t^2\over \vert
\overrightarrow{p_g}\vert \vert \overrightarrow{p_b}\vert}(x_Ey -
x_E-y+s).$$
The limits on the two variables are (see also the notation
introduced in \eq{notations}):
\begin{eqnarray}
&& \sqrt{b} \leq  x_E \leq s,\nonumber\\
&& y_{-} \leq  y \leq y_{+} ~;~~ y_\pm = {s-x_E\over 1-x_E \mp
\sqrt{x_E^2-b}}~.\nonumber
\end{eqnarray}
To obtain the desired differential distribution $d\Gamma/dx_b$ one
must integrate over $y$ and use $x_b=x_E/s$ as implied by \eq{xb}.
In the integration over $y$ the so-called plus prescription
appears. It is discussed in \refsec{secplus}.

\section{Spence Functions}

We present some relations between the Spence functions (or
di-logarithms) that we have used. Most are taken from \cite{dilog}
where more details about the properties of those functions can be
found.

That function is defined through the integral:
\begin{equation}
{\mathrm{Li}}_2 (z)=-\int_0^z {{{dt}\over t} \ln (1-t)},
\end{equation}
and has a cut from $z=1$ to $\infty$. Some particular values are:
\begin{equation}
\Sp{1} = {\pi^2\over 6}~,~~ \Sp{-1} = -{\pi^2\over 12}.
\end{equation}

The Spence function satisfies the following relations
\cite{dilog}:
\begin{eqnarray}
\Sp{-{1\over x}} &=& - \Sp{-x} -{1\over 2}\ln^2(x) -{\pi^2\over
6},\nonumber\\
\Sp{x} &=&  -\Sp{{-x\over 1-x}} -{1\over 2} \ln^2(1-x),\nonumber\\
{1\over 2} \Sp{x^2} &=& \Sp{x} +\Sp{-x},\nonumber\\
\Sp{{x\over 1-x}{y\over 1-y}} &=& \Sp{{x\over 1-y}} + \Sp{{y\over
1-x}} \nonumber\\
&-& \Sp{x}-\Sp{y} -\ln(1-x)\ln(1-y),\nonumber\\
\Sp{{y(1-x)\over x(1-y)}} &=& \Sp{x} -\Sp{y} +\Sp{{y\over x}} +
\Sp{{1-x\over 1-y}} \nonumber\\
&-&{\pi^2\over 6} +\ln(x)\Ln{{1-x\over 1-y}}. \nonumber
\end{eqnarray}
Combining the above relations one can derive other relations:
\begin{eqnarray}
\Sp{x} &=& -\Sp{1-x} + {\pi^2\over 6} -\ln(x)\ln(1-x), \nonumber\\
\Sp{x} &=& \Sp{{1\over 1-x}} +\ln(1-x)\Ln{{1-x\over -x}} -{1\over
2}\ln^2(1-x) -{\pi^2\over 6},\nonumber\\
\Sp{{1-x\over -x}} &=& 2\Sp{{-x\over 1-x}} - 3\Sp{1-x} -2
\ln(x)\Ln{1-x} \nonumber\\
&+& \ln^2(1-x) - {1\over 2}\ln^2(x) + {\pi^2\over 3}.
\end{eqnarray}

\section{Plus Prescription}\label{secplus}

Here we discuss some properties of the so-called plus prescription
$[f(x)]_+$ defined through:
\begin{equation}
\int_0^1[f(x)]_+ g(x) dx = \int_0^1 f(x)\left(g(x)-g(1)\right) dx
\label{plus def}
\end{equation}
where $g$ is a sufficiently regular function, while the function
$f$ typically is non-integrable in the point $x=1$. The plus
prescriptions are strictly distributions; for $x<1$ they can be
thought as the function itself i.e.:
$$ [f(x)]_+ = f(x)\ \ {\rm for}\ \  x<1. $$

Those distributions appear through the following identity:
\begin{equation}
\lim_{\epsilon \to 0} {\left(x-x_{min}\right)^{2\epsilon}\over
(1-x)^{1+2\epsilon}}~ =~ -{1\over 2\epsilon} \delta(1-x) + {1\over
(1-x)_+} + {\mathcal{O}}(\epsilon). \label{plus series}
\end{equation}
One can prove it by multiplying both sides with a test function,
use the definition \eq{plus def} and then compare the terms
multiplying the different powers of $\epsilon$. The lower limit
can be any number $x_{min}:\ 0\leq x_{min}<1$. In the massless
case, $m_b=0$, one needs the identity \eq{plus series} including
the term linear in $\epsilon$, the latter being:
$$ \left({\ln(1-x)\over1-x}\right)_+ - {\ln(x)\over 1-x} ~.  $$

Using Eq.\eq{plus def}, one can easily prove the following
important distributional identities:
\begin{eqnarray}
\left[f(x)\right]_+g(x) &=& f(x)g(x) - \left(g(1) \int_0^1
f(y)dy\right) \delta(1-x),\nonumber\\
\left[f(x)g(x)\right]_+ &=& \left[f(x)\right]_+g(x) -
\left(\int_0^1 \left[f(y)\right]_+g(y)dy\right) \delta(1-x),
\nonumber\\
\left[f(x)\right]_+g(x) &=& \left[f(x)\right]_+g(1) +
f(x)\left(g(x)-g(1)\right). \label{plus identities}
\end{eqnarray}
All of these identities are applicable to functions $f,g$ for
which the identities themselves make sense. As an example, using
the above relations, one can write the splitting function
$P_{qq}^{(0)}(x)$ in Eq.\eq{AP functions} as:
\begin{eqnarray}
P_{qq}^{(0)}(x) &=& C_F\left( {1+x^2\over 1-x}\right)_+ =~
C_F\left( {2\over (1-x)_+} - (1+x) +{3\over
2}\delta(1-x)\right).\label{Pqqsplit}
\end{eqnarray}

\section{Mellin-Space Results}\label{secMellin}

The Mellin transformation is defined through Eq.\eq{Mellin}. That
transformation has the factorization property \eq{Mellin fac}. The
inverse Mellin transform is given in Eq.\eq{Mellin inv}. The
following functions are often encountered:
\begin{eqnarray}
\int_0^1{dz~z^{N-1}{1\over{(1-z)_+}}}&=&-S_1(N-1),\nonumber\\
\int_0^1{dz~z^{N-1}{{\ln z}\over{(1-z)_+}}}&=&-\psi_1(N),\nonumber\\
\int_0^1{dz~z^{N-1}\left[{\ln(1-z)\over{1-z}}\right]_+ }&=&{1\over
2} \left[S_1^2(N-1)+S_2(N-1)\right],\nonumber\\
\int_0^1 dz~z^{N-1}\ln(1-x) &=&-{S_1(N)\over N},\label{s1s2}
\end{eqnarray}
where (we correct a typo in that definition in Eq.(B.2) in
\cite{cormit}):
\begin{equation}
\psi_k(x)={{d^{k+1}\ln\Gamma(x)}\over {dx^{k+1}}}
\end{equation}
and
\begin{eqnarray}
S_1(N)& \equiv &\psi_0(N+1)-\psi_0(1) = S_1(N-1)+{1\over N},\nonumber\\
S_2(N)& \equiv &-\psi_1(N+1)+\psi_1(1) = S_2(N-1)+{1\over
N^2},\nonumber\\
S_3(N)& \equiv &{1\over 2} \left[\psi_2(N+1)-\psi_2(1)\right].\nonumber\\
\end{eqnarray}
The large $N$ behavior of some of the often encountered singular
functions is \cite{CT}:
\begin{eqnarray}
\int_0^1{dz~z^{N-1}{1\over{(1-z)_+}}}&=& -\ln(N)-\gamma_E
+ {\mathcal{O}}(1/N),\nonumber\\
\int_0^1{dz~z^{N-1}\left[{\ln(1-z)\over{1-z}}\right]_+ }&=&
{1\over 2}\ln^2(N) + \gamma_E\ln(N) \nonumber\\
&+& {1\over 2}(\gamma_E^2+\zeta(2)) + {\mathcal{O}}(1/N).
\label{large N}
\end{eqnarray}

\section{$N$-space Result for the Coefficient Function and $ D_q^{ini}$}

The $N$-space expression for the coefficient function \eq{diff} is
\cite{cormit}:
\begin{eqnarray}
\hat\Gamma_N
(m_t,m_W,\mu,\mu_F)&=&1+{{\alpha_S(\mu)C_F}\over{2\pi}}
\left\{\ln{{m_t^2}\over{\mu_F^2}}\left[{1\over{N(N+1)}}-2S_1(N)+
{3\over 2}\right]\right.\nonumber\\
&+&[1+2\ln(1-w)]{1\over{N(N+1)}}-2\psi_1(N)
-2\psi_1(N+2)\nonumber\\
&+&{{4w(1-w)}\over{1+2w}}
\left[{{_2F_1(1,N+1,N+2,1-w)}\over{N+1}}\right.\nonumber\\
&-&\left. {{_2F_1(1,N+2,N+3,1-w)}\over{N+2}}\right]
+S_1^2(N-1)+S_1^2(N+1)\nonumber\\
&+&S_2(N-1)+S_2(N+1)+2[1-2\ln(1-w)]S_1(N)\nonumber\\
&+&2\ln w\ln(1-w)-2{{1-w}\over{1+2w}}\ln(1-w)
-{{2w}\over{1-w}}\ln w\nonumber\\
&+&4{\mathrm{Li}}_2(1-w)-6-{{2\pi^2}\over 3}\Bigg\}.\label{coeff
function N}
\end{eqnarray}
The $N$-space expression for the quark initiated initial condition
\eq{d01 gen} for the perturbative fragmentation function is
\cite{Mele Nason}:
\begin{eqnarray}
D_{q,N}^{ini}(\mu_0,m) &=& 1+{\as(\mu_0^2)C_F\over 2\pi}\left[
\Ln{{\mu_0^2\over m^2}}\left({3\over 2}+{1\over N(N+1)}
-2S_1(N)\right)\right. \nonumber\\
&-& 2S_1^2(N) +{2\over N(N+1)}S_1(N) - {2\over (N+1)^2}
\nonumber\\
&-& \left. 2S_2(N) +2 -{1\over N(N+1)}+2S_1(N)\right] \label{D ini
N}
\end{eqnarray}
One can derive those transformations with the use of the results
in \refsec{secMellin}.

\section{$N$-space Expressions for the Kernels $P_N^{(0,1)}$ }

Applying the results in \refsec{secMellin} one has the following
$N$-space expression for the leading order splitting function:
\begin{equation}
P_N^{(0)} = C_F\left({3\over 2} + {1\over N(N+1)}-2S_1(N)\right).
\label{P0 N}
\end{equation}
The result for the non-singlet NLO kernel is reported in
\cite{Mele Nason} and reads:
\begin{equation}
P_N^{(1)} = C_F^2\left[ P_F(N) + \Delta(N)\right] + {1\over 2}
C_FC_A P_G(N)+{1\over 2}C_Fn_fP_{NF}(N), \label{P1 N}
\end{equation}
where $C_F$ and $C_A$ are given in \eq{CFCA}, $n_f$ is the number
of the active flavors, and
\begin{eqnarray}
P_F(N) &=& \left(2S_1(N) -{1\over N(N+1)}\right) (
2S_2(N)-2\zeta(2)) -{2(2N+1)\over N^2(N+1)^2}S_1(N)
\nonumber\\
&+& 4S_3(N) - 3S_2(N)+3\zeta(2) + {3N^3+N^2-1\over N^3(N+1)^3}
-{23\over 8}, \nonumber\\
P_{NF}(N) &=& {20\over 9}S_1(N) - {4\over 3}S_2(N) -{1\over
6}-2{11N^2 +5N-3\over 9N^2(N+1)^2},\nonumber\\
P_G(N) &=& -P_F(N) +S_1(N)\left(-{134\over 9} -2 {2N+1\over
N^2(N+1)^2}\right) +4S_1(N)S_2(N)\nonumber\\
&+& S_2(N)\left({13\over 3} -{2\over N(N+1)}\right) + {43\over 24}
+ {151N^4 +263N^3 +97N^2 +3N +9\over 9N^3(N+1)^3}, \nonumber\\
\Delta(N) &=& 2\left(-2S_1(N) +{3\over 2} + {1\over N(N+1)}\right)
\left(2S_2(N) -{1\over 3}\pi^2 - {2N+1\over N^2(N+1)^2}\right).
\nonumber
\end{eqnarray}

\section{Derivation of the Factors $K_i$ in Eqns.\eq{Kfact} and
\eq{Kfactor-zero}}

Here we present in some detail the derivation of the factor $K_i$
given in Eqns.\eq{Kfact} and \eq{Kfactor-zero}. We introduce a
functional $\Delta$, such that to a function $f(N)$:
$$ \lim_{N\to \infty} f(N) =\sum_{i>0}\ln^i(N)+f_0 +{\mathcal{O}}(1/N),$$
it assigns the constant $f_0$, i.e.:
$$\Delta[f(N)]=f_0.$$
Then the desired coefficient $K_i$ is defined as:
$$\Delta\left[ H^q_i(z,\mu_F^2,\lambda)\right] = C_FK_i(\mu_F^2,\lambda).$$
To calculate the action of $\Delta$, one first has to extract the
singular dependence of a function using the results in \eq{plus
identities}. For example, making use of the relations \eq{large
N}, one gets for the leading order splitting function in
Eq.\eq{Pqqsplit}:
$$\Delta\left[P^{(0)}_{qq}(z)\right] = C_F\left({3\over 2} -2\gamma_E\right).$$
We next calculate explicitly one of the more complicated terms in
the coefficient function $H_i^q$ that can be found in \cite{kr}:
\begin{eqnarray}
\Delta\left[\left(f_\lambda(z)\right)_+\right]=\left\{%
\begin{array}{ll}
{1\over\lambda^2}\left(\lambda+\ln(1-\lambda)\right)
                                 ,&\hbox{$\lambda<1$;}\\
    -\gamma_E, & \hbox{$\lambda=1$,} \\
\end{array}%
\right.
\end{eqnarray}
where
$$f_\lambda(z) = {1-z\over (1-\lambda z)^2} .$$
To demonstrate that result, one needs to separately consider the
cases $\lambda<1$ and $\lambda=1$. The action of $\Delta$ in the
latter case is obvious. When $\lambda<1$, $f_\lambda(z)$ is
integrable and we extract the $z\to 1$ singular contribution using
\eq{plus identities}:
\begin{equation}
[f_\lambda(z)]_+ = f_\lambda(z) + {1\over \lambda^2}\left(\lambda
+ \ln(1-\lambda)\right)~\delta(1-z)~,~~\lambda<1. \label{flz}
\end{equation}
However, one must ensure that the first term in \eq{flz} vanishes
in the large $N$ limit. One can show that this term is indeed
$1/N$ suppressed and does not contribute. However, for large but
fixed $N$, in the limit $\lambda\to 1$:
$$f_\lambda(z) \simeq {\ln(1-\lambda)\over N}, $$
i.e. although the first term in \eq{flz} is formally $1/N$
suppressed, it is no longer small in the limit of vanishing quark
mass.

\end{document}